\documentclass{emulateapj} 
\usepackage{apjfonts}
\usepackage{psfig}
\usepackage{mathrsfs}
\slugcomment{PASP, in press}

\def\ngc#1{\hbox{NGC$\,$#1}}

\def\etal{{et~al.}}
\def\ie{{\it i.e.}}
\def\eg{{\it e.g.}}
\def\cf{{\it cf.}}
\def\rnum#1{{\uppercase\expandafter{\romannumeral#1}}}

\def\hst{{\it HST\/}}

\def\hr{${}^{\rm h}\,$}
\def\mn{${}^{\rm m}\,$}
\def\sc{${}^{\rm s}$}
\def\Sc{${}^{\rm s}$\llap{.}}

\def\Min{${}^{\prime}$\llap{.}}
\def\Sec{${}^{\prime\prime}$\llap{.}}
\def\deg{${}^\circ\,$}
\def\min{${}^{\prime}\,$}
\def\sec{${}^{\prime\prime}$}
\def\ltsim{ \,{}^<_\sim\, }
\def\gtsim{ \,{}^>_\sim\, }

\def\vmi{\hbox{\it V--I\/}}
\def\vmr{\hbox{\it V--R\/}}
\def\rmi{\hbox{\it R--I\/}}
\def\bmv{\hbox{\it B--V\/}}

\def\bmi{\hbox{\it B--I\/}}
\def\bmr{\hbox{\it B--R\/}}
\def\imj{\hbox{\it I--J\/}}

\def\vmk{\hbox{\it V--K\/}}

\def\sharp{\hbox{\it sharp\/}}

\def\today{\number\year\space \ifcase\month\or
  January\or February\or March\or April\or May\or June\or
  July\or August\or September\or October\or November\or December\fi
  \space\number\day}
\def\now{\number\year\space \ifcase\month\or
  January\or February\or March\or April\or May\or June\or
  July\or August\or September\or October\or November\or December\fi
  \space\number\day .\number\time}

\begin{document}


\slugcomment{PASP, in press}
\shorttitle{\ngc{4147}}
\shortauthors{Stetson, Catelan, \& Smith}

\title{Homogeneous Photometry V:  
The Globular Cluster \ngc{4147}\footnotemark[1]${}^,$\footnotemark[2]}
\footnotetext[1]{Based in part on archival observations made with ESO 
Telescopes at the La Silla and Paranal Observatory under programme ID 
60.A-9050(A).}
\footnotetext[2]{This publication makes use of data products from the 
Two Micron All Sky Survey, which is a joint project of the University 
of Massachusetts and the Infrared Processing and Analysis Center/California 
Institute of Technology, funded by the National Aeronautics and Space 
Administration and the National Science Foundation.}

\author{Peter B. Stetson\altaffilmark{3,}\altaffilmark{4}}
\affil{Dominion Astrophysical Observatory, Herzberg Institute of Astrophysics,\\
National Research Council, 5071 West Saanich Road, Victoria, BC V9E 2E7, Canada;
Peter.Stetson@nrc.gc.ca}

\author{M.~Catelan}
\affil{Departamento de Astronom\'ia y Astrof\'isica, Pontificia Universidad
Cat\'olica de Chile, Avenida Vicu\~na Mackenna 7860, 782-0436 Macul, Santiago,
Chile; mcatelan@astro.puc.cl}

\author{Horace~A.~Smith}
\affil{Department of Physics and Astronomy, Michigan State University,
East Lansing, Michigan~48824; smith@pa.msu.edu}

\altaffiltext{3}{Guest Investigator of the UK Astronomy Data Centre.}

\altaffiltext{4}{Guest User, Canadian Astronomy Data Centre, which is operated
by the Herzberg Institute of Astrophysics, National Research Council of Canada.}

\received{}
\revised{}
\accepted{}

\begin{abstract}

New {\it BVRI\/} broad-band photometry and astrometry are presented for the
globular cluster \ngc{4147}, based upon measurements derived from 524
ground-based CCD images mostly either donated by colleagues or retrieved from
public archives.  We have also reanalysed five exposures of the cluster obtained
with WFPC2 on the {\it Hubble Space Telescope\/} in the F439W and F555W ($B$ and
$V$) filters.  We present calibrated color-magnitude and color-color diagrams. 
Analysis of the color-magnitude diagram reveals morphogical properties generally
consistent with published metal-abundance estimates for the cluster, and an age
typical of other Galactic globular clusters of similar metallicity.  We have
also redetermined the periods and mean magnitudes for the RR Lyrae variables,
including a new c-type variable reported here for the first time.  Our data do
not show clear evidence for photometric variability in candidate V18, recently
reported by Arellano~Ferro \etal\ (2004 RMxAA 40, 209).  These observations also
support the non-variable status of candidates V5, V9, and V15. 

The union of our light-curve data with those of Newburn (1957 AJ 62, 197),
Mannino (1957 MmSAI 28, 285) and Arellano~Ferro \etal\ (op. cit.) permits the
derivation of significantly improved periods.  The mean periods and the
Bailey period-amplitude diagrams support the classification of the cluster
as Oosterhoff~I despite its predominantly blue horizontal branch.  The number
ratio of c- to ab-type RR~Lyrae stars, on the other hand, is unusually
high for an Oosterhoff~I cluster.

The calibrated results have been made available through the first author's web
site.  

\end{abstract}


\section{INTRODUCTION}

Many images of the globular cluster \ngc{4147} = C1207+188 exist in public
astronomical data archives around the world, primarily because it is included
among the six photometric calibration fields defined by the ``KPNO Video
Camera/CCD Standards Consortium'' (Christian \etal\ 1985).  Yet, surprisingly
few detailed studies of its color-magnitude diagram and variable-star properties
are available in the literature.

According to the summary of cluster properties compiled by
Harris (1996)\footnote{ http://physwww.physics.mcmaster.ca/\%7Eharris/mwgc.dat}
(Revision: February 2003), the cluster lies at the position
$\alpha\,=\,12$\hr10\mn06\Sc2, $\delta\,=\,+18$\deg32\min31\sec (J2000),
$l^{II}\,=\,$253\deg, $b^{II}=+77$\deg, thus near the boundary between the third
and fourth Galactic quadrants and not far from the north Galactic Pole.  Its
foreground reddening is accordingly quite small, $E(\bmv) \approx 0.02$. 
\ngc{4147} lies some 19$\,$kpc from the Sun and 21$\,$kpc from the Galactic
center, making it clearly a member of the halo rather than the disk
subpopulation of globular clusters.  In fact, it lies pretty much within the
transition zone between the inner and outer components of the Galactic halo
(\eg, Carney \etal\ 1990, 1991).  The cluster metallicity is listed as
[Fe/H]$\,=\,$--1.8; apparently nothing is known about its [$\alpha$/Fe] ratio.

NGC$\,$4147 is intrinsically rather small:  among the 146 Galactic globular
clusters with estimated absolute visual magnitudes in Harris's compilation, with
$M_V = -6.2$ \ngc{4147} ranks 112$^{\footnotesize th}$ in total intrinsic
luminosity, comparable to notoriously sparse clusters like Palomar~4.  However,
the cluster's half-light radius is estimated at 2.4~pc (van~den~Bergh \& Mackey
2004), which is more typical of inner-halo clusters than outer-halo ones. 
Djorgovski \& King (1986) list \ngc{4147} as {\it possibly\/} being among the
$\sim\,$20\% of Galactic globular clusters with central density cusps believed
to be the result of gravothermal core collapse, although Auri\`ere \& Lauzeral
(1991) suggest that the central brightness cusp might be explained by the
presence of a mere three bright giants in the inner 4\sec\ of the cluster.
\ngc{4147} does not seem to be a candidate post-core-collapse cluster either in
the surface photometry of Trager \etal\ (1993) or in the
velocity-dispersion data of Pryor \& Meylan (1993).  

An early photometric study of \ngc{4147} was published by Sandage \& Walker
(1955), based upon both photoelectric and photographic measurements from
Mount Wilson Observatory and Palomar Observatory.  They regarded the cluster as
an important test of the modern theory of stellar evolution, which has since
become universally accepted but was then quite new.  They also reported the
discovery of ten new variable stars, presumably of the RR Lyrae type, which were
assigned the designations V5--V14; one variable star had previously been
discovered by Davis (1917), and three more had been reported by Baade (1930). 

More recent work on \ngc{4147} has been rather sparse, perhaps in part because
of its relatively great distance and isolation in a direction well apart from
most other globular clusters.  Auri\`ere \& Lauzeral (1991) published $B,V$ CCD
photometry of a 100\sec$\times$160\sec\ region around the center of the cluster,
based upon a night of observations in 1\Sec0--1\Sec2 seeing conditions at the
Observatoire du Pic du Midi.  They found a fairly steep red giant branch (RGB)
typical of metal-poor globular clusters, along with a prominent blue horizontal
branch (HB) not unlike that of NGC~288 (except for the clear presence
of a more substantial RR Lyrae component).  However, their photometry did not go
deep enough to reach the main-sequence turnoff of the cluster.  Since then, Wang
et al. (2000) have also presented color-magnitude diagrams for this cluster,
although they barely reached the HB level.  Piotto \etal\ (2002)
reported on HST photometry for the innermost regions of the cluster, which
although revealing the main-sequence turnoff point for the first time, appeared
to extend $\ltsim 1$~mag below it. This, along with the fact that the HB of the
cluster was not particularly well defined in the HST study, especially at its
``horizontal'' level (\ie, around the RR Lyrae region), have led to few attempts
to utilize these data for reliable age dating of the cluster.

The first detailed study of the variable-star population in \ngc{4147} was
carried out by Newburn (1957), who added three more entries to the catalog of
variable-star candidates in the cluster.  In this paper, Newburn also retracted
his earlier claim that candidate V9 was a variable, which he had made in a
private communication to A.~R.~Sandage.  Another study of six of the best
cluster RR~Lyrae candidates was carried out at about the same time by Mannino
(1957).  As of May 2005 Christine Clement's web site\footnote{
http://www.astro.utoronto.ca/$\sim$cclement/read.html} (Clement \etal\ 2001) still
lists those 17 objects as the only known or suspected photometric variables in
the field of \ngc{4147}; among them, the periods listed for seven are flagged as
dubious and V9 is indicated as ``probably not var.'' 

On the basis of these data, \ngc{4147}\ has stood out from other Galactic
globular clusters due to its reportedly unusual RR Lyrae star properties. In
particular, Castellani \& Quarta (1987) classified this cluster as belonging to
Oosterhoff (1939) Type I (Oo~I) despite the fact that it possesses a low metal
abundance and a blue HB.  Such a classification seems
inconsistent with the scenario---which gained significant impetus in the
late-1980's/early-1990's with the work by Lee \etal\ (1990), and later
on by Clement \& Shelton (1999)---whereby RR~Lyrae stars in Oo~I globular clusters
(predominantly red or intermediate HBs) are relatively unevolved objects,
whereas those in Oo~II globulars (predominantly blue HBs) are evolved from a
position on the blue zero-age HB (ZAHB). However, when this potential conflict
with the evolutionary interpretation was identified,  the possibility was soon
raised that at least some of the RR Lyrae periods reported in the literature
were in fact incorrect.  For this reason, Clement (2000) stressed the need for
additional work on the variable stars in this cluster. 

Arellano~Ferro \etal\ (2004; hereinafter AF04) have recently provided new
periods and light curves in the $V$ and $R$ photometric bandpasses for the 17
previously known \ngc{4147}\ variable candidates, and for an eighteenth
candidate variable that they identified in the cluster field.  The work is
derived from 23 nights of observations from three observatories sampling a total
range of 171.6 days in 2003.  The number of individual magnitudes reported by
AF04 for any given star ranged from a minimum of 68 (V18) to a maximum of 551
(V9).  This work confirmed that some of the older periods for the variable stars
were incorrect, but nevertheless the cluster's Oo~I classification was also
confirmed.  AF04 did not provide a color-magnitude diagram for \ngc{4147}.

In the remainder of this paper we present the results of our analysis of 524
ground-based images of \ngc{4147} as well as five exposures obtained with the
WFPC2 camera on the {\it Hubble Space Telescope\/}.  These data span the period
1983--2003, and are independent of those employed by AF04.
Section 2 below describes the nature and the provenance of our CCD images. 
Section 3 presents some details of the methodology by which we determined
fundamental positions and magnitudes for stars in the cluster field.  Section 4
discusses new color-magnitude and color-color diagrams for the cluster. In
Section 5 we present new period determinations and inferred physical properties
for the known variables in the cluster field based upon a combination of our own
data with the independent data of Newburn, Mannino,  and AF04.  Here we also
include results for a new c-type RR Lyrae star that is identified here for
the first time.  Finally, we present a brief discussion of the significance of the
present results.

\begin{deluxetable*}{llllrrrrrr}
\footnotesize
\tablecaption{Photometric data sets for \ngc{4147}}
\tablecolumns{10}
\tablenum{1}
\tablehead{
\colhead{Observing run} &\colhead{Telescope} &\colhead{Detector} 
&\colhead{Year/Month} &\colhead{Clr} &\colhead{Cld} &\colhead{$B$} 
&\colhead{$V$} &\colhead{$R$} &\colhead{$I$}
}
\startdata
nbs          & CTIO 4m        & RCA1         & 1983 Jan      &  4 & -- &  5 &  6 &  3 &  2 \\
jvw          & INT 2.5m       & RCA          & 1986 Mar/Apr  & -- &  1 &  4 &  4 &  4 &  3 \\
igs          & INT 2.5m       & GEC4         & 1989 Mar/Apr  & -- &  1 &  3 &  6 &  7 &  6 \\
c90ic17      & CFHT 3.6m      & RCA4         & 1990 May      & -- &  1 & -- &  3 & -- &  3 \\
c90ic02      & CFHT 3.6m      & RCA4         & 1990 May      & -- &  1 &  2 &  2 & -- &  2 \\
rdj          & JKT 1.0m       & GEC3         & 1991 Apr      & -- &  1 &  5 &  5 &  5 &  4 \\
rld          & JKT 1.0m       & GEC6         & 1991 May      & -- &  1 &  3 & -- &  6 & -- \\
rdj2         & JKT 1.0m       & GEC3         & 1992 Mar      & -- &  1 &  5 &  3 &  5 &  5 \\
psb          & INT 2.5m       & EEV5         & 1992 Mar      &  3 &  1 &  5 &  5 &  5 & -- \\
c92ic34      & CFHT 3.6m      & Lick2        & 1992 Mar      & -- &  1 & -- & -- &  1 &  1 \\
dhpj         & INT 2.5m       & GEC6         & 1992 Apr      & -- &  1 & -- & -- & 57 & 16 \\
rjt          & INT 2.5m       & EEV5         & 1992 Apr/May  & -- &  1 &  4 &  4 &  5 &  4 \\
saic         & CFHT 3.6m      & HRCam/saic1  & 1992 May/Jun  & -- &  1 & -- &  4 &  8 & -- \\
c92ic05      & CFHT 3.6m      & Lick2        & 1992 Jun      & -- &  1 & -- &  1 &  1 &  2 \\
h92iic22     & CFHT 3.6m      & HRCam/saic1  & 1992 Jul      & -- &  1 & -- &  2 &  4 &  1 \\
bolte        & KPNO 2.1m      & t1ka         & 1994 Apr      &  1 & -- & -- &  2 & -- &  2 \\
pwm          & JKT 1.0m       & EEV7         & 1994 Apr      & -- &  1 & -- &  1 & -- &  1 \\
siv          & INT 2.5m       & EEV5         & 1994 Apr/May  &  1 &  1 & -- & 18 & -- & 17 \\
itp          & JKT 1.0m       & EEV7         & 1994 May      & -- &  1 &  4 &  3 &  4 &  2 \\
smh2         & INT 2.5m       & TEK3         & 1995 Jan      & -- &  1 &  1 &  2 &  1 &  2 \\
mxt          & INT 2.5m       & TEK3         & 1995 Apr      &  4 &  1 & 20 & 17 & 17 & 16 \\
rr           & INT 2.5m       & TEK1         & 1996 May      & -- &  1 & -- & 43 & -- & -- \\
bond9        & KPNO 0.9m      & t2ka         & 1997 May      &  1 &  1 &  3 &  3 & -- &  3 \\
n4147        & {\it hst\/}    & WFPC2        & 1999 Jun      &  1 &  1 &  3 &  2 & -- & -- \\
jun00        & CFHT 3.6m      & CFH12k       & 2000 Jun      &  1 & -- & -- &  2 & -- & -- \\
bono         & MPG-ESO 2.2m   & WFI          & 2002 Feb      &  1 & -- &  5 & -- & -- & -- \\
hannah       & JKT 1.0m       & SIT2         & 2002 Mar      & -- &  1 &  3 &  3 &  3 & -- \\
arg02        & JKT 1.0m       & SITe2        & 2002 May      &  1 & -- &  2 &  2 &  2 & -- \\
vimos1       & VLT Melipal 8m & VIMOS        & 2003 Apr      &  1 & -- &  1 &  1 &  1 &  1 \\
alf03        & JKT 1.0m       & SITe1        & 2003 May      &  1 & -- &  1 &  1 &  1 &  1 \\
\enddata
\end{deluxetable*}

\section{DATA}

Table~1 details the CCD images that are available in the $B$, $V$, $R$,
and $I$ filters for the globular cluster \ngc{4147}.  All observing runs but one
have been taken from public archives or have been donated to the cause from
private collections.  The first four columns of the table list, respectively,
the arbitrary name that we have assigned to a given observing run, the
telescope, the camera system, and the approximate dates of the observations. 
The columns labeled ``Clr'' and ``Cld'' represent the number of {\it data sets
per CCD\/} that were analysed under the assumption that they had been obtained
under, respectively, photometric and non-photometric conditions.  (See Stetson
2005 for the sense in which we use the terms ``observing run,'' ``data set,''
``photometric,'' and ``non-photometric.'') Each number in the column ``Clr''
represents the number of photometric nights during the observing run upon which
images of \ngc{4147} were obtained.  A ``1'' under the heading ``Cld'' indicates
that one or more nights of non-photometric data were bundled into a single
photometric reduction.  Finally, the columns labeled ``$B$,'' ``$V$,'' ``$R$,''
and ``$I$'' represent the number of individual exposures that were obtained in
those filters during each observing run.  Note that the VIMOS, WFI, and CFH12k
cameras contain, respectively, four, eight, and twelve CCDs.  A single
exposure with one of these cameras therefore contributes that number of
independent CCD images to the total set of data.  However, since in each case
the separate images are non-overlapping, the number of exposures represents the
maximum number of CCD images in which any given star can appear.  The
CFHT+CFH12k run labeled ``jun00'' is the one set of observations that we
obtained ourselves.  

The WFPC2 images of \ngc{4147} were obtained with the {\it Hubble Space
Telescope\/} in the course of observing program 7470, ``A Snapshot Survey of
Galactic Globular Clusters,'' I.~R.~King, PI.  We obtained the on-the-fly
recalibrated copies of these images through the services of the Canadian
Astronomy Data Centre.  

In addition to the 452 ground-based exposures and five WFPC2 exposures listed in
Table~1, which produced a total of 534 distinct CCD images, we also had seven
ground-based exposures (ten images) in one or another flavor of the $U$ bandpass.
These were included in the ALLFRAME reductions in order to extract whatever
information they might contribute to the completeness of the star list and the
precision of the astrometry, but no attempt was made to calibrate or employ them
photometrically.  Among the 524 ground-based images of \ngc{4147}, the best
seeing achieved was 0\Sec37, 25$^{th}$ percentile 0\Sec97, median 1\Sec24,
75$^{th}$ percentile 1\Sec6, and worst 5\sec.

For each observing run obtained from the archives, we requested {\it all\/}
the CCD images obtained during the course of the run, including such bias frames
and flat-field frames as were available, as well as any images of other
astronomical targets in case they might be---or might someday become---secondary
standard fields (see Stetson 2000).  Mean bias, flat-field, and, when
necessary, fringe frames were constructed in accordance with procedures that
have by now become well established, and the images of science targets were
corrected for these instrumental signatures in the usual way.  Those data sets
that were contributed from private collections (``nbs,'' ``bolte,'' ``bond9,''
``arg02,'' and ``alf03'') had been corrected for bias and flat-field structure
before the images were passed on to us and, again, we tried to make sure that we
had copies of {\it all\/} images of science targets from those observing runs.

The total body of imagery for \ngc{4147} spans an area of roughly 71\min\
east-west versus 39\min\ north-south, centered on $\alpha\,=\,12$\hr10\mn12\Sc4,
$\delta\,=\,+18$\deg36\min38\sec\ (J2000).  However, the outer limits of this
field are defined by the CFH12k data, which exist only for the $V$ filter.
Those stars for which $B$, $V$, {\it and\/} $I$ photometry (at least) are
available are contained within the bounds 12\hr08\mn34\Sc4$\ltsim \alpha 
\ltsim$12\hr10\mn55\Sc8, +18\deg20\min47\sec$\ltsim \delta
\ltsim$+18\deg47\min07\sec (J2000).  A congeries of images representing this
34\min$\times$26\min\ area of sky is presented here as Fig.~1.

\begin{figure*}[t]
  \figurenum{1}
  \plotone{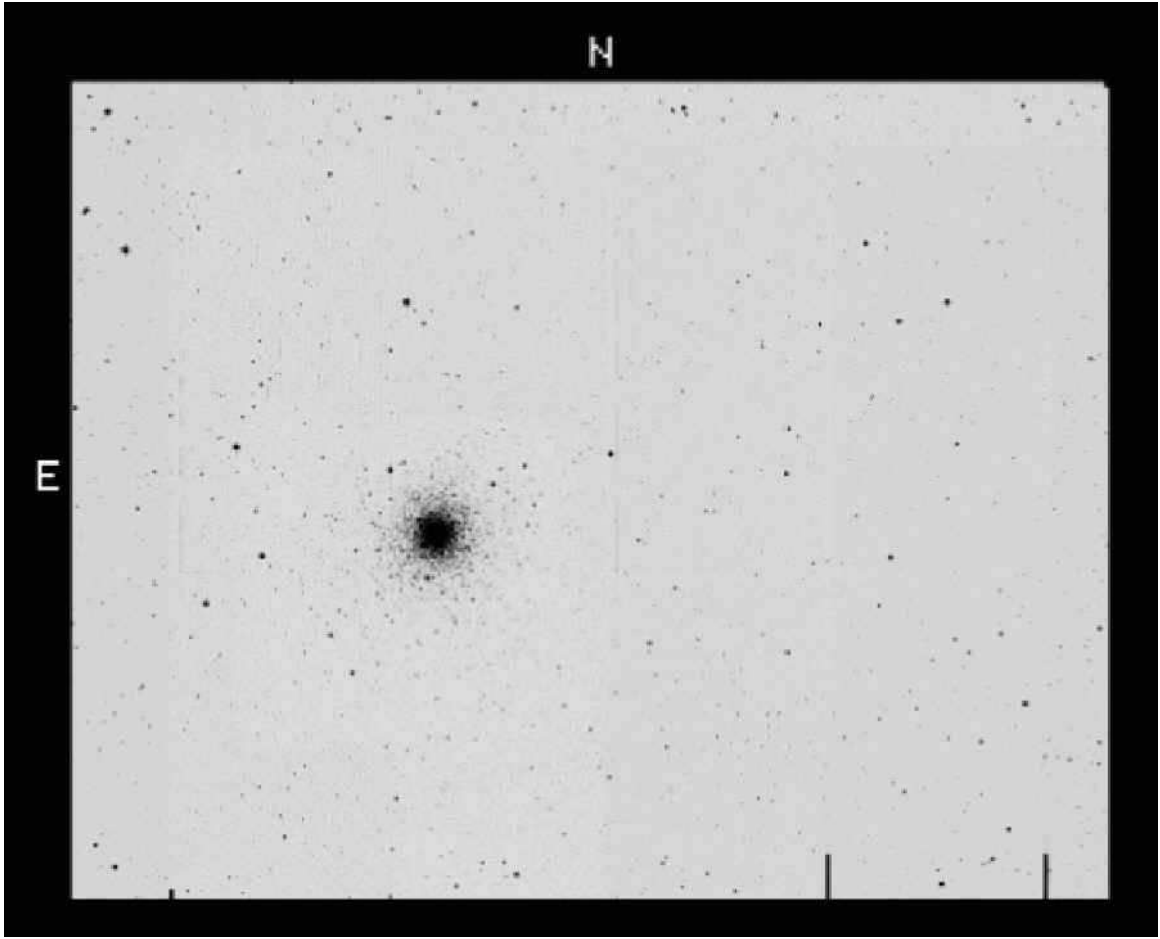}
\caption{A digital stack of our CCD images representing the
34\min$\times$26\min\ area of sky surrounding \ngc{4147} where we were able to
derive calibrated photometry in at least the $B$, $V$, and $I$ photometric
bandpasses.  The axes of the image are accurately aligned with the north-south
and east-west directions.  North and east are indicated.}
      \label{Fig01}
\end{figure*}

\section{ANALYSIS}

The ground-based and \hst\ observations of \ngc{4147} were analysed separately,
as experimental reductions of ground-based and \hst\ data together have so far
proven disappointing.  One would think that we could use the \hst\ imagery to 
establish an ironclad star list and set of centroid positions, which could
then be imposed as prior conditions on the analysis of ground-based images in a
solution for photometric parameters alone within the area of overlap.  We have
not yet stumbled upon an effective way to make this work.  It is often the case
that a bright star is seen to have numerous fainter companions in the \hst\ 
images.  However, when this information is used in the analysis of the
ground-based data, only the stars' relative {\it positions\/} are carried along
from the \hst\ to the ground-based reductions, not their relative {\it
magnitudes\/}; otherwise, it would be impossible to fairly treat stars of
differing colors, or stars whose brightnesses vary with time.  It seems that
under these circumstances---when a given blob of light in the ground-based data
encompasses several distinct but unequal detections in the \hst\ imagery---the
software at present has too much freedom to distribute the ground-based photons
among the various \hst\ detections as it attempts to optimally model the
detailed distribution of light in the observed blob.  The net result is a
compact clump of objects that individually appear to brighten and dim
spasmodically in response to the varying distribution of noise in the object's
profile as recorded in the different ground-based images.  In most cases it
seems preferable, at least with the current generation of software, to ignore
the fact that brighter stars may have fainter companions visible in the \hst\
images.  Instead, we reduce the ground-based data independently of the \hst\
star lists, and cross-identify the stars {\it ex post facto\/}.  It will
generally be obvious from the WFPC2 data which stars sufficiently dominate their
companions that such a comparison will be meaningful, and which stars will be so
badly blended in the ground-based data that no reasonable comparison is
possible.  

\subsection{Astrometry}

Employing the services of the Canadian Astronomy Data Centre, we extracted from
the U.~S.~Naval Observatory ``USNO-A2.0'' Guide-Star Catalog (Monet \etal\ 1998)
all 7,941 sources within a square box 120\Min0 on a side centered on coordinates
$\alpha\,=\,12$\hr10\mn13\Sc79, $\delta\,=\,+18$\deg31\min22\Sec4 (J2000). 
These coordinates represent the origin of the differential $(X,Y)$ coordinate
system that we will henceforth employ for identifying detected objects.  Also 
through the Data Centre, we extracted images 80\min\ on a side, centered on the
same coordinates, from the STScI Digitized Sky Survey~1 ``O'' plate, and the
Digitized Sky Survey~2 ``B,'' ``R,'' and ``I'' plates.  These were analyzed
with a modernized version of the Stetson (1979) software.  The program DAOMASTER
(Stetson 1993) was then used  to transform the data from these star lists and
from our own ALLFRAME (Stetson 1994) analysis of the ground-based CCD images
to a common reference system based upon the USNO2 coordinates.  Ten-parameter
cubic fits in $X$ and $Y$ were used to effect the transformations.  The $(X,Y)$
coordinates in our composite star list should now be accurately aligned with the
cardinal directions, with $X$ increasing east and $Y$ increasing north. 
Positions are expressed in units of arcseconds with the origin of the coordinate
system at the celestial coordinates given above.  Measurements from the WFPC2
images were subsequently transformed to the same system by comparing the
coordinates of detected objects to positions derived from the ground-based CCD
images, again employing a 10-parameter cubic transformation for each of the two
spatial dimensions.  

The precision at the present epoch of the USNO2 positions is generally in the
range 0\Sec2--0\Sec4 per detection, and appears to be dominated by the proper
motions of nearby stars and the difficulty of unambiguously centroiding extended
objects like stellar blends and galaxies.  Our positional system as a whole
should therefore be the same as the USNO system with an accuracy $\sim$
0\Sec4$/\sqrt{7900} \sim 0$\Sec01, but we can provide no independent estimate
of the absolute accuracy of the USNO system itself.  The precision of the
position of any one star relative to the others in our catalog is probably never
better than a few $\times\,0$\Sec01, and will be much worse than this for faint
or crowded stars, and for non-stellar detections.

\subsection{Ground-based Photometry}

Profile-fitting and concentric-aperture photometry were obtained for
all images of science targets with the DAOPHOT-ALLSTAR-ALLFRAME-DAOGROW-$\ldots$
software packages following commonly understood reduction procedures (\eg,
Stetson 1987, 1990, 1994).  The corpus of ground-based instrumental magnitudes
measured in the various {\it bvri\/} systems was then transformed to Stetson's
current best approximation of Landolt's (1992) {\it BVRI\/} system via the
CCDSTD-CCDAVE-NEWTRIAL (Stetson 1993) software packages.  The way in which the
instrumental magnitudes from the various observing runs are transformed to a
common standard system duplicating that of Landolt as closely as possible has
been discussed in some detail recently (Stetson 2005).  

In brief, for each {\it photometric\/} night, observations of large numbers of
primary and secondary standards are used to determine the photometric
zero points, extinction coefficients, and polynomial-approximation
color-transformation coefficients relating instrumental to standard magnitudes
for that specific telescope/filter/detector combination.  These quantitative
transforming relationships are used to convert the instrumental magnitudes for
hand-selected stars in the \ngc{4147}\ field to the standard system.  The
totality of calibrated data for each of these stars from all photometric nights
is robustly averaged to define a local sequence of secondary standards in the
\ngc{4147}\ field itself.  

For a {\it non-photometric\/} data set, the data for all celestial fields that
contain at least two standard stars spanning some range of color are used to
determine the corrections for bandpass-mismatch as functions of the standard
color; during this analysis the photometric zero point of each individual CCD
image is allowed to float.  In calibrating an individual image of \ngc{4147}\
from a non-photometric data set, the color transformation derived from all
standard-star observations included in that data set is imposed, but the
photometric zero point of each image is determined only from the {\it local\/}
secondary standards contained within that image itself.

In the field of \ngc{4147}, the first author has identified 712 stars that 
appear to be both bright and isolated enough to be potentially useful as
photometric standard stars.  Among these, 412 stars have been sufficiently
well observed that they are listed on his web site (as of January 2005) as
potential secondary standards for the calibration of other science targets:
these are defined as those stars having at least five observations on
photometric nights {\it and\/} standard errors\footnote{For a single measurement
of a star in a single image, the standard error of the measurement is based
upon a compromise between two considerations:  first, the readout noise and
Poisson noise of star plus sky in each pixel, and second the observed scatter of
the residuals of the individual pixels from the best-fitting point-spread
function.  When the number of pixels contained within the star image is small,
the former consideration dominates, when the stellar image is spread over many
pixels, the latter dominates.  The standard error of the average of many
measurements of a particular star in a given filter is similarly based on a
compromise of two considerations: first, the above-described standard errors of
the individual measurements; second, the actual observation-to-observation
agreement of the measured magnitudes, with more weight being accorded to the
latter as the number of independent observations grows.} 
of the mean calibrated magnitude
$\leq 0.02\,$mag in at least two of the {\it BVRI\/} filters, and no indication
of intrinsic variability greater than 0.05$\,$mag, root-mean-square, when data
from all filters are considered together.  For our present purposes, we will use
the local reference stars solely to redetermine the photometric zero points of
the individual CCD images to place them all on as internally consistent a system
as possible.  All other color terms, extinction coefficients, and spatially
dependent corrections will be imposed as known quantities from previous
calibration stages.  Since the zero point of any given CCD image is now the only
unknown quantity, for our present purposes we have slightly relaxed the
aforementioned criteria and adopted a local reference sequence consisting of
531 stars that were observed on at least three photometric occasions and have
standard errors of the mean magnitude $\leq 0.04\,$mag in at least two of the
four principal filters, and have no evidence of variability in excess of
0.05$\,$mag, root-mean-square.  The minimum, median, and maximum number of these
local reference stars in any individual CCD image were, respectively, 1, 159,
and 479.

Our experience is that the aggregate of CCD data for any given astronomical field
obtained on any given photometric night can typically be calibrated to the standard
magnitude system with an external accuracy $\sim\,$0.02$\,$mag, root-mean-square
(this is a {\it very\/} crude generalization).  Some vague sense of the likely
external accuracy of our photometry can therefore be obtained from the data in
Table~1.  For instance, at least some data in the $V$ band were obtained for the
\ngc{4147}\ field on 18 photometric nights, so the absolute $V$ magnitude scale
has probably been established with an accuracy no better than
$\sim0.02/\sqrt{18} = 0.005\,$mag.  A pessimist might say that the 0.02$\,$mag
figure actually applies to the accuracy possible from a given {\it run\/},
rather than {\it night\/}.  In this case, the external accuracy would be
guesstimated at $0.02/\sqrt{8} = 0.007\,$mag.  In either case, the absolute
accuracy in $B$, $R$, and $I$ would be even poorer than these estimates, since
these filters were less commonly used than $V$.  

\subsection{WFPC2 Data}

The instrumental magnitudes for stars in the WFPC2 observations of \ngc{4147}
were extracted from the images in pretty much the same way as was done for the
ground-based data (see, \eg, Stetson \etal\ 1998 for more details).  That done,
the WFPC2 star lists were searched for cases of individual bright stars that,
while not being saturated, nevertheless sufficiently dominated their fainter
neighbors that one might be able to relate their WFPC2 instrumental magnitudes
to their magnitudes on the fundamental Landolt (1992) photometric system via
the ground-based observations of the cluster.  Table~2 and
Table~3, respectively, list the positions and photometric results for
the 91 stars hand-selected for this purpose.

\begin{figure}[t]
  \figurenum{2}
  \plotone{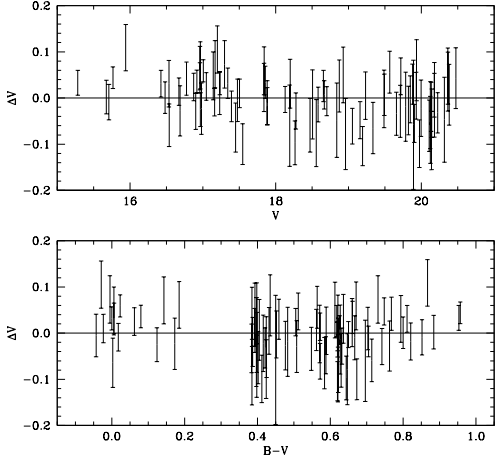}
\caption{Photometric residuals, in the sense (ground-based
{\it minus\/} WFPC2), between our calibrated ground-based $V$-band photometry
and our WFPC2 $V$-band photometry calibrated with the use of the
color-transformation coefficients of Holtzman \etal\ \ Each star in the
transfer sequence is represented by a $\pm1\sigma$ error bar.}
      \label{Fig02}
\end{figure}

\begin{figure}[t]
  \figurenum{3}
  \plotone{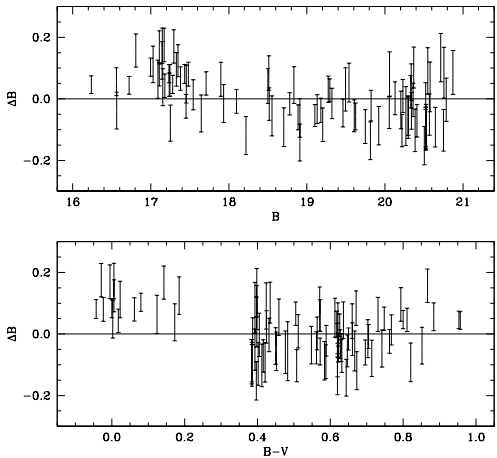}
\caption{As in Fig.~2, photometric residuals, in the sense
(ground-based {\it minus\/} WFPC2), between our calibrated ground-based $B$-band
photometry and our WFPC2 $B$-band photometry when the color-tansformation
coefficients of Holtzman \etal\ are used.  Each star is represented by a
$\pm1\sigma$ error bar.} 
      \label{Fig03}
\end{figure}

In calibrating these data to the ground-based photometric system, we found that
the color-transformation coefficients of Holtzman \etal\ (1995) for the $V$
filter were satisfactory: 
$$\hbox{\rm F555W} = V + \hbox{\rm constant} +
0.060\,(\bmv) - 0.033\,(\bmv)^2$$
left no serious residual trends with color or magnitude, as we illustrate in
Fig.~2.  However, Holtzman's $B$ transformation, 
$$\hbox{\rm F439W} = B + \hbox{\rm constant} - 0.003(\bmv) + 0.088(\bmv)^2,$$
did not seem to work for the \ngc{4147}\ data.  When we imposed this color
transformation on the data from the four WFPC2 chips and solved only for a
photometric zero point for each image, we obtained the transformation residuals
shown in Fig.~3.  Not only is there a systematic curvature of the
fitting residuals with color, but since the bluest stars all have similar
apparent magnitudes---they are on the blue horizontal branch---the poor color
transformation appears as a magnitude nonlinearity as well.  We therefore used
these \ngc{4147} data to redetermine the color transformations for the $B$
filter, and arrived at the following relationship: 
$$\hbox{\rm F439W} = B + \hbox{\rm constant} + 0.54(\bmv) - 0.42(\bmv)^2.$$
This represents the weighted average of the transformations derived
independently from the four CCDs, and this average transformation was imposed
equally on the data from the four chips in the final reduction.  The
star-by-star residuals from this best fit are illustrated in Fig.~4. 
The data shown in the lower panel suggest that a cubic color term might be
called for, as the bluest stars still tend to have positive fitting residuals,
and the two reddest stars---which have very small uncertainties---have negative
residuals.  However, in view of the number and quality of the calibrating stars,
we judge a cubic transformation to be too extreme to be attempted in this case. 
The formal uncertainties of the linear and quadratic coefficients are already
$\pm 0.04$ and $\pm 0.21$, respectively.  

Our $B$-band color-transformation coefficients are painfully huge, and very
different from the published values.  We note that in a previous analysis of
these same data by Piotto \etal\ (2002), the Holtzman color corrections were
adopted.  Some indications of a problem with the standard F439W calibration were
noted by Bedin \etal\ (2000) in their study of \ngc{2808} = C0911-646, which
they dealt with by means of an empirical {\it linear\/} adjustment of the
WFPC2-based \bmv\ colors to ground-based values for stars in the cluster field. 
We can offer no explanation for the difference between Holtzman's calibration
and ours, and fall back on the feeble justification that this transformation
appears to be necessary to make our analysis of the WFPC2 images accord with our
ground-based results.

\begin{figure}[t]
  \figurenum{4}
  \plotone{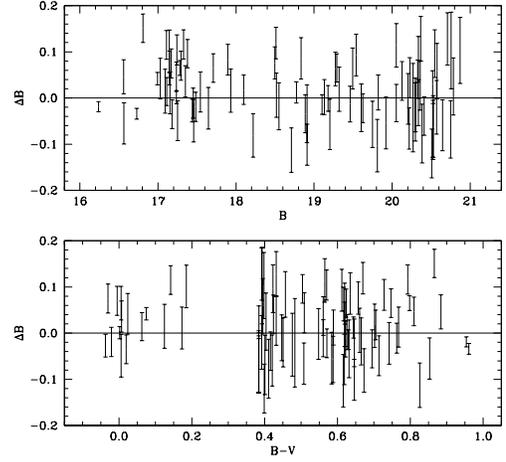}
\caption{As in Fig.~3, photometric residuals, in the sense
(ground-based {\it minus\/} WFPC2), between our calibrated ground-based $B$-band
photometry and our WFPC2 $B$-band photometry except that here we have used
color-transformation coefficients determined directly from the stars in the
transfer sequence, rather than those of Holtzman \etal}
      \label{Fig04}
\end{figure}

\begin{figure}[t]
  \figurenum{5}
  \plotone{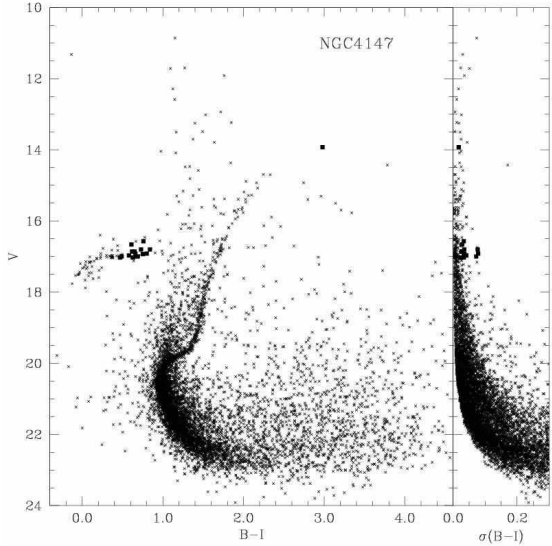}
\caption{A (\bmi,$V$) color-magnitude diagram (left panel)
including all the stars in our photometric survey area having measurements in at
least the $B$, $V$, and $I$ filters, provided $\sigma(\bmi) < 0.30\,$mag.
A few stars lie outside the limit of this figure.  The right-hand panel plots
$V$ and $\sigma(\bmi)$ for the same stars.  In each case, catalogued variable
stars have been indicated by filled squares.}
      \label{Fig05}
\end{figure}

\section{COLOR-MAGNITUDE AND COLOR-COLOR DIAGRAMS}
\subsection{Color-Magnitude Diagrams for \ngc{4147}}

The left panel of Fig.~5 is a ground-based $V$ versus \bmi\
color-magnitude diagram (CMD) for every detection in the field of \ngc{4147} for
which we have photometry in at least the $B$, $V$, and $I$ filters, and
$\sigma(\bmi) < 0.30\,$mag, within the indicated magnitude and color limits. 
The right panel plots the value of $\sigma(\bmi)$ for the same detections
against the visual magnitude.  Note that the horizontal scales of the two parts
of this diagram are not the same.  Solid squares in this plot represent average
photometric indices for the 15 previously identified variable
candidates\footnote{At this point, we do not consider V5, V9, or V15 to be
likely variable stars; see \S5.3 below.} in the cluster field, plus one
additional variable candidate that we have identified in these data.  The
variable candidate at $(\bmi,V) = (2.98,13.93)$ is V18, discovered by
AF04,  which they tentatively identify as a foreground RR
Lyrae star.  For those variable candidates for which we were able to estimate
periods and produce reasonable light curves (see next section), we have derived
mean magnitudes by converting the fitted light curves to flux units, integrating
over one cycle, and converting the results back to a magnitude scale.  In the
case of V18---for which we were unable to find a period---the mean photometric
indices are only robust averages of the individual magnitudes we have in hand. 
There will be more discussion of these stars in \S5 below.  

The CMD clearly shows the presence of foreground stars and
background galaxies, especially those with red colors and faint magnitudes.  The
presence of stellar blends in the crowded cluster center is also evident in the
broad lump of stars above the subgiant branch.  The fact that $\sigma(\bmi)$
tends to values larger than 0.1$\,$mag primarily for stars considerably fainter
than $V=20$ suggests that our photometry is reasonably complete to at least this
limit\footnote{Note that $\sigma(\bmi)\sim 0.3$ implies, in the worst case,
$\sigma(B) \sim \sigma(I) \sim 0.2$, or a signal-to-noise ratio of about 5 in
each filter; if the S/N ratio is slightly poorer than this in one filter,
then it will be much better than this in the other.  Therefore a star with
$\sigma(\bmi) \sim 0.3$ is at least a $7\sigma$ detection in $B$ and $I$
considered together, and is probably a $10\sigma$ detection when the $V$ filter
is added, since that bandpass is close to the peak quantum efficiency of
standard CCDs.}, which is close to the main-sequence turnoff (TO) of the
cluster.  

We therefore estimated the astrometric position of the cluster center as
follows: a virtual circular aperture of some specified radius was scanned over
the catalog of detections with calibrated positions and magnitudes.  This
aperture was judged to be centered on the cluster when the median $X$- and
median $Y$-position of all objects with $V\leq20.0$ contained within the
aperture coincided with the center of the aperture itself.  For instance, when a
virtual aperture of radius 5\min\ was concentric with the cluster, it contained
938 catalog entries, and the one-dimensional root-mean-square width of the
distribution of objects about the centroid was $\sigma = 68$\sec, so the
precision of this cluster centroid may be estimated at $\sigma/\sqrt{N-1}
=$~68\sec/$\sqrt{937} \approx 2$\sec\ in each direction.  We presume that any
residual incompleteness of stars with $V \ltsim 20$ at small cluster radii will
be more or less independent of position angle, so this should not systematically
affect our estimate of the cluster centroid position.

Repeat experiments with virtual-aperture radii from 9\min\ down to 10\sec\
produced ratios $\sigma/\sqrt{N-1}$ that continued to decrease monotonically
for smaller aperture radii.  During these experiments the derived cluster
centroid position drifted over an {\it extreme\/} range of 2\Sec9 in right
ascension and 2\Sec4 in declination.  The center of these two ranges corresponds
to the position 
$\alpha\,=\,12$\hr10\mn06\Sc34, $\delta\,=\,+18$\deg32\min33\Sec4 (J2000), 
which is very close to the estimate
obtained with a virtual-aperture radius of 60\sec, enclosing 741 objects with a
positional root-mean-square dispersion of 27\sec.  This estimated position,
which we believe to be accurate to $\sim1$\Sec0 in each coordinate, lies about
3\sec\ northeast of the value tabulated by Harris (1996).

\begin{figure}[t]
  \figurenum{6}
  \plotone{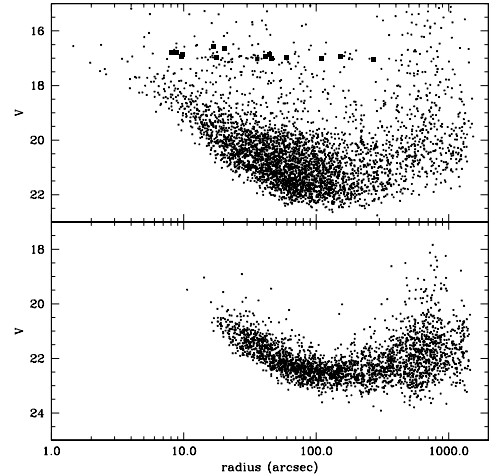}
\caption{(Upper panel) A plot of apparent visual
magnitude $V$ against radial distance from the cluster center for all stars
with color uncertainties $\sigma(\bmi) < 0.10\,$mag); (lower panel)
the same for all stars with $0.10 \leq \sigma(\bmi) < 0.30\,$mag.  This
illustrates the level of completeness expected at any given distance from
the cluster center.  Obviously,  crowding errors near the center of the
cluster mean that precise photometry is possible only for relatively bright
stars.  The error-curves bend upward slightly at large distances from the
cluster because these parts of the astronomical field were recorded only
in a comparatively small number of CCD images.  Catalogued variables have
been indicated by filled squares.}
      \label{Fig06}
\end{figure}

The upper panel of Fig.~6 shows the radial distances of all stars with
$\sigma(\bmi) \leq 0.10\,$mag from this adopted cluster centroid position,
plotted against their apparent visual magnitudes.  The filled squares mark 
15 of the 16 variable-star candidates (V18 now lies above the upper
edge of the diagram).  The lower panel shows $V$ magnitude against radius for
all stars with $0.10 < \sigma(\bmi) \leq 0.30\,$mag.  The rising lower
envelopes for $r < 100$\sec\ result from the increasing effect of crowding
for faint stars at small radii.  Taken together, these plots
suggest that serious incompleteness for stars with $V\sim20$
probably sets in only for radii less than 10\sec; the detection limit improves
from $V \sim 21$ at 20\sec\ radius to $V\sim 23$ at 100\sec, and remains {\it
roughly\/} constant at that level as far as the remotest corner of the field
some 20\min\ from the cluster center.  The slight upturn of the error--magnitude
curves at larger radii is due to the fact that the outer parts of the field are
contained in fewer CCD images than the cluster center.  Our star list is
probably close to complete to $V\sim18$ at all radii down to the cluster center. 
The cluster HB, indicated by the variable stars and an
overdensity of constant stars near $V = 17$, can be clearly traced out to a
radius of 200\sec, with one probable RR~Lyrae variable (the one discovered here)
lying nearly 5\min\ from the cluster center.  In the upper panel there also
appears to be a vague edge to the distribution of upper-main-sequence stars with
$20\ltsim V\ltsim22$ near a radius of 200\sec.

\begin{figure}[t]
  \figurenum{7}
  \plotone{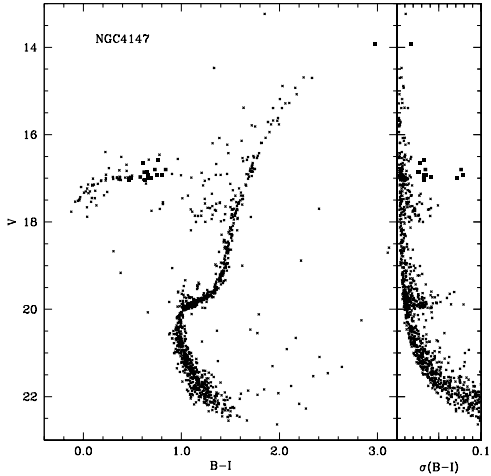}
\caption{A cleaner $(\bmi,V)$
color-magnitude diagram for \ngc{4147} made by adopting progressively higher
magnitude limits in radial zones closer to the cluster center.  Catalogued
variable stars have been indicated by filled squares.}
      \label{Fig07}
\end{figure}

Fig.~5 represents every star with reliable {\it BVI\/} photometry in our
sample.  We now want to produce a cleaner CMD in order
to better define the morphology of the cluster fiducial sequence. 
Fig.~7 is an attempt to produce a cleaner diagram by plotting only the
catalog entries with the following properties: $\sigma(\bmi) \leq 0.10\,$mag,
and

\begin{center}
$\begin{array}{rlcrl}
             & r \leq 200^{\prime\prime} & \hbox{\rm and} & & V \leq 18.0;\\
 30^{\prime\prime} \leq & r \leq 200^{\prime\prime} & \hbox{\rm and} & 18.0 < & V \leq 20.0;\\
100^{\prime\prime} \leq & r \leq 200^{\prime\prime} & \hbox{\rm and} & 20.0 < & V.
\end{array}$
\end{center}

\noindent These selection criteria have been applied only to the putatively
constant stars; all the variable candidates have been plotted as filled squares
regardless of position or magnitude. 

It is clear from the figure that even these criteria include a few stars of
larger photometric uncertainty, especially among stars slightly brighter than
the boundaries of our imposed magnitude zones at $V\ltsim18.0$ and
$V\ltsim20.0$; presumably these stars with larger color errors are
concentrated in the innermost parts of their radial zones as well.  This
illustrates a known shortcoming of the ALLFRAME analysis package: when
photometric errors are dominated by {\it crowding\/}, rather than {\it noise\/},
ALLFRAME systematically underestimates the uncertainties of the derived
photometric indices.  This is shown most clearly by the group of stars to the
blue of the RGB and below the HB, with $17.5 \ltsim V
\ltsim 18.0$ and $1.0 \ltsim \bmi \ltsim 1.4$.  These are mostly cluster giants
that have been scattered by up to 0.4$\,$mag in \bmi\ color by crowding errors,
despite the fact that their $\sigma(\bmi)$ color uncertainties have been
estimated to be less than 0.10$\,$mag.  This tendency of the software results
from the fact that those photometric uncertainties are estimated from the known
readout noise and Poissonian photon statistics in the digital image as well as
the size of the individual-pixel residuals from the profile fits.  However,
when a given blob of light is being modeled by many overlapping profiles each of
which has an amplitude and a centroid in two dimensions that may be freely
adjusted, many different combinations of fitted parameter values may produce
fits of nearly the same quality.  As a result, the adopted solutions may in fact
be rather farther from the ``true'' solution than the fitting residuals would
seem to imply.  The stars here are affected most strongly in the $B$ filter
because at shorter wavelengths the brightness contrast between the giant-branch
stars and the fainter, bluer subgiants and turnoff stars is smaller than in $V$
or $I$, which makes the crowding effects more severe.

\begin{figure}[t]
  \figurenum{8}
  \plotone{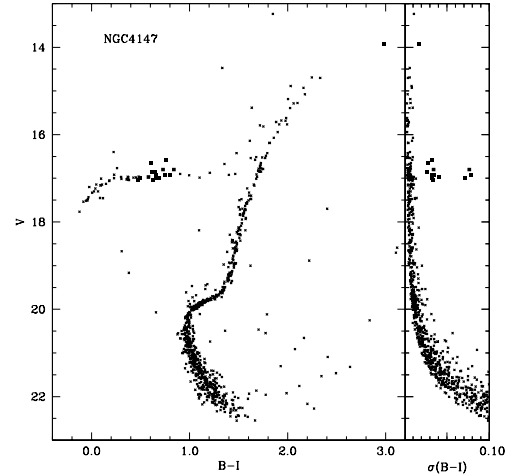}
\caption{A still cleaner $(\bmi,V$) color-magnitude diagram
for \ngc{4147} has been produced by augmenting the radially graduated
magnitude limits with a separation-based acceptance criterion applied to
the individual stars.  As before, all catalogued variables have been
indicated by filled squares.}
      \label{Fig08}
\end{figure}

To reduce the influence of stars whose photometric errors have been
underestimated due to crowding, one can employ a separation index such as that
defined by Stetson \etal\ (2003; see their Sec.~4.1) to supplement
the $\sigma$-based selection criterion.  Fig.~8 here results from the
same selection criteria as Fig.~7 augmented by $\hbox{\it sep} \geq 3.0$ for an
assumed seeing of 1\Sec0, which means that a star must be at least sixteen times
brighter than the summed contribution of all other stellar profiles at its
position---\ie, its light must be contaminated by no more than 6\% by known
companions---when the seeing is one arcsecond.  As one can see, this additional
acceptance criterion completely removes the slight haze of stars below the
nominal position of the red horizontal branch, and reduces the number of stars
above the main-sequence turnoff and subgiant branch.

\begin{figure}[t]
  \figurenum{9}
  \plotone{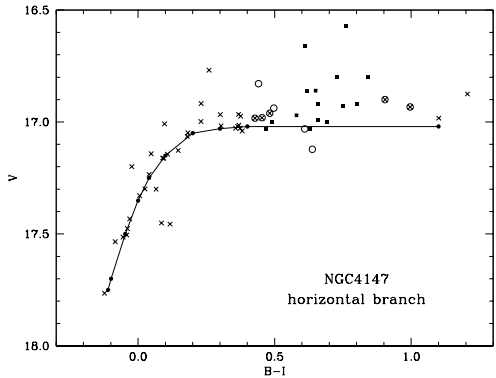}
\caption{An enlargement of the
horizontal-branch region of the $(\bmi,V)$ color-magnitude diagram of
\ngc{4147}.  As before, filled squares represent the catalogued variable
stars in the cluster.  Crosses represent constant stars taken from
Fig.~8.  Open circles represent the apparently constant stars near the
instability strip discussed in the text, and listed in Tables~7 and 14.
Those open circles that do not contain crosses represent those
constant stars that failed to pass the $\hbox{\it sep} > 3$ acceptance
criterion for Fig.~8.  The solid curve represents our hand-drawn
estimate of the ZAHB; it has been arbitrarily extended redward from
\bmi = 0.40 at a fixed level of $V = 17.02$ to indicate where
we believe red horizontal-branch stars should be, if any were present.
A few stars do lie near this locus, but their membership status is
not otherwise known.}
      \label{Fig09}
\end{figure}

\begin{figure}[t]
  \figurenum{10}
  \epsscale{1.25}
  \plotone{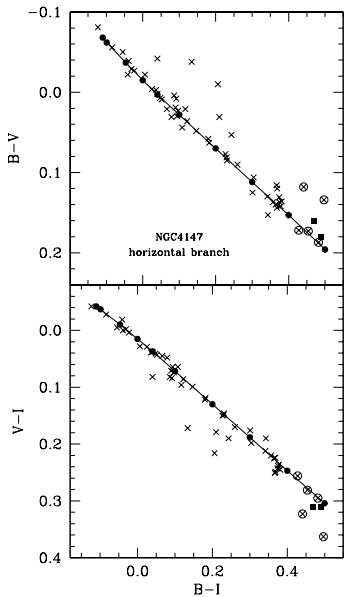}
\caption{Two examples of the color-color diagrams for
horizontal-branch stars in \ngc{4147}\ that we used to transform the ZAHB
from the $(\bmi,V)$ plane to other colors.  (Upper) The
(\bmi,\bmv) plane; (lower) the (\bmi,\vmi) plane.  We also produced and
employed a (\bmi,\vmr) diagram (not illustrated).  Point types are as in Fig.~9,
except that in this case apparently constant stars failing the $\hbox{\it
sep} > 3$ acceptance criterion have not been plotted.}
      \label{Fig10}
\end{figure}

\begin{figure}[t]
  \figurenum{11}
  \plotone{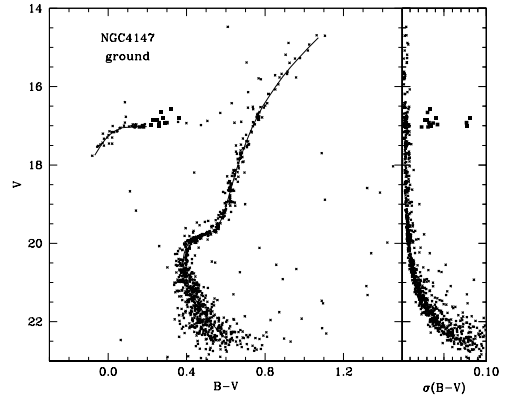}
\caption{A ground-based $(\bmv,V)$ color-magnitude
diagram for \ngc{4147}, cleaned as described in the text.  Catalogued
variable stars have been plotted as filled squares, and the solid curve
represents our hand-drawn fiducial sequence for the cluster.  This figure
is intended for direct comparison with Fig.~12.}
      \label{Fig11}
\end{figure}

In Table~4 we present our derived fiducial sequences for \ngc{4147}. 
The principal sequences were derived from hand-sketched curves drawn on
large-scale plots of cleaned CMDs with $V$ plotted against,
separately, \bmi, \bmv, \vmr, and \vmi.  Once normal points had been read out
with a ruler, we considered the first and second differences between the
tabulated points, and made minor adjustments to the normal points to produce
reasonable smoothness.  The final curves were visually verified by digitally
overplotting the final normal points on the CMDs.  In the
case of the horizontal branch, we estimated the locus from a large-scale plot of
the $(\bmi,V)$ CMD (Fig.~9), because
this diagram offers the most favorable ratio of photometric uncertainty to color
range.  The hand-drawn curve skirts the lower envelope of the horizontal branch
where it is nearly flat---and therefore is expected to represent the 
ZAHB---and continues through the greatest density of points
where it bends faintward at the blue end.  This adopted curve has been
transcribed to the other filters via color--color plots, like the ones shown as
Fig.~10 here.  

Fig.~11 shows our ground-based $(\bmv,V)$ CMD for
the \ngc{4147} stars selected by the criteria listed above, with our derived
fiducial sequences superimposed.  For comparison, we show in Fig.~12 the
$(\bmv,V)$ CMD for all stars with $V$ and $B$ magnitudes
derived from the \hst\ observations.  Here we have made no selection on radius or
crowding; every point is plotted provided only that $\sigma(\bmv) < 0.30\,$mag,
and the derived color and magnitude fall within the limits of the diagram.  The
RR Lyrae candidates that fell within the WFPC2 field have again been plotted as
filled squares, but since all the \hst\ images were obtained within an interval
of about 20 minutes, these represent instantaneous photometric quantities, and
the estimated photometric errors do not stand out from those of the other stars. 
The same fiducial sequences as in Fig.~11 are reproduced here; they indicate
that our calibration has done a reasonable job of referring the instrumental
F439W and F555W magnitudes to the standard {\it BV\/} system.  Only the bluest
horizontal-branch stars appear to have been measured a bit too red, and the 
reddest giants have been measured a bit too blue.  This may be a reflection
of the third-order color calibration that we eschewed in the previous section.

\begin{figure}[t]
  \figurenum{12}
  \plotone{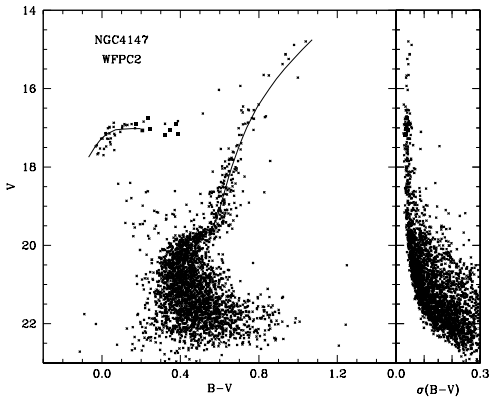}
\caption{Color-magnitude diagram derived from the
WFPC2 observations of \ngc{4147}, where measurements in the F439W and
F555W filters have been transformed to the $B,V$ photometric system of
Landolt as described in the text.  This figure is intended for direct
comparison to Fig.~11, and the same fiducial sequence as in that figure
has been plotted here.  Instantaneous photometric indices for the variable
stars that fell within the WFPC2 coverage have been indicated by filled
squares.}
      \label{Fig12}
\end{figure}

\subsection{Distribution of Blue Stragglers}

A comparison of Figs.~11 and 12 suggests that a population of blue 
stragglers---stars brighter than the TO and bluer than the lower giant
branch---is much more prominent in the WFPC2 data than in the ground-based data. 
To investigate whether there is a significant reality behind this appearance, we
have estimated the relative frequency of stars in the blue straggler region
of the CMD in various subsamples of the data.  We must warn the reader, however,
that this is not a fully rigorous experiment:  there are complicating factors
beyond our control.  In particular, \ngc{4147}\ was approximately centered on
the PC chip of WFPC2.  Readers will be familiar with the peculiar ``E''-shaped
footprint of the WFPC2 on the sky.  With the large inter-chip dead zones (we do
not attempt to calibrate photometry from positions $x < 75\,$px or $y < 75\,$px
on the WFC chips, or $x < 100\,$px or $y < 100\,$px on PC; see Stetson 1998),
gaps in the WFPC2 coverage extend to as close as 16\sec\ from the cluster
center.  The farthest corner of the WFC field lies some 90\sec\ from the cluster
center.  The ground-based coverage of the cluster, of course becomes confused
and imprecise as the center is approached (Fig.~6 above).  Furthermore, the
photometric boxes we will use here are comparatively crude, so we will not be
measuring a specific blue-straggler frequency in the sense of Bolte \etal\
(1993), for instance.

We define a blue-straggler box in the CMD by the limits $0.00 < \bmv \leq 0.40$
and $18.0 < V < 19.5$; similarly we define a lower giant-branch box by $0.40 <
\bmv \leq 0.80$ and $18.0 < V < 19.5$ (\cf\ Figs.~11 and 12).  We have not
carried out formal artificial-star tests on these images.  However, we note that
Fig.~6 has shown that the $\sigma(\bmi)$ photometric uncertainties of stars
brighter than $V = 19.5$ exceed 0.10$\,$mag only for stars within 8\sec\ or
9\sec\ of the center of the cluster.  Our past experience is that the detection
completeness is not worse than 90\% when the photometry achieves these levels of
precision.  Especially with the broad wavelength difference between the $B$ and
$I$ filters enhancing the detectability of, respectively, blue and red stars, we
expect that any residual incompleteness or large photometric errors due to
blending will be negligible for these fairly closely matched photometric boxes
at radii greater than 20\sec.

\begin{deluxetable}{rrrrr}
\footnotesize
\tablecaption{Fiducial sequences for \ngc{4147}}
\tablecolumns{5}
\tablenum{4}
\tablehead{
\colhead{$V$} &\colhead{\bmi} &\colhead{\bmv} 
&\colhead{\vmr} &\colhead{\vmi} 
}
\startdata
\multicolumn{5}{c}{Giant branch} \\
14.75 & 2.275 & 1.072 & 0.623 & 1.203 \\
15.00 & 2.168 & 1.022 & 0.594 & 1.146 \\
15.25 & 2.079 & 0.975 & 0.572 & 1.104 \\
15.50 & 1.998 & 0.931 & 0.551 & 1.067 \\
15.75 & 1.927 & 0.890 & 0.534 & 1.037 \\
16.00 & 1.865 & 0.854 & 0.517 & 1.011 \\
16.25 & 1.811 & 0.820 & 0.503 & 0.991 \\
16.50 & 1.760 & 0.789 & 0.492 & 0.971 \\
16.75 & 1.715 & 0.763 & 0.482 & 0.952 \\
17.00 & 1.671 & 0.738 & 0.473 & 0.933 \\
17.25 & 1.631 & 0.717 & 0.465 & 0.914 \\
17.50 & 1.594 & 0.697 & 0.457 & 0.897 \\
17.75 & 1.560 & 0.679 & 0.449 & 0.881 \\
18.00 & 1.531 & 0.662 & 0.442 & 0.869 \\
18.25 & 1.507 & 0.646 & 0.436 & 0.861 \\
18.50 & 1.484 & 0.632 & 0.430 & 0.852 \\
18.75 & 1.460 & 0.618 & 0.425 & 0.842 \\
19.00 & 1.438 & 0.604 & 0.421 & 0.834 \\
19.25 & 1.411 & 0.590 & 0.417 & 0.821 \\
\multicolumn{5}{c}{subgiant branch} \\
19.50 & 1.369 & 0.567 & 0.413 & 0.802 \\
19.60 & 1.339 & 0.557 & 0.404 & 0.782 \\
19.65 & 1.314 & 0.548 & 0.395 & 0.766 \\
19.70 & 1.277 & 0.533 & 0.383 & 0.744 \\
19.75 & 1.229 & 0.510 & 0.370 & 0.719 \\
19.80 & 1.176 & 0.478 & 0.355 & 0.698 \\
19.85 & 1.127 & 0.455 & 0.340 & 0.672 \\
19.90 & 1.083 & 0.435 & 0.325 & 0.648 \\
19.95 & 1.052 & 0.421 & 0.312 & 0.631 \\
20.00 & 1.026 & 0.410 & 0.302 & 0.616 \\
20.10 & 0.994 & 0.398 & 0.290 & 0.596 \\
\multicolumn{5}{c}{main sequence} \\
20.25 & 0.975 & 0.388 & 0.286 & 0.587 \\
20.50 & 0.965 & 0.380 & 0.286 & 0.585 \\
20.75 & 0.979 & 0.388 & 0.290 & 0.591 \\
21.00 & 1.011 & 0.404 & 0.298 & 0.607 \\
21.25 & 1.061 & 0.426 & 0.310 & 0.635 \\
21.50 & 1.127 & 0.453 & 0.326 & 0.674 \\
21.75 & 1.205 & 0.484 & 0.346 & 0.721 \\
22.00 & 1.296 & 0.520 & 0.370 & 0.776 \\
22.25 & 1.408 & 0.565 & 0.397 & 0.843 \\
22.50 & 1.532 & 0.615 & 0.426 & 0.917 \\
\multicolumn{5}{c}{Horizontal branch} \\
17.75 & --0.110 & --0.068 & --0.032 & --0.042 \\
17.70 & --0.099 & --0.062 & --0.031 & --0.037 \\
17.50 & --0.047 & --0.037 & --0.023 & --0.010 \\
17.35 & 0.000 & --0.015 & --0.015 & 0.015 \\
17.25 & 0.040 & 0.003 & --0.008 & 0.037 \\
17.15 & 0.100 & 0.028 & 0.006 & 0.072 \\
17.05 & 0.200 & 0.070 & 0.032 & 0.130 \\
17.03 & 0.300 & 0.112 & 0.063 & 0.188 \\
17.02 & $\geq$0.400 & $\geq$0.153 & $\geq$0.100 & $\geq$0.247 \\
\enddata
\end{deluxetable}

If we consider the entire {\it BVI\/} survey area {\it
outside\/} the 200\sec\ limit, we count 12 stars in the blue-straggler box and
54 stars in the lower-giant box, contained within an area of some 890$\,$sq.\
arcmin; we take these as representative of the field population.  If we now
consider the ground-based data for the annulus 20\sec--200\sec, we count 9 blue
stragglers and 153 faint giants in an area of some 35$\,$sq.\ arcmin.  If we
subtract the scaled field component, the blue straggler:giant ratio becomes
9:151 for the cluster rounded to nearest whole numbers, or 0.06.  Now if we
consider the WFPC2 photometry (recognizing that a portion of the WFPC2 field
overlaps with the 20\sec-200\sec annulus) we count 23 blue stragglers and 137
faint giants (ratio = 0.17) in a total area of some 4.7$\,$sq.\ arcmin.  Field
corrections to these latter numbers are negligible: less than a quarter of a
star.  Incompleteness due to crowding is also a non-issue in the WFPC2 data.
From this, it is already clear that the blue stragglers in \ngc{4147}\ are
strikingly more centrally concentrated than the faint giants.  This conclusion
is reinforced by the observation that, within the WFPC2 coverage, the rms
distance of the blue stragglers from our adopted cluster center is 12\sec, that
of the faint giants is 35\sec.

For further discussion of the evidence and implications of radial gradients in
blue straggler populations in globular clusters, the reader is referred to the
recent papers by Piotto \etal\ (2004) and Sabbi \etal\ (2004).  The high
central concentration of blue stragglers is generally held to be consistent
with the idea that they were formed by stellar collisions or binary mergers in
a past episode of core collapse.

\begin{deluxetable*}{ccccccc}
\footnotesize
\tablecaption{NGC~4147: main-sequence turnoff in various colors}
\tablecolumns{7}
\tablenum{5}
\tablehead{
\colhead{$V_{TO}$} 
&\colhead{$( \bmi )_{TO}$} 
&\colhead{$( \bmr )_{TO}$} 
&\colhead{$( \vmi )_{TO}$} 
&\colhead{$( \bmv )_{TO}$} 
&\colhead{$( \rmi )_{TO}$} 
&\colhead{$( \vmr )_{TO}$} 
}
\startdata
\multicolumn{7}{c}{Ground-based data} \\
20.48 & 0.971 & --- & --- & --- & --- & --- \\
20.50 & --- & 0.677 & --- & --- & --- & --- \\
20.43 & --- & --- & 0.580 & --- & --- & --- \\
20.55 & --- & --- & --- & 0.390 & --- & --- \\
20.49 & --- & --- & --- & --- & 0.293 & --- \\
20.51 & --- & --- & --- & --- & --- & 0.284 \\
\multicolumn{7}{c}{WFPC2 data} \\
20.51 & --- & --- & --- & 0.398 & --- & --- \\
\enddata
\end{deluxetable*}

\subsection{Comparison with Other Clusters---Relative Abundances and Ages}

The data in Table~4 suggest that the TO of \ngc{4147}, defined as the bluest
point on the fiducial sequence, lies at $V\approx20.5$.  A more sensitive
analysis, involving the fitting of a parabola to the actual stellar photometry
(\ie, not the normal points) in a restricted magnitude range symmetric about the
TO (see Stetson \etal\ 1999), leads to the results shown in Table~5. 
Each separate CMD indicates both a color and a visual
magnitude corresponding to the bluest point on the stellar sequence.  A straight
unweighted average of the seven values of $V_{TO}$ is 20.50, but we regard the
determinations near the top of the table as the strongest and those near the
bottom as the weakest, so we adopt 20.48--20.49 as our best guess at the turnoff
magnitude.  With the flat part of the lower envelope of the horizontal branch
(HB) quite well constrained at a value near $V = 17.02$, this implies a TO--ZAHB
magnitude difference $\Delta V\,=\,3.46$ or 3.47$\,$mag.  Allowing some
0.06$\,$mag for the typical difference between the ZAHB and the mean HB
(\eg, Catelan 1992; Cassisi and Salaris 1997), this still puts \ngc{4147}
squarely within the band of normal globular clusters in the plot of Rosenberg
\etal\ (1999, their Fig.~3) that relates $\Delta V_{TO}^{HB}$ to [Fe/H],
implying a completely normal age compared to other globular clusters included in
their analysis.

\begin{figure}[t]
  \figurenum{13}
  \plotone{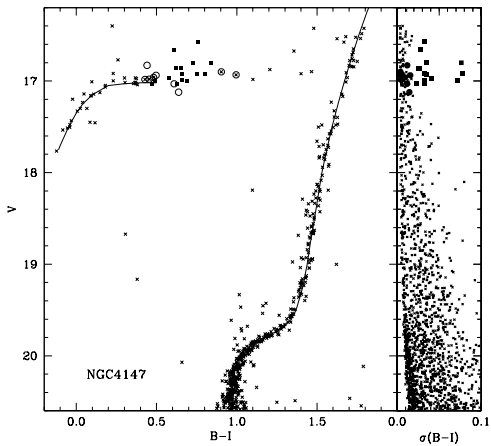}
\caption{An enlargement of the 
subgiant-to-horizontal-branch region of our $(\bmi,V)$ color-magnitude diagram
for \ngc{4147} intended for comparison to Figs.~14 and 15---the corresponding
diagrams for the globular clusters M$\,$3 and M$\,$55.  Our hand drawn fiducial
sequences have been superimposed, and point types have the same significance as
in Fig.~9.}
      \label{Fig13}
\end{figure}

\begin{figure}[t]
  \figurenum{14}
  \plotone{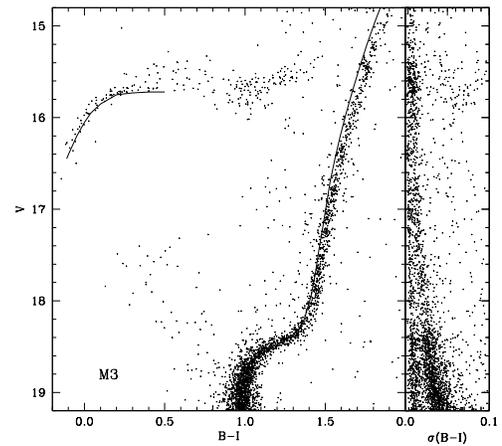}
\caption{Plot of the subgiant-to-horizontal-branch
region of our $(\bmi,V)$ color-magnitude diagram for M$\,$3 
intended for comparison to Fig.~13.  The solid curve represents our
adopted fiducial sequence for \ngc{4147} shifted vertically upward 
by 1.30$\,$mag to provide an optimum match to M$\,$3 in the turnoff/subgiant
and blue horizontal-branch regions of the diagram.  No horizontal shift
has been applied.  The perceptible mismatch between the two clusters'
giant branches suggests that M$\,$3 is more metal-rich than \ngc{4147}.}
      \label{Fig14}
\end{figure}

The normalcy of NGC4147 is further illustrated by the comparison between
Fig.~13 and Fig.~14: the former shows the (\bmi,$V$)
CMD of \ngc{4147} from just below the
turnoff to just above the horizontal branch, with our fiducial sequences
superimposed.  The latter diagram shows our unpublished data for the cluster
M$\,$3 = \ngc{5272} = C1339+286, which we have collected and processed in the
same way as for \ngc{4147} and many other targets.  Here we have not imposed a
radial selection, but have plotted only stars with $\sigma(B-I) < 0.10$ and
$\hbox{\it sep\/} > 5$.  The solid curve is our hand-fitted fiducial sequence
for \ngc{4147}, shifted brightward by 1.30$\,$mag and with no horizontal shift. 
The agreement of the colors of the horizontal branches indicate that the
reddening difference between the clusters is effectively zero (Harris 1996 gives
E(\bmv) = 0.02 for \ngc{4147}, 0.01 for M$\,$3).  The slight displacement and
relative tilt of the giant branches indicates that M$\,$3 is by a small amount
the more metal-rich of the two clusters, an inference consistent with the data
in Harris's compilation catalog:  he lists [Fe/H]~=~--1.83 for \ngc{4147} and
--1.57 for M$\,$3; these are evidently on the metallicity scale of Zinn \& West
(1984).  The agreement of the luminosity of the subgiant branches indicates that
the ages of the two clusters are indistinguishable with the present data.

\begin{figure}[t]
  \figurenum{15}
  \plotone{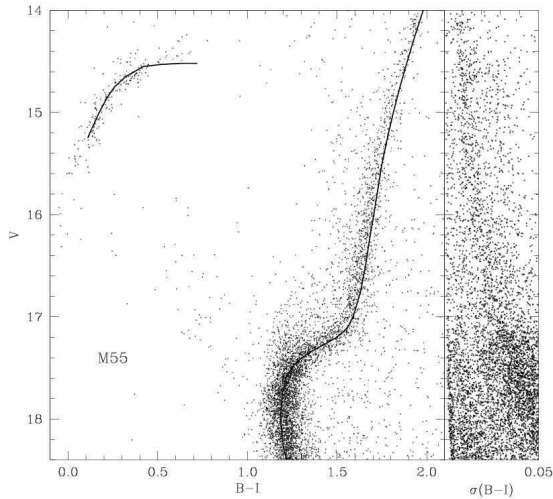}
\caption{Plot of the subgiant-to-horizontal-branch region of
our $(\bmi,V)$ color-magnitude diagram for M$\,$55 intended for comparison to
Fig.~13.  The solid curve represents our adopted fiducial sequence for
\ngc{4147} shifted vertically upward by 2.50$\,$mag and horizontally to the
right by 0.22$\,$mag to provide an optimum match to M$\,$3 in the
turnoff/subgiant/red-giant and blue horizontal-branch regions of the diagram.}
      \label{Fig15}
\end{figure}

Fig.~15 compares our fiducial sequences for \ngc{4147}\ to our unpublished
photometric results for the globular cluster M$\,$55 = \ngc{6809} = C1936-310. 
This comparison is particularly interesting because the two clusters appear to
have the same chemical abundances to within the precision with which they can be
determined:  Harris (1996) lists [Fe/H]~=~--1.81 for M$\,$55.  As is usually the case,
the original CCD images are a mix of data we have taken ourselves and data we
have requested from archives or received from colleagues with the intention of
creating a large, homogeneous database of photometry for star clusters and
resolved galaxies.  Unlike \ngc{4147} and M$\,$3, M$\,$55 is a southern cluster
and has literally no observing runs in common with the other two.  A
quantitative comparison, then, relies heavily on the validity of the
standard-star system used to calibrate the data.  

The principal sequences in our CMDs for M$\,$55 are
perceptibly broadened by amounts in excess of our expected photometric errors. 
This is in contrast to the photometry of Mandushev \etal\ (1996) who found a
tight main sequence with $\sigma(\vmi) \sim 0.015\,$mag at and just below the
turnoff.  We have attempted to compensate by plotting only the stars with the
very best data: $\sigma(\bmi) < 0.05\,$mag, but this has not reduced the
scatter.  There are a number of possible explanations.  First, part of the
greater width of our principal sequences is undoubtedly due to the fact that
many the images available to us have quite short exposures:  their comparatively
large values of $\sigma(\bmi)$ are visible in the right-hand panel of Fig.~15.  
However, the perceived scatter does not decrease for the brighter stars as
rapidly as one would expect.  Second, our 30\min$\times$30\min\ field includes
the cluster center, while the 4\min$\times$4\min\ field of Mandushev \etal\ was
some 7\min\ (=~2.4 core radii) from the center.  Our sample therefore presumably
includes more stars whose photometry is adversely affected by crowding, although
plots of our color residual from M55's giant branch versus position do not
support the notion that---at least on the giant branch---the photometric errors
decrease with increasing distance from the cluster center.  Third, we believe it
is possible that M55 is subject to differential reddening; its Galactic
coordinates are $l\,=\,$9\deg, $b\,=\,-23$\deg, and the aforementioned
plots of color residual versus position suggest that the reddening E(\bmi) may
be increasing toward the east.  We should also note here that our main-sequence
ridge line for M$\,$55 in the $(\vmi,V)$ color magnitude diagram lies
0.02$\,$mag to the blue of that of Mandushev \etal\ over the magnitude
range common to the two studies.  

We find a best overall match between \ngc{4147}\ and M$\,55$ when the fiducial
sequences for the former are shifted by +0.22$\,$mag in \bmi\ and by
--2.50$\,$mag in $V$.  A quantitative determination of M$\,$55's turnoff
magnitude by the fit of a parabola to more than 3,000 stars within
$\pm\,0.4\,$mag of the turnoff in the (\bmi,$V$) CMD yields
$(\bmi)_{TO} = 1.212$ and $V_{TO}=17.99$, comparable to the values of 1.19 and
17.98 inferred from our manual shift of the \ngc{4147} sequences---including the
giant branch and horizontal branch---to those of M$\,$55.  Harris's (1996)
compilation catalog lists $V_{HB} = 17.01$ for \ngc{4147}---extremely close to
the value of 17.02 that we have determined here---and 14.40 for M$\,$55.  The
inferred difference in apparent visual distance modulus, --2.61, is not
compatible with our data.  

The perceived horizontal shift between the clusters is presumably due primarily
to a difference in reddening.  If E(\vmi)$\,\approx\,1.3\,$E(\bmv), then
E(\bmi)$\,\approx\,2.3\,$E(\bmv), and our adopted shift of +0.22$\,$mag in \bmi\
corresponds to $\Delta\,$E(\bmv)~=~+0.10$\,$mag.  Yet Harris lists
E(\bmv)~=~0.02 for \ngc{4147} and 0.08 for M$\,$55.  We note that the reddening
maps of Schlegel \etal\ (1998) predict E(\bmv)$\,=\,$0.026 for
\ngc{4147}\ and 0.135 for M$\,$55; the implied difference of 0.11$\,$mag is less
dissimilar to what we find than Harris's tabulated values.  

Closer investigation of Fig.~15 reveals that our shifted \ngc{4147}\ fiducial
falls perceptibly to the blue of the center of M$\,$55's main-sequence band,
while \ngc{4147}'s giant branch is toward the red side of M$\,$55's.  We have
already noted that a quantitative determination of the color difference between
the turnoffs of M$\,$55 and \ngc{4147}, $\Delta(\bmi)_{TO} = 1.212 - 0.971
=0.24\,$mag, is slightly greater than the 0.22$\,$mag adopted as the best
compromise shift between the fiducial sequences overall.  The extent of the
color difference between a cluster's main-sequence turnoff and its lower giant
branch is affected by both age and abundance, with a smaller difference
indicating either a greater age or a lower metallicity.  By this measure M$\,$55
is then either older or more metal-poor, since for this assumed difference in the
clusters' reddening values its main sequence is redder and its giant branch
is bluer than in \ngc{4147}.  However, an age difference would also affect
the difference in apparent magnitude between the flat part of the horizontal
branch and the turnoff or the nearly flat part of the subgiant branch.  As
discussed above, this age indicator is very nearly the same in the two clusters;
if anything, in Fig.~15 M$\,$55's subgiant branch is skewed slightly to the
blue/bright side of that of \ngc{4147}, and the stars on the sloping part of
M$\,$55's HB are skewed to the red/faint side of \ngc{4147}'s.  If we accept that
these minor differences are statistically significant, this would indicate a
{\it younger\/} age for M$\,$55, in conflict with the age implications of the
color differences between turnoff and giant branch.  We therefore suspect, on
the basis of these photometric indicators, that M$\,$55 may actually be slightly
more metal-poor than \ngc{4147}.  If this inference is correct, then the
dereddened main sequence of M$\,$55 should lie slightly to the {\it
blue\/}---not to the red---of that of \ngc{4147}, and we conclude that the
reddening toward M$\,$55 must be still higher than our previous estimate by at
least another 0.02$\,$mag in $\Delta$E(\bmi), \ie, about 0.26---or about 0.11 in
$\Delta$E(\bmv)---which brings it into still closer agreement with the Schlegel
\etal\ predictions.  Note that the redder measured colors of Mandushev \etal\
(1996) would imply an even higher reddening for M$\,$55 if its metal abundance
is similar to that of \ngc{4147}.  If the reddening of M$\,$55 is as high as
E(\bmv)$\,\approx\,$0.12--0.15$\,$mag, variability in the reddening at the level
of 0.02--0.03$\,$mag (rms) in E(\bmv), or $\approx\,$0.05--0.07$\,$mag in
E(\bmi), would not be surprising.  This may be part of the explanation for
the slightly broadened principal sequences in Fig.~15.

Even with all these caveats and conditions, the vertical magnitude difference
between the clusters' horizontal branches and subgiant branches, $\Delta
V_{TO}^{ZAHB} \approx 3.46$ (\ngc{4147}), $\sim 17.99 - 14.52 = 3.47$ (M$\,$55),
indicates that their ages are the same to within the precision of these data. 
We note that VandenBerg (2000) has listed a value $\Delta V_{TO}^{HB} = 3.65$ for
M$\,$55, completely at odds with our estimate even once allowance is made for
the magnitude difference between the ZAHB and the mean HB.  VandenBerg cites
Chaboyer \etal\ (1996) as his source for this measurement. 
They in turn cite Buonanno \etal\ (1989), who derived $V_{TO} =
17.90 \pm 0.07$ for M$\,$55 based upon their own CCD measurements calibrated
to agree with the photographic photometry of Alcaino (1975), and $V_{HB} =
14.35 \pm 0.07$ from the photographic study of Lee (1977).  For comparison,
Buonanno \etal\ also cite $V_{TO} = 18.03$ and $V_{HB} = 14.33$ for the same
cluster from a literature survey by Peterson (1986), but these values do not
appear to have been used in their analysis.  The original estimate of
E(\bmv)$\,=\,$0.08$\,$mag for M$\,$55 also appears to have come from the
photographic study by Lee.  We believe the present work supersedes these values.  
The result $\Delta V_{TO}^{HB} = 3.50\pm0.12$ found by Rosenberg \etal\ (1999)
accords well with ours, whether we apply the $\sim 0.06\,$mag correction
for the ZAHB {\it minus\/} mean HB magnitude difference or not.

\subsection{Morphological Parameters of the Evolved Sequences}

The morphology of the giant branch of a globular cluster's CMD
has long been used as indicative of its metallicity.  Ferraro \etal\
(1999) produced quantitative equations relating various measurable quantitities
in the ($V$,\bmv) plane to [Fe/H] on the scale of Carretta \& Gratton (1997;
CG97).  Fig.~16 shows our dereddened $(\bmv,V)$ CMD
for \ngc{4147} with the various parameters illustrated.  The \bmv\ colors have
been dereddened assuming $E(\bmv) = 0.02$.  In this diagram we have extrapolated
the giant-branch fiducial sequence (heavy long-dashed curve) in accordance with
the notion that the star designated V18 by AF04 may be in fact a normal
giant-branch member of the cluster, presumably near the giant branch
tip\footnote{We note, however, that our star catalog for the field includes a
companion star lying some 0\Sec2 from V18, whose presence is inferred from the
best-seeing images.  V18 is present in the PC images, but in the extreme
corner where the point-spread function is distorted and we do not attempt a
calibration (the spherically aberrated image of the star falls on a vertex of
the pyramid mirror before being corrected, resulting in lost light), and is
badly saturated or close to saturation in the F555W exposures, so it is
difficult to use them to confirm or refute the reality of the companion.  If
this detection is false, then we may have significantly underestimated the
apparent brightness of V18, and the metallicities inferred here from the
extrapolated-giant-branch morphology indices should be revised downward. 
Auri\`ere \& Lauzeral (1991) noted the presence of a bright star 25\sec\ from
the cluster center with $V=13.91$ and $\bmv = 1.56$, which agrees reasonably
well with our measurements---$V=13.93$ and $\bmv=1.44$---even though they made
no mention of such a companion and probably could not have seen one with their
1\sec\ seeing.}.  We have also projected the horizontal branch across the
diagram to the giant branch (heavy dashed horizontal line) at the level $V =
17.02$.

\begin{figure}[t]
  \figurenum{16}
  \plotone{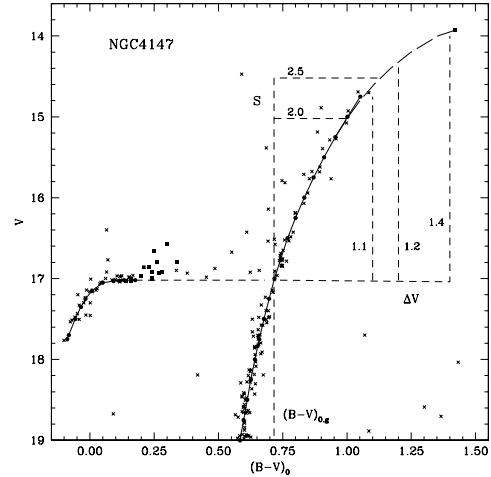}
\caption{Figure illustrating the various giant-branch
morphology parameters that have been developed to gauge a globular cluster's
metal content, as applied to our $(\bmv,V)$ color-magnitude diagram for
\ngc{4147}.  The solid curves represent our adopted cluster fiducial
sequences; the long-dashed curve is a notional extrapolation stemming from the
postulate that star V18 lies near the giant-branch tip.  Short-dashed curves
are the geometric representations of the various morphology metrics.
Apparently constant stars in \ngc{4147} are represented by small crosses,
and catalogued variables by filled squares.}
      \label{Fig16}
\end{figure}

Table~6 presents our quantitative results for the different morphology
parameters and their corresponding values of [Fe/H]${}_{\hbox{\footnotesize
CG97}}$ according to the equations given in Table~4 of Ferraro \etal\ \  We have
marked with colons those results that are particularly dubious because they rely
on our extrapolation of the giant branch to V18.  We note that the $S_{2.0}$
index betokens a rather high metallicity for \ngc{4147}: higher than the
other photometric indices, and high compared to published abundance estimates
for the cluster.  We do not know why this index stands out.  In Ferraro's
compilation the various relations between the photometric indices and
metallicity exhibit scatter of order 0.12$\,$dex--0.20$\,$dex in terms of
abundance; the dispersion for $S_{2.0}$ is given as 0.18$\,$dex.  The
uncertainty of these estimators for \ngc{4147} may well be greater than for the
average cluster, due to the paucity of bright giants.  The metallicity implied
by this one index is therefore anomalous at not more than a 2$\sigma$ level. 
This is probably not extreme enough to require a special explanation.  For
comparison, Ferraro \etal\ listed the following parameters for \ngc{4147} based
upon the CMD of Sandage \& Walker (1955): $(B-V)_{\hbox{\footnotesize 0,g}} =
0.76$, $\Delta V_{1.1} = 2.17$, $S_{2.0} = 6.63$, implying
[Fe/H]${}_{\hbox{\footnotesize CG97}}$ = --1.53, --1.50, and --1.19.  On the
CG97 scale the adopted metallicities of \ngc{4147}, M$\,$3, and M$\,$55 are,
respectively, --1.58, --1.34, and --1.61.  On the metallicity scale of Zinn \&
West (1984) all these abundances would be about 0.2$\,$dex lower.  Our inferred
metallicity estimates for \ngc{4147} based upon the $(\bmv,V)$ giant-branch
morphogy parameters are consistent with the claims we made above: namely that it
is perceptibly more metal-poor than M$\,$3 and similar to or marginally more
metal-rich than M$\,$55 based on their (\bmi,$V$) CMDs.

\begin{deluxetable}{cll}
\footnotesize
\tablecaption{NGC~4147: Color-magnitude diagram morphology and metallicity}
\tablecolumns{3}
\tablenum{6}
\tablehead{
\colhead{Parameter} &\colhead{Value} &\colhead{[Fe/H]${}_{\hbox{\footnotesize CG97}}$} 
}
\startdata
(\bmv)${}_{\hbox{\footnotesize 0,g}}$ & 0.716 & --1.80 \\
$\Delta V_{1.1}$        & 2.40  & --1.75 \\
$\Delta V_{1.2}$        & 2.70$\,${\bf :} & --1.73$\,${\bf :} \\
$\Delta V_{1.4}$        & 3.07$\,${\bf :} & --1.56$\,${\bf :} \\
$S_{2.0}$               & 7.02  & --1.31 \\
$S_{2.5}$               & 6.02$\,${\bf :} & --1.64$\,${\bf :} \\
\enddata
\end{deluxetable}

The horizontal-branch morphology of a globular cluster is of particular
interest, as it is generally correlated with the metal abundance of the cluster,
but not perfectly so.  These departures from a simple one-to-one mapping of
metal abundance onto horizontal-branch shape are known as the ``second-parameter
problem''---a riddle of some forty years' standing that has been intensively
debated, but is still without a unique, generally accepted solution.  It is not
our intent here to pursue this debate, and so we omit a discussion of the
literature, but see Catelan (2005) for a recent discussion and references. 
However, it is still useful to quantify the morphology of the
horizontal branch in \ngc{4147}\ so that it may take its place in future
discussions.  In the absence of notable clumps, gaps, extensions, or bimodality,
the HB of \ngc{4147}\ can be adequately characterized by the canonical ratio
$(B-R)/(B+V+R)$ where $B$, $V$, and $R$ are, respectively, the number of stars
on the blue HB, the number of RR~Lyrae variables, and the number of stars on the
red HB.  For this estimate we consider all stars within 200\sec\ of the cluster
center without regard to photometric uncertainty or separation index.  We define
the blue HB as all stars with $16.3 < V < 18.0$ and $-0.40 < \bmi < 0.65$ (\cf\
Fig.~5); the red HB as all stars with $16.3 < V < 17.3$ and $0.65 \leq \bmi
< 1.40$; and the RR~Lyraes are those discussed in \S5 except V19, which lies
beyond 200\sec\ from the cluster center.  With these definitions, the counts
are $B:V:R = 56:14:13$.  The number of stars on the red HB is possibly
overestimated, because at these colors the inclusion of a field star or two is a
possibility; furthermore, it is possible that one or two giants have been
scattered into the zone by photometric errors or crowding.  Assuming that the
red HB stars have been overcounted by some number in the range 0--4, then $+0.52
\ltsim (B-R)/(B+V+R) \ltsim +0.57$.  This revises slightly downward the HB type
recently provided for the cluster by Mackey \& van den Bergh (2005), namely
+0.66, but is equal to the value tabulated in the Harris (1996) catalog,
+0.55.

\begin{figure}[t]
  \figurenum{17}
  \plotone{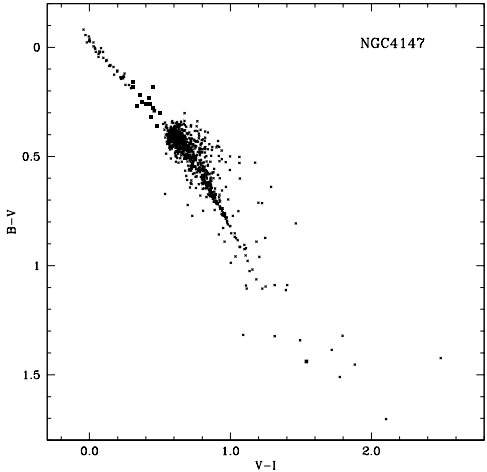}
\caption{$(\vmi,\bmv)$ color-color diagram for
\ngc{4147}, for stars accepted according to the same criteria as Fig.~8.
As before, catalogued variable stars are indicated by filled squares.}
      \label{Fig17}
\end{figure}

\begin{figure}[t]
  \figurenum{18}
  \plotone{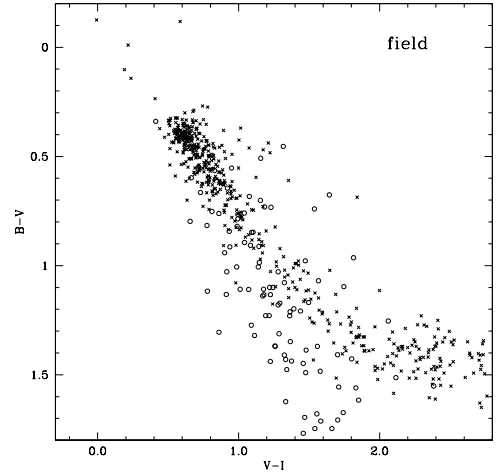}
\caption{$(\vmi,\bmv)$ color-color diagram for
stars in our photometric survey area lying more than 200\sec\ from the
center of \ngc{4147}.  Detections with a \sharp\ index greater than
0.5, indicated perceptibly broadened profiles, have been designated by
empty circles;  these are mostly background galaxies.}
      \label{Fig18}
\end{figure}

There are other parameters that provide more detailed information on the density
and extent of the blue HB ``tail,'' such as $(B2-R)/(B+V+R)$ (where $B$, $V$,
and $R$ are the same as before, but $B2$ is the number of blue HB stars bluer
than $(\bmv)_0 = -0.02$~mag), and $\Delta V_{tail}$ (the difference in magnitude
between the 10\% faintest and the 10\% brightest blue HB stars). These indices
are defined and their interpretation is discussed in 
Buonanno (1993), Buonanno \etal\ (1997) and Catelan \etal\ (2001).  From the same
data set as before, $B2 = 12$ and $(B2-R)/(B+V+R) = -0.01$, and 
$\Delta V_{tail} = 17.69 - 16.64 = 1.05$.

\subsection{Comparison of the Cluster and Field Populations}

Fig.~17 presents the \bmv\ versus \vmi\ color-color diagram for likely
cluster members with the same selection as before:

\begin{center}
$\begin{array}{rlcrl}
             & r \leq 200^{\prime\prime} & \hbox{\rm and} & & V \leq 18.0;\\
 30^{\prime\prime} \leq & r \leq 200^{\prime\prime} & \hbox{\rm and} & 18.0 < & V \leq 20.0;\\
100^{\prime\prime} \leq & r \leq 200^{\prime\prime} & \hbox{\rm and} & 20.0 < & V.
\end{array}$
\end{center}

\noindent plus the standard error of each color $< 0.10\,$mag, and $\hbox{\it
sep\/} > 3$.  Fig.~18 presents the same color-color diagram for the
likely field population, defined as all stars more distant than 200\sec\ from our
adopted cluster center, with color errors $< 0.10\,$mag and $\hbox{\it sep\/} >
3$.  As a further wrinkle, here we have plotted detections with a sharpness
index $< 0.50$ as crosses, and those with \sharp\ $\geq 0.50$ as empty
circles.  These latter should be almost exclusively background galaxies.  Here
we see that the color-color sequence for the field dwarf stars bends in a
direction quite different from that followed by the cluster giants
(Fig.~17).  This effect has been noted before (\eg, Stetson \etal\ 2003), and it
provides a possible means of distinguishing foreground dwarfs from cluster
giants, at least for the reddest stars.  The possibility of mistaking a field
galaxy for a cluster giant on the basis of this two-color diagram appears to be
somewhat greater.  It is worth noting that there are a few blue stars in the
field that could conceivably be cluster blue horizontal-branch stars at large
radii.  As noted previously (\eg, Stetson \etal\ 2004), the blue
star that lies well off the mean color-color ridge line at $(\vmi,\bmv) \sim
(0.6,-0.1)$ is most likely a photometric mistake or, possibly, a non-stellar
object.  Note also that variable candidate V18 falls well to the left of the
main band of field stars at the same \bmv\ color.  This strengthens the case for
it being a giant-star member of the cluster.

\begin{figure}[t]
  \figurenum{19}
  \plotone{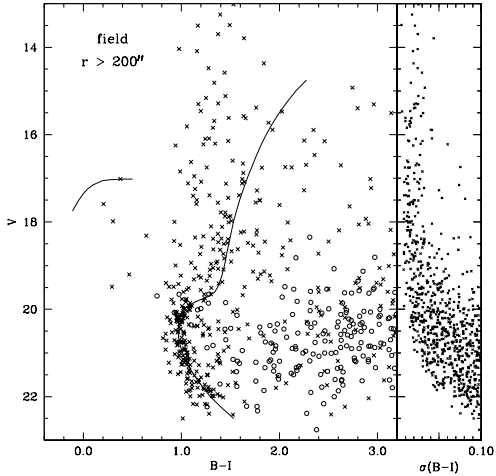}
\caption{$(\bmi,V)$ color-magnitude diagram representing
the stars in our photometric survey area lying more than 200\sec\ from
the cluster center.  Star-like detections (\sharp\ less than 0.5)
are plotted as crosses, and apparently extended detections (\ie, likely
galaxies; \sharp\ greater than 0.5) are shown as empty circles.
Our adopted fiducial sequences for \ngc{4147} have been indicated as
solid curves.}
      \label{Fig19}
\end{figure}

We adopted 200\sec\ as the outer radius for our cluster sample only because
Fig.~6 seemed to indicate a break in the density profile of the cluster at
that distance.  The Harris (1996) compilation indicates an estimated tidal
radius slightly in excess of 6\min\ for \ngc{4147}.  Fig.~19 is a
CMD for the same objects as in the field color-color
diagram, Fig.~18 above.  As before, probable non-stellar objects with $\hbox{\it
sharp\/} \geq 0.5$ are plotted as empty circles.  Here it is clear that the
cluster main sequence is still populated beyond a radius of 200\sec, and one of
the blue stars in the field has the right apparent magnitude for a position on
the cluster's ZAHB.  Even at distances greater than 6\min,
it is still possible to see an apparently significant overdensity of stars near
the location of the cluster turnoff (Fig.~20).  At these radii we are
in the regime where the detection limit is becoming brighter again, due to the
smaller number of images covering this part of the field.  This may explain the
paucity of main-sequence stars below the turnoff.  However, in the absence
of radial velocities, proper motions, or other supporting measurements, we are
unwilling to press this as evidence for an extra-tidal population at this time.  
We can also see from this figure that none of the other blue stars in the field
appears to belong to the cluster horizontal branch.  We have also searched for
evidence for more short-period variables in the \ngc{4147} field and have found
only one:  the candidate we have provisionally named V19 at some 4\Min5
from the cluster center.

\begin{figure}[t]
  \figurenum{20}
  \plotone{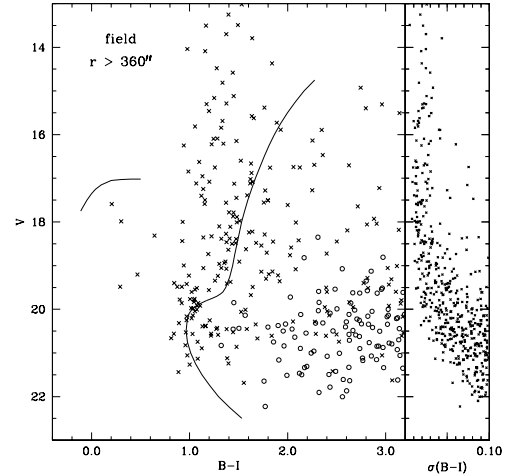}
\caption{The same as Fig.~19, except for those detections
lying more than 6\min\ from the center of \ngc{4147}.}
      \label{Fig20}
\end{figure}

\section{VARIABLE STARS}

\subsection{Astrometry}

Table~7 lists astrometric positions for the variable-star candidates in
\ngc{4147}.  The first column is the star identification, followed by the star's
right ascension and declination as of equinox J2000.0.  Then comes the star's
rectilinear coordinates, in units of arcseconds with $X$ increasing east and $Y$
increasing north, relative to our adopted reference point of
$\alpha\,=\,12$\hr10\mn13\Sc79, $\delta\,=\,+18$\deg31\min22\Sec4 (J2000).  The
last two columns give the star's differential position with respect to our best
estimate of the cluster center:  (--106.0,+71.0) in our $(X,Y)$ coordinate
system, or 
$\alpha\,=\,12$\hr10\mn06\Sc34, $\delta\,=\,+18$\deg32\min33\Sec4 (J2000)
in celestial coordinates.  Candidates V1--V18 have been discussed previously in
the literature, as mentioned in the Introduction.  We have discovered a new
probable c-type RR~Lyrae star in the field of \ngc{4147}, which we
provisionally name V19, pending approval by higher authorities.  This star is
marked by the uppermost, leftmost circle in Fig.~21.  The star is also
visible in Plate~I of Sandage \& Walker (1955):  it is the black dot about
27$\,$mm to the right and 9$\,$mm above star I-3.  Our best light curve for the
star, phased according to a derived period of 0.273$\,$933$\,$d, is shown here
as Fig.~22.

\begin{deluxetable*}{lccrrrr}
\footnotesize
\tablecaption{NGC~4147:  Astrometry for variable candidates
and possIble constant stars near the instability strip}
\tablecolumns{7}
\tablenum{7}
\tablehead{
\colhead{ID} 
&\colhead{RA} 
&\colhead{Dec} 
&\colhead{\ $X$} 
&\colhead{\ $Y$} 
&\colhead{\ $\Delta X$} 
&\colhead{\ $\Delta Y$} \\
& \colhead{2000.0} & \colhead{2000.0}
}
\startdata
V1  & 12~09~59.38 & +18~31~48.4 & --204.9 &  +26.1 &  --98.9 &  --44.9 \\
V2  & 12~10~04.96 & +18~32~04.5 & --125.6 &  +42.1 &  --19.6 &  --28.9 \\
V3  & 12~10~04.38 & +18~31~58.4 & --133.9 &  +36.0 &  --27.9 &  --35.0 \\
V4  & 12~10~06.39 & +18~32~50.1 & --105.2 &  +87.7 &    +0.8 &   +16.7 \\
V5  & 12~10~07.38 & +18~32~35.1 &  --91.2 &  +72.7 &   +14.8 &    +1.7 \\
V6  & 12~10~08.51 & +18~33~00.0 &  --75.2 &  +97.6 &   +30.8 &   +26.6 \\
V7  & 12~10~06.66 & +18~32~39.7 & --101.5 &  +77.3 &    +4.5 &    +6.3 \\
V8  & 12~10~06.95 & +18~32~34.8 &  --97.3 &  +72.4 &    +8.7 &    +1.4 \\
V9  & 12~10~08.24 & +18~32~59.7 &  --78.9 &  +97.3 &   +27.1 &   +26.3 \\
V10 & 12~10~03.74 & +18~31~48.4 & --142.9 &  +26.0 &  --36.9 &  --45.0 \\
V11 & 12~10~05.53 & +18~31~51.8 & --117.4 &  +29.4 &  --11.4 &  --41.6 \\
V12 & 12~10~06.70 & +18~32~28.4 & --100.9 &  +66.0 &    +5.1 &   --5.0 \\
V13 & 12~10~06.37 & +18~32~14.1 & --105.5 &  +51.7 &    +0.5 &  --19.3 \\
V14 & 12~10~06.94 & +18~32~32.4 &  --97.4 &  +70.0 &    +8.6 &   --1.0 \\
V15 & 12~10~07.00 & +18~32~24.8 &  --96.6 &  +62.4 &    +9.4 &   --8.6 \\
V16 & 12~10~07.35 & +18~32~40.1 &  --91.6 &  +77.7 &   +14.4 &    +6.7 \\
V17 & 12~10~10.67 & +18~34~51.2 &  --44.4 & +208.8 &   +61.6 &  +137.8 \\
V18 & 12~10~05.63 & +18~32~11.6 & --116.1 &  +49.2 &  --10.1 &  --21.8 \\
V19 & 12~10~21.98 & +18~35~02.0 &  +116.4 & +219.6 &  +222.4 &  +148.6 \\
C1  & 12~10~01.77 & +18~32~00.5 & --170.9 &  +38.1 &  --64.9 &  --32.9 \\
C2  & 12~10~04.11 & +18~32~38.9 & --137.6 &  +76.5 &  --31.6 &    +5.5 \\
C3  & 12~10~05.88 & +18~32~34.0 & --112.5 &  +71.6 &   --6.5 &    +0.6 \\
C4  & 12~10~06.17 & +18~32~37.6 & --108.4 &  +75.2 &   --2.4 &    +4.2 \\
C5  & 12~10~06.38 & +18~32~36.1 & --105.4 &  +73.7 &    +0.6 &    +2.7 \\
C6  & 12~10~06.51 & +18~32~35.2 & --103.5 &  +72.8 &    +2.5 &    +1.8 \\
C7  & 12~10~08.11 & +18~32~11.1 &  --80.9 &  +48.7 &   +25.1 &  --22.3 \\
\enddata
\end{deluxetable*}

\subsection{Periods}

We summarize the published and present best periods for each of the variable
candidates in Table~8.  As always with RR Lyrae stars, there is some
ambiguity in the period determinations due to the uncertain number of cycles
that take place in the dark ages between successive observing seasons.
For instance, the difference between Newburn's (1957) period of 0.492$\,$39$\,$d
for V2 and Mannino's (1957) period of 0.493$\,$06$\,$d for the same star
represents the difference between 741.7 cycles per year and 740.7 cycles per
year.  We believe that the combination of AF04's dense string of observations in
the first half of 2003 and our own sparser string of observations spanning
1983--2003 (but mostly 1992--2003) offers the best available chance of resolving
those difficulties.  Accordingly, in the penultimate column of Table~8 we list
the best---in our judgment---``modern'' periods for the candidates, based upon
our analysis of the union of the AF04 data with our own.  Similarly, for
completeness we list ancient periods for the same stars based upon our
reanalysis of the union of Newburn's data with Mannino's, where possible, or
Newburn's data alone for stars not observed by Mannino.  In obtaining the
ancient periods, we chose the cycle count that implied a period closest to the
modern period.  In most cases, that turned out to be the same best period as was
found in a blind search, although not always.

\begin{figure*}[t]
  \figurenum{21}
  \epsscale{0.87}
  \plotone{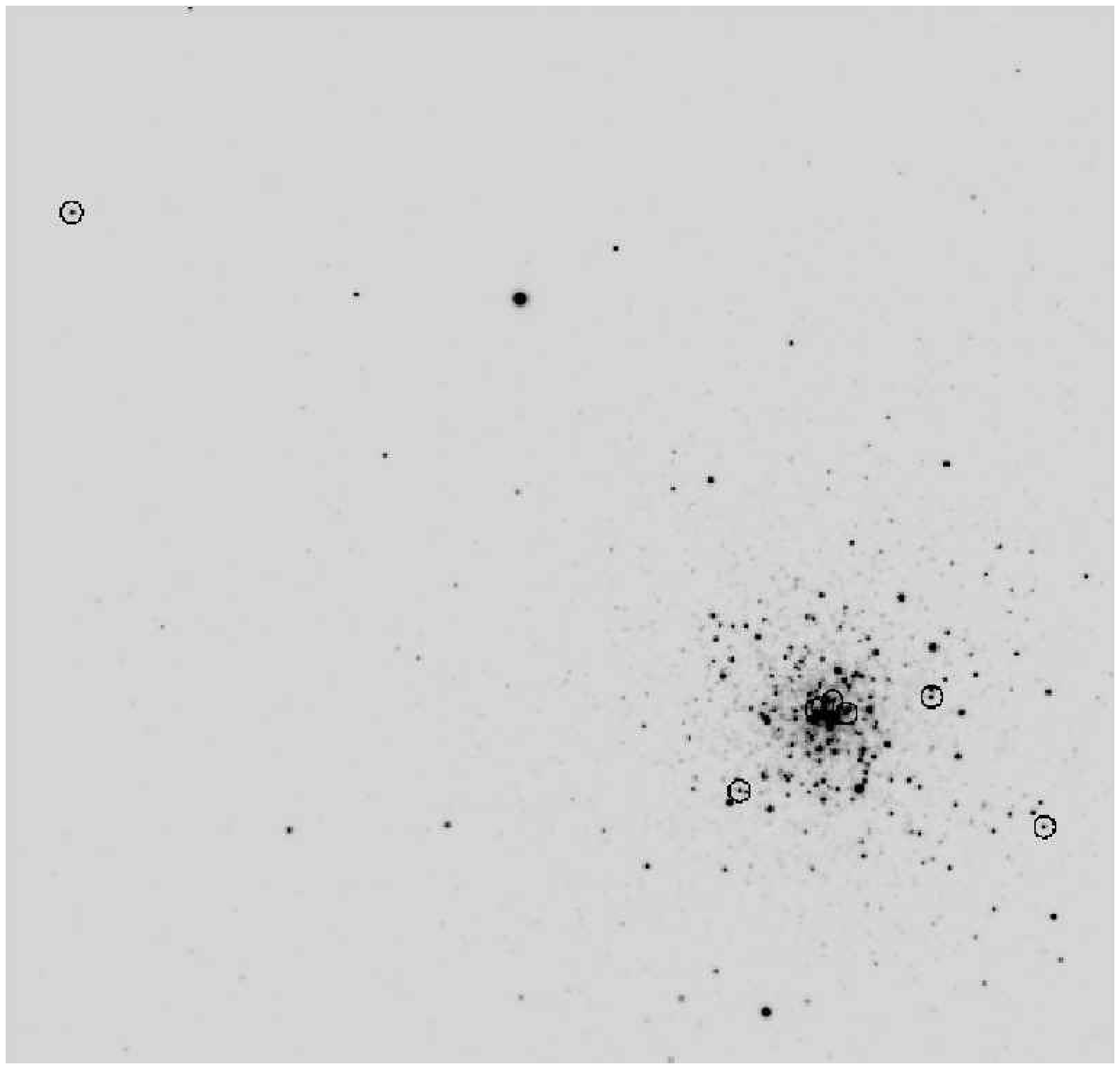}
\caption{Finding chart for newly detected variable candidate,
V19, indicated by the circled dot nearest the upper left corner of the
figure.  Circled dots in the lower right quadrant represent apparently constant
stars near the instability strip, as listed in Tables~7 and 14.  The orientation
is the same as Fig.~1.}
      \label{Fig21}
\end{figure*}

\subsection{Comments on Individual Stars}

A few of the variables represent special problems, and we discuss them
individually here.

{\bf V2:}  The Newburn and Mannino data do not phase well together for any
period in the range 0.20--0.80$\,$d.  The period indicated here is the least bad.
The modern data agree better, but with some indication of a varying amplitude
such as is typical of the Blazhko effect (see AF04, Fig.~2).

\begin{deluxetable*}{lllllll}
\footnotesize
\tablecaption{NGC~4147:  Best periods for variable candidates}
\tablecolumns{7}
\tablenum{8}
\tablehead{
\colhead{ID} 
&\colhead{Newburn} 
&\colhead{Mannino} 
&\colhead{Arellano~Ferro} 
&\colhead{This study} & modern data & older data \\
& \ \ (1957) & \ \ \ (1957) & \ \ \etal\ (2004)
}
\startdata
V1 & 0.500 38            & 0.500 3860 & 0.500 38 & 0.500 403   & 0.500 399   & 0.500 38 \\
V2 & 0.492 39            & 0.493 06   & 0.493 25 & 0.493 180   & 0.493 182   & 0.493 12 \\
V3 & 0.281 58            & 0.280 542  & 0.280 58 & 0.280 542 7 & 0.280 542 9 & 0.280 54 \\
V4 & 0.300 97            &    ---     & 0.299 22 & 0.300 031   & 0.300 033   & 0.300 03 \\
V5 & 0.341 25$\,${\bf :} &    ---     & not var. & not var.    & not var.    & not var. \\
V6 & 0.618 60            &    ---     & 0.609 75 & 0.609 730   & 0.609 732   & 0.609 69 \\
V7 & 0.512 94            &    ---     & 0.514 39 & 0.514 245   & 0.514 243   & 0.514 31 \\
V8 & 0.389 7$\,${\bf :}  &    ---     & 0.278 61 & 0.278 652   & 0.278 651   & 0.278 64 \\
V10 & 0.351 98            & 0.352 314  & 0.352 33 & 0.352 301   & 0.352 300 7 & 0.352 31 \\
V11 & 0.387 36            & 0.387 39   & 0.387 45 & 0.387 419   & 0.387 431   & 0.387 39 \\
V12 & 0.5$\,${\bf :}      &    ---     & 0.504 61 & 0.504 700   & 0.504 701   & 0.504 67 \\
V13 & 0.375 9$\,${\bf :}  &    ---     & 0.408 13 & 0.408 320   & 0.408 318   & 0.408 29 \\
V14 & 0.525 5$\,${\bf :}  &    ---     & 0.259 50 & 0.356 376   & 0.356 372   & 0.356 38 \\
V15 & 0.335 4$\,${\bf :}  &    ---     & not var. & not var.    & not var.    & not var. \\
V16 & 0.277 5$\,${\bf :}  &    ---     & 0.369 4  & 0.372 259   & 0.372 261   & 0.372 09 \\
V17 & 0.375 86            & 0.374 73   & 0.374 94 & 0.371 229   & 0.374 952   & 0.374 84 \\
V18 &    ---              &    ---     & 0.492 05 & not var.    & not var.    &  --- \\
V19 &    ---              &    ---     &    ---   & 0.273 933   &   ---       &  --- \\

\enddata
\end{deluxetable*}

{\bf V4:}  AF04 report a best period of 0.299$\,$22$\,$d for this star, revised
from the value 0.300$\,$97$\,$d of Newburn.  Our own reanalysis of the AF04 $V$-
and $R$-band data considered in isolation derives a best period of
0.300$\,$04$\,$d, closer to the value 0.300$\,$031$\,$d that we derive from an
analysis of our data alone, and to 0.300$\,$033$\,$d that we derive from the
merger of our data with those of AF04.  In all cases, however, there is
appreciable dispersion in magnitude at all phases of the light curve, suggesting
some wander in the epoch of zero phase.

{\bf V5, V9, V15:}  We concur with the judgment of Newburn (V9) and AF04 that
these stars do not appear to be varying.  The heading of Mannino's Tabella~II
states that candidate V15 was one of the stars he studied.  This must be a
typographical error:  the body of the text, Tabella~I, and Mannino's derived
period all indicate that candidate V17 was the one he investigated.

\begin{figure}[t]
  \figurenum{22}
  \plotone{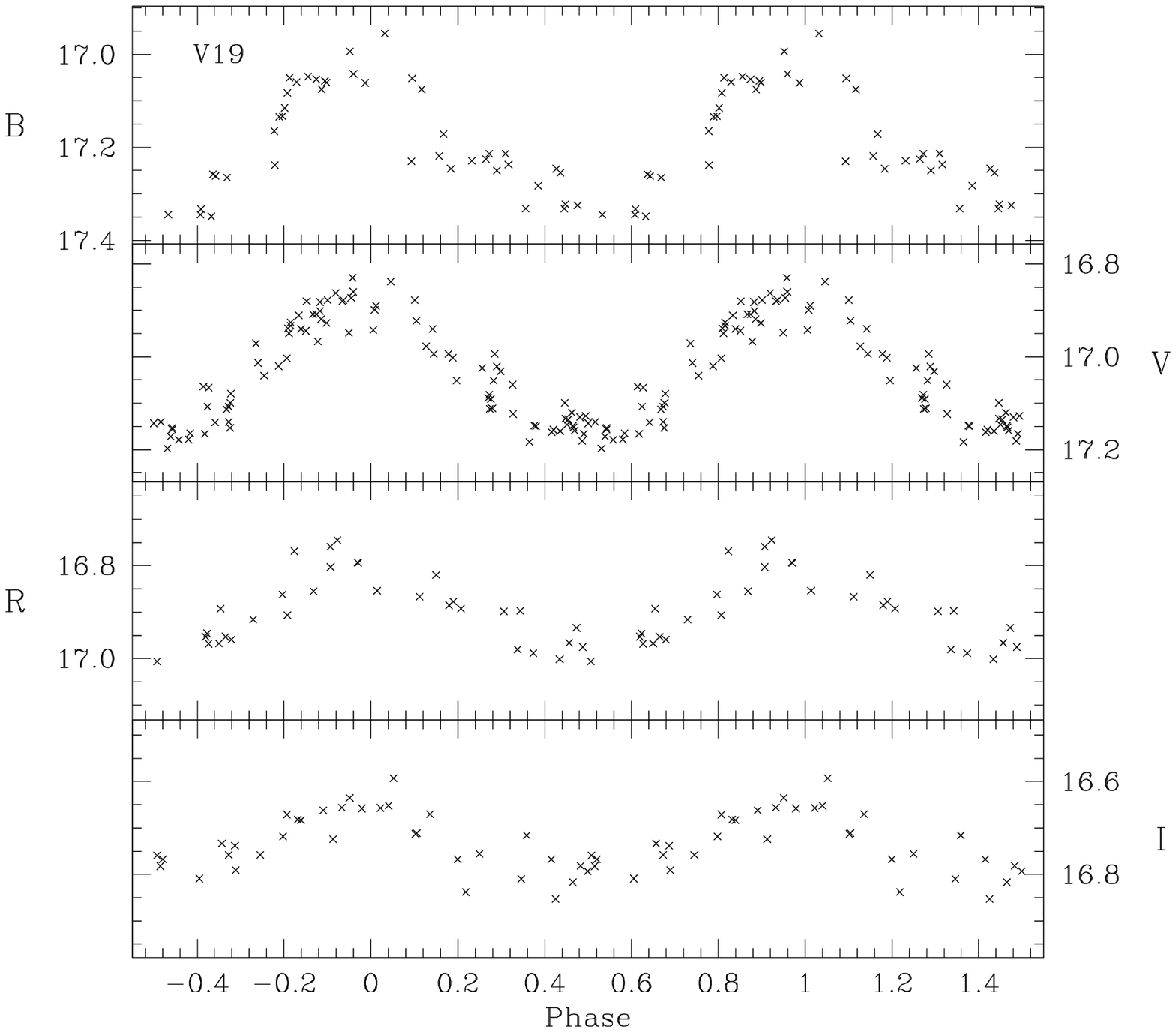}
\caption{Our light curves for the newly identified
variable-star candidate V19, in the field of \ngc{4147}\ in the
(from top to bottom) $B$, $V$, $R$, and $I$ photometric bandpasses.
Twenty years' worth of data have been phased on the period 0.273$\,$933$\,$d.}
      \label{Fig22}
\end{figure}

{\bf V6:}  The merest hint of the striking amplitude variations exhibited in the
AF04 light curve is seen in our data.

{\bf V8:}  There is a systematic magnitude offset between our data and those of
AF04:  we measure it 0.10$\,$mag brighter in $V$ and 0.14$\,$mag brighter in
$R$ than they.  We have applied these offsets to their photometry before 
combining it with ours to derive the modern period listed in Table~8 and the
Fourier components discussed below.

\begin{figure}[t]
  \figurenum{23}
  \plotone{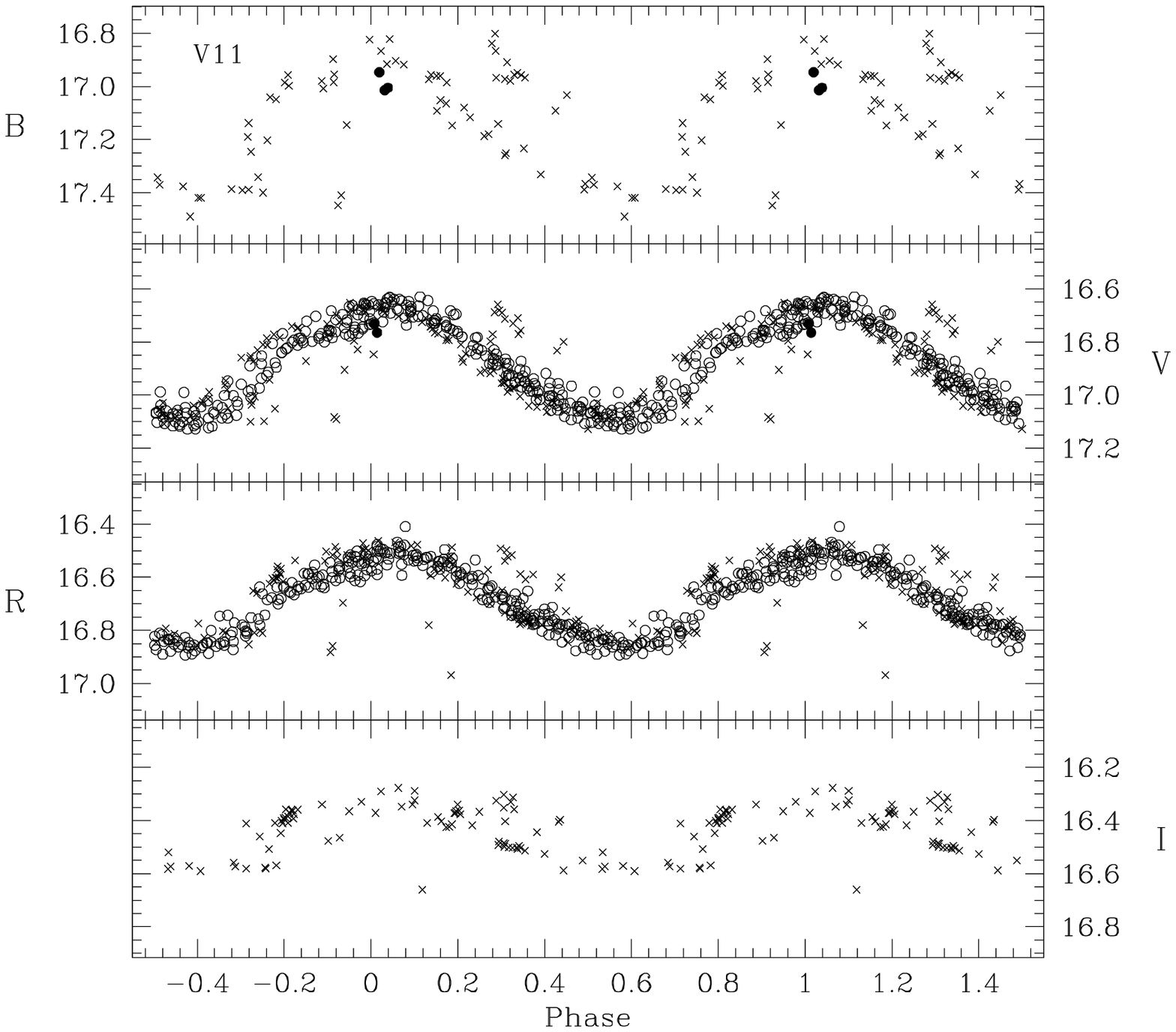}
\caption{Our best phased light curve for variable candidate
V11 in the field of \ngc{4147}.  Our data (crosses for our ground-based data
in $B$, $V$, $R$ and $I$, filled circles for WFPC2 measurements in $B$ and $V$
only), and the measurements of AF04 (empty circles, $V$ and $R$ bands only)
have been phased on a common period of 0.387$\,$431$\,$d.  A significant
number of points fail to match this light curve, as discussed in the text.}
      \label{Fig23}
\end{figure}

{\bf V11:}  A significant number of data points in $B$, $V$ and $R$ do not
phase up with the rest.  In Fig.~23 these are visible as the
striking arc of points above the main body of the light curve between phases
of 0.2 and 0.4.  The slope is right but the phase is wrong.  There is even a
hint of something similar happening at the same phase in the $I$-band light
curve.  (The small scatter of points below the light curves are the sort of
behavior one sees when the star falls very near the edge of the CCD or on a
cosmetic blemish in individual images.  These anomalies are less likely to be
astrophysically significant, and our robust fitting teechniques effectively
ignore them.)  Upon closer investigation, we found that these discrepant points
come from two periods of time:  (a)~before HJD 2,448,500 (\ie, observing runs
``nbs,'' ``igs,'' ``c90c17,'' ``c90ic02, and ``rdj,'' which took place up to and
including April 1991; V11 did not fall within the field covered during the
``rld'' observing run in May 1991, and the ``jvw'' run in March/April 1996 did
not produce any observations in the crucial phase range 0.2--0.4); and
(b)~observing run ```arg02,'' which took place in May 2002 (observing runs
``bono'' and ``hannah,'' which took place during the same observing season, did
not have any coverage in the phase range 0.2--0.4).  There is no other
evidence for this anomaly after  the summer of 1991.

\begin{figure}[t]
  \figurenum{24}
  \plotone{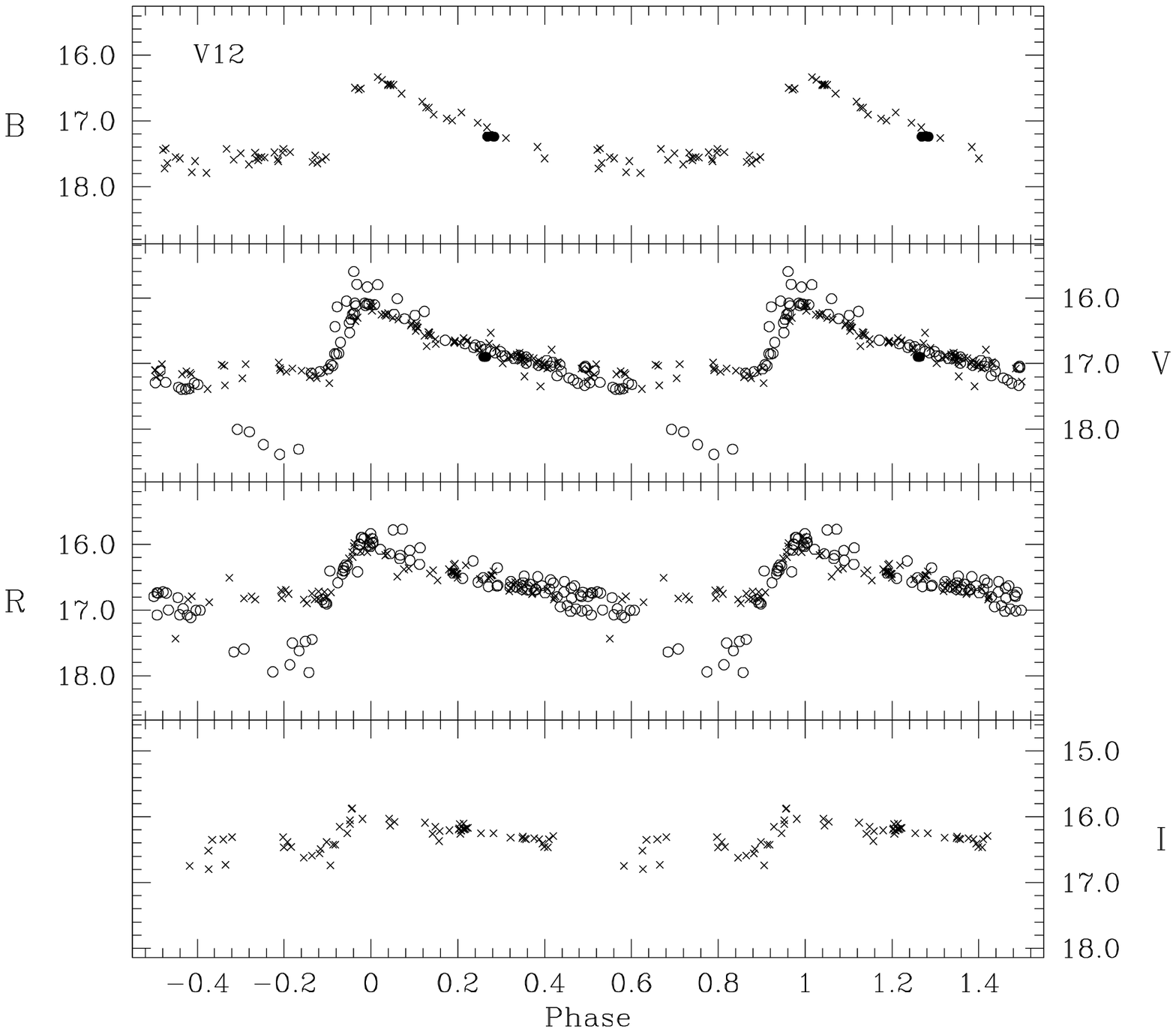}
\caption{Our best phased light curve for variable candidate
V12 in the field of \ngc{4147}.  Our data and the measurements of AF04 
(symbol types as in Fig.~22) have been phased on a common period of
0.504$\,$701$\,$d. Our data trace a significantly shallower and flatter
minimum than those of AF04, suggesting that our light curve is contaminated
by the light of an undetected companion.}
      \label{Fig24}
\end{figure}

{\bf V12:}  We have a serious disagreement with AF04 concerning the amplitude of
variation in $V$ and $R$ (Fig.~24).  The peculiar flat-bottomed
appearance of our light curves strongly suggests that we have confounded V12
with a companion that AF04 successfully distinguished from the variable.

{\bf V13:}  We are not able to phase our data with a period near that of AF04. 
From our data we get reasonable, although somewhat noisy, light curves for a
period of 0.408$\,$320$\,$d, in contrast to their period of 0.408$\,$13$\,$d. 
Our second-best period in the range 0.2--0.408$\,$2$\,$d is 0.407$\,$866$\,$d,
and it is appreciably worse than the other one.  Conversely, the AF04 data do
not phase at all well for any period in the range 0.408$\,$2--0.8$\,$d.  

{\bf V16:}  Our data in the $B$, $R$, and $I$ filters phase reasonably well
for a period near that found by AF04, but the data in the $V$ filter do not.
By experimenting, we found that by considering only the data taken before
HJD 2,450,000 (October 9, 1995), we were able to get reasonable light curves
in all filters for a period of 0.372$\,$25$\,$d, which is close to our overall
best period.  Conversely, for the data taken since then, including the AF04
data, the best period is 0.372$\,$88$\,$d, with a small amount of enhanced
scatter in the $V$-band light curve.

{\bf V17:}  Like the case with V11 and V16, in $B$ and $R$ our data phase well
with AF04's for a period very near their derived value, but a small percentage
of the $V$ data are out of phase.  (The $I$-band data for this star are
comparatively poor, but they appear to dislike this period, too.) Here, the
optimum slice point seemed to be near HJD 2,451,000 (July 5, 1998).  For the
data from earlier than that date, quite nice light curves were obtained for a
period of 0.371$\,$222$\,$d; the later data, which are dominated by AF04,
indicate a best period of 0.374$\,$944$\,$d with amplitude variations.  There is
also a slightly more complicated scenario.  (1)~If we take {\it our\/} data from
before HJD 2,450,000, they can be reasonably well fit with a period of
0.374$\,$8$\,$d.  A period of 0.371$\,$2$\,$d is not as good.  (2)~In the data
taken between 2,450,000 and 2,450,999 there are a number of reasonable periods: 
one of them is around 0.374$\,$8$\,$d and, again, 0.371$\,$2$\,$d is not as
good.  (3)~But if we try to phase these two groups together, there is no single
period that works well.  (4)~If we try to phase all our data from after
2,450,000, but not the AF04 data, the comparatively few points obtained after
2,451,000 could not be well phased with those obtained earlier at any period. 
Conclusion: It seems that each major group (the Mannino data, the Newburn data,
our pre-2,450,000 data, our 2,450,000-999 data, the AF04 data), taken by itself,
can be well fit with a period in the range 0.374$\,$8$\,$d to 0.374$\,$9$\,$d. 
It is only when we seek to combine the data sets over a long time interval
that a 0.371$\,$2$\,$d period appears.  This suggests that we have some sort of
phase jumps or short-lived period changes that we do not record well, but that
the main period does not really change very much.

\begin{figure}[t]
  \figurenum{25}
  \plotone{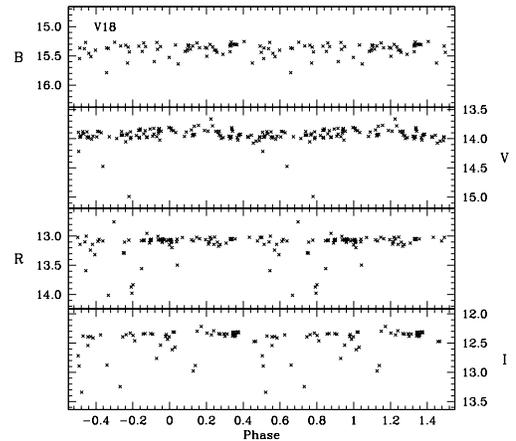}
\caption{Our best phased light curve for variable candidate
V18 based upon our data alone, where the search was restricted to periods
in the range 0.2--0.8$\,$d.  The data have been folded on the best period,
which was 0.324$\,$183$\,$d.  When the period search is extended to 1000$\,$d,
a slightly more significant period---statistically speaking---is found at
32.99$\,$d, but the resulting light curve is no more convincing than this one.}
      \label{Fig25}
\end{figure}

{\bf V18:}  Candidate variable V18 was identified by AF04 and was thought by
them to be a likely foreground RR~Lyrae variable projected onto the cluster
core.  We used our software to examine our data for any plausible periodicities
in the range 0.20$\,$d to 0.80$\,$d.  Fig.~25 presents the light curve
corresponding to the most plausible periodic variation found, with a period of
0.324$\,183\,$d.  In presenting this figure, we have resisted the temptation to
edit out by hand those magnitude measurements that are obviously discrepant.  As
the CMDs of the previous section have shown, this star is
by a significant margin the brightest one in the direction of the main body of
the cluster.  It is very likely that this star is near or above the saturation
level of the detector in some of the long-exposure images, especially for the
longer-wavelength filters.  It may also have fallen near the edge of the CCD or
on a cosmetic blemish in some of the exposures.  The star is close to the
center of the cluster, and our reconstruction of the scene splits the image of
V18 into three components, with companions separated by 0\Sec2 and 0\Sec9 from
the primary star.  The reality of the second and third components of the blend
may reasonably be doubted, but their presence and astrometric
positions---inferred from the best-seeing images available---have been uniformly
assumed in poorer-seeing images as well.  Differing seeing conditions can quite
easily lead to the attribution of different fractions of the photons to
different components of the optical blend in different images.  In any long list
of automatically derived photometry there will always be some corrupt
measurements.  However, our robust software has been designed to reduce the
weight of grossly discrepant data points unless it can find some way to phase
them up into the reasonable semblance of a light curve.  A periodicity search
extended out to a possible period as large as 1000$\,$d turned up a
statistically slightly more significant periodicity of 32.99$\,$d for V18, but
that light curve was no more visually satisfying than Fig.~25.  Our conclusion
is that the star may or may not be be varying (we suspect it isn't), but we
certainly can not infer the physical nature of any intrinsic variation from the
data in hand.  

{\bf V19:}  This star was not previously recognized as a variable candidate,
so it has no previous time-resolved photometry in the literature.  The light
curve fit to our data is noisy, and can be slightly improved if we assume two
periodicities rather than one or, alternatively, a change of period or
discontinuous jump in phase during the time span covered by our observations. 
The available data do not permit a definitive conclusion, and the period that we
have tabulated seems to be a reasonable compromise.

It is probably noteworthy that the four most anomalous variables, V11, V13, V16,
and V17, have best periods of 0.39, 0.41, 0.37, and 0.37$\,$d, respectively.  
These relatively long RRc periods suggest that the erratic phasing may have
something to do with a blending of the fundamental and first overtone pulsation
modes.  However, although the fits to the light curves of these stars might
be marginally improved by double-mode solutions, in no case do we see conclusive
evidence for genuine double-mode behavior.

Comparison of the older and modern periods for the \ngc{4147} RR~Lyrae
variables (Table~8) indicates that some of the variables have undergone real
period changes over the past half century.  Neglecting V17, for which
the long-term period behavior is perhaps uncertain, the mean period change
for the thirteen remaining variables is +0.5 days per million years.
This is a large average period change compared to theoretical
expectations (Catelan 2005 and references therein).
However, RR~Lyrae stars are known to exhibit ``noisy" period changes
as well as the longer-term period changes expected from evolution
(\eg, Rathbun \& Smith 1997).  It is likely that
observations spanning half a century do not yet reveal the true
evolutionary period changes of the \ngc{4147} RR~Lyrae stars.

\begin{figure}[t]
  \figurenum{26}
  \epsscale{0.7}
  \plotone{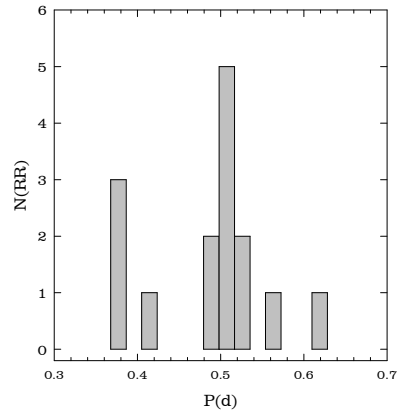}
\caption{Fundamentalized period histogram for \ngc{4147} RR~Lyrae
variables. Note the sharp peak around $P_{\rm f} \simeq 0.51$~d, which resembles
what is seen for M3 and other globular clusters (Rood \& Crocker 1989; Catelan
2004a). The three variables with the shortest periods, which in principle could
be second-overtone (RRe) variables, are more readily classified as
short-period RRc stars (see text).}
      \label{Fig26}
\end{figure}

\subsection{RR Lyrae Specific Frequency and Period Distribution}

The revised number of RR~Lyrae stars in \ngc{4147} also leads to a revision of
its specific RR~Lyrae fraction, defined as $S_{RR} = N_{RR} \times
10^{0.4\,(7.5 + M_V)}$, where $M_V = -6.16$ (Harris 1996) is the cluster's
integrated absolute magnitude in $V$.  With $N_{RR} = 15$, we find $S_{RR} =
51.5$, compared to 30.8 in the current version of the Harris catalog. With this
result, \ngc{4147} now has one of the highest $S_{RR}$ values known, falling
behind only six other Galactic globular clusters.

In Fig.~26 we show a histogram of the RR~Lyrae pulsation periods in
\ngc{4147}.  The RRc periods were ``fundamentalized'' by adding 0.128 to
their $\log\,P$ values. As can be clearly seen, there is a sharp peak in the
period distribution at a period around $P_{\rm f} \approx 0.51$~d. Such a
peaked period distribution is not uncommon among Galactic globular clusters, as
discussed by Rood \& Crocker (1989) and Catelan (2004a). The origin of this
phenomenon is unclear at present, but it does suggest that the full width of the
instability strip is not uniformly populated.  The reader is referred to the
quoted papers, as well as to Castellani \etal\ (2005), for
critical discussions.

\begin{figure*}
  \figurenum{27}
  \epsscale{0.65}
  \plotone{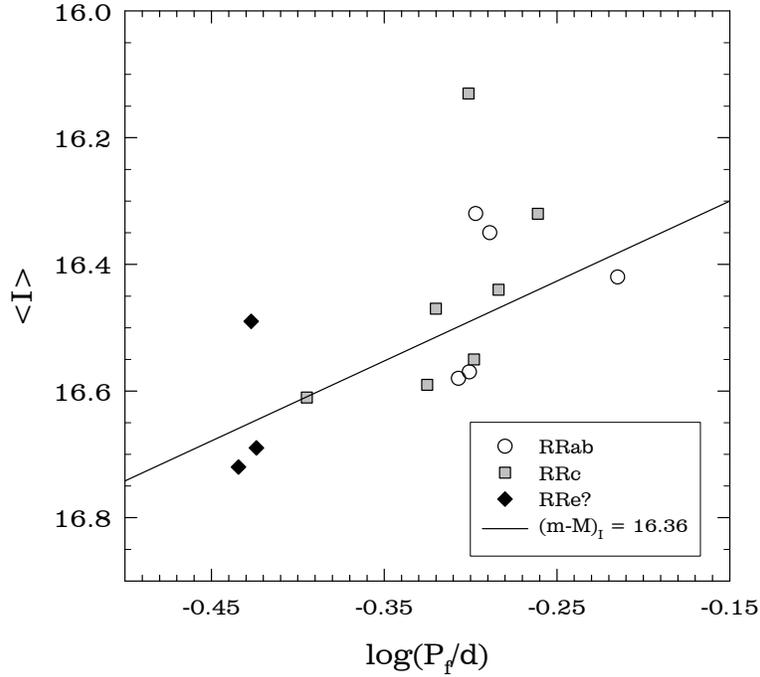}
\caption{Period-luminosity relation in the $I$ band for \ngc{4147} 
RR~Lyrae stars.  RRab stars are shown as open circles, RRc stars as filled gray
squares, and candidate RRe's as black losanges. Periods of RRc and
(candidate) RRe stars were fundamentalized in the same way, by adding +0.128 to their
$\log\,P$ values. The line is the theoretical relation by Catelan \etal\ (2004)
for a metallicity $Z = 0.0005$, an HB type $(B-R)/(B+V+R) = +0.55$, and a
distance modulus $(m-M)_I = 16.36\,$mag, or $(m-M)_V = 16.39\,$mag.}
      \label{Fig27}
\end{figure*}

\begin{deluxetable*}{lcccccccc}
\footnotesize
\tablecaption{NGC~4147:  Mean magnitudes and colors for variable candidates}
\tablecolumns{9}
\tablenum{9}
\tablehead{
\colhead{ID} 
&\colhead{$\left< B\right>$} 
&\colhead{$\left< V\right>$} 
&\colhead{$\left< R\right>$} 
&\colhead{$\left< I\right>$} 
&\colhead{$\left< B\right>-\left< I\right>$} 
&\colhead{$\left< B\right>-\left< V\right>$} 
&\colhead{$\left< V\right>-\left< R\right>$} 
&\colhead{$\left< V\right>-\left< I\right>$} 
}
\startdata
V1  & 17.26 & 17.00 & 16.80 & 16.57 & 0.69 & 0.26 & 0.20 & 0.43 \\
V2  & 17.21 & 17.03 & 16.81 & 16.58 & 0.63 & 0.17 & 0.22 & 0.45 \\
V3  & 17.18 & 17.00 & 16.86 & 16.69 & 0.49 & 0.18 & 0.14 & 0.31 \\
V4  & 17.19 & 16.97 & 16.80 & 16.61 & 0.58 & 0.22 & 0.17 & 0.36 \\
V6  & 17.22 & 16.92 & 16.66 & 16.42 & 0.80 & 0.30 & 0.26 & 0.50 \\
V7  & 17.08 & 16.80 & 16.59 & 16.35 & 0.73 & 0.28 & 0.21 & 0.45 \\
V8  & 17.11 & 16.86 & 16.68 & 16.49 & 0.62 & 0.25 & 0.18 & 0.37 \\
V10 & 17.25 & 16.99 & 16.81 & 16.59 & 0.66 & 0.26 & 0.17 & 0.40 \\
V11 & 17.09 & 16.86 & 16.64 & 16.44 & 0.65 & 0.23 & 0.22 & 0.42 \\
V12 & 17.16 & 16.80 & 16.51 & 16.32 & 0.84 & 0.36 & 0.29 & 0.48 \\
V13 & 16.93 & 16.66 & 16.52 & 16.32 & 0.61 & 0.27 & 0.14 & 0.34 \\
V14 & 17.22 & 16.94 & 16.72 & 16.47 & 0.75 & 0.28 & 0.22 & 0.47 \\
V16 & 16.89 & 16.57 & 16.38 & 16.13 & 0.76 & 0.32 & 0.19 & 0.44 \\
V17 & 17.16 & 16.93 & 16.75 & 16.55 & 0.61 & 0.23 & 0.18 & 0.38 \\
V19 & 17.19 & 17.03 & 16.88 & 16.72 & 0.47 & 0.16 & 0.15 & 0.31 \\
\enddata
\end{deluxetable*}

\subsection{Physical Properties of the Variable Stars}

Table~9 contains the flux-weighted mean magnitudes for the variable
candidates that we obtained by integrating over the fitted light curves.  Here
we have omitted candidates V5, V9, V15, and V18, because we were unable to fit
reasonable light curves to the available data.  For those variables with
ambiguous periods, we have used the light curve corresponding to the best single
period that characterized all our observations, without reference to AF04 or any
other sources of data.  Similarly, Table~10 lists our derived total
amplitudes for the perceived variation in the four photometric bandpasses for
each of the variable stars.  For stars with Blazhko-like symptoms, these
represent {\it average\/} amplitudes; we feel that longer and more nearly
continuous strings of observations would be needed before extremes of amplitude
could be reliably determined.  It is important to note that the mean magnitudes
(Table~9) are based upon a light-curve analysis of {\it our data only\/}, while
the amplitudes (Table~10) are based on a similar light-curve analysis of {\it
our data merged with those of AF04\/}.  We chose to do it this way so that our
mean magnitudes for the variable candidates would be on exactly the same
photometric system as our results for the rest of the stars in the cluster
field.  Since the amplitude determination is fundamentally differential in
nature, it would be less affected if the photometry of AF04 is on a system that
differs from ours by a few hundredths of a magnitude, and would also benefit
from the more densely populated light curves.

\begin{deluxetable}{lllll}
\footnotesize
\tablecaption{NGC~4147:  Pulsation amplitudes for variable candidates}
\tablecolumns{5}
\tablenum{10}
\tablehead{
\colhead{ID} 
&\colhead{$\left< B\right>$} 
&\colhead{$\left< V\right>$} 
&\colhead{$\left< R\right>$} 
&\colhead{$\left< I\right>$} 
}
\startdata
V1  & 1.37 & 1.15 & 0.95 & 0.70 \\
V2  & 1.21 & 1.02 & 0.81 & 0.65 \\
V3  & 0.62 & 0.51 & 0.39 & 0.31 \\
V4  & 0.56 & 0.45 & 0.38 & 0.33 \\
V6  & 1.50 & 0.85 & 0.86 & 0.63 \\
V7  & 1.19 & 1.06 & 0.91 & 0.39 \\
V8  & 0.52 & 0.40 & 0.29 & 0.24 \\
V10 & 0.55 & 0.40 & 0.29 & 0.24 \\
V11 & 0.60 & 0.41 & 0.34 & 0.25 \\
V12 & 1.33 & 1.12 & 0.95 & 0.79 \\
V13 & 0.50 & 0.45 & 0.35 & 0.34 \\
V14 & 0.45 & 0.37 & 0.26 & 0.23 \\
V16 & 0.49 & 0.21 {\bf :} & 0.32 & 0.23 \\
V17 & 0.50 & 0.34 & 0.29 & 0.22 \\
V19 & 0.31 & 0.28 & 0.21 & 0.14 \\
\enddata
\end{deluxetable}

Table~11 lists selected Fourier parameters extracted from our fitted
light curves.  Following traditional practise in the field, we have performed a
sine decomposition for the ab-type variables (P$\,>\,$0.45$\,$d, in this
case), and a cosine decomposition for the c-type variables (see, \eg, Corwin
\etal\ 2003).  All Fourier components up to and including the eighth were
computed for the ab-type variables, and up to and including the sixth
for the shorter-period c-type variables.  However, for compactness here we
list only the astrophysically most interesting parameters.  Table~12
lists physical parameters of the ab-type variables, as based on the
empirical analyses by Jurcsik \& Kov\'acs (1996) and others (see Corwin \etal\
for extensive references).  Table~13 presents physical properties of the 
c-type variables, as inferred from comparison to Fourier analyses of
theoretical light curves extracted from hydrodynamic pulsation models (\eg,
Simon \& Clement 1993).  These have been computed as explained by Corwin
\etal\ \  Note that validity of the Simon \& Clement results for RRc variables
has recently been questioned by Catelan (2004b).  Also, for the RRab
variables, the physical parameters should be taken seriously only for those with
the Jurcsik \& Kov\'acs compatibility parameter $D_m < 5$.

\begin{deluxetable*}{lcccccccl}
\footnotesize
\tablecaption{NGC~4147:  Fourier parameters for suspected variables}
\tablecolumns{9}
\tablenum{11}
\tablehead{
\colhead{ID} 
&\colhead{$A_{1}$}
&\colhead{$A_{21}$}
&\colhead{$A_{31}$}
&\colhead{$A_{41}$}
&\colhead{$\phi_{21}$}
&\colhead{$\phi_{31}$}
&\colhead{$\phi_{41}$}
&\colhead{Note}
}
\startdata
\multicolumn{9}{c}{ab-type variables} \\
V1  & 0.438 & 0.400 & 0.294 & 0.194 & 2.22 & 4.70 & 0.99 \\
V2  & 0.384 & 0.564 & 0.254 & 0.107 & 2.09 & 4.01 & 5.77 \\
V6  & 0.318 & 0.475 & 0.305 & 0.182 & 2.46 & 5.23 & 1.66 \\
V7  & 0.364 & 0.499 & 0.367 & 0.272 & 2.33 & 4.81 & 1.12 \\
V12 & 0.428 & 0.378 & 0.291 & 0.198 & 2.44 & 4.81 & 0.94 & all modern data \\
V12 & 0.786 & 0.616 & 0.282 & 0.090 & 3.34 & 0.46 & 4.27 & AF04 data only \\
\multicolumn{9}{c}{c-type variables} \\
V3  & 0.241 & 0.224 & 0.061 & 0.023 & 4.63 & $2.80\,\pm\,0.14$ & 1.16 \\
V4  & 0.214 & 0.185 & 0.062 & 0.067 & 4.55 & $2.58\,\pm\,0.48$ & 5.09 \\
V8  & 0.189 & 0.199 & 0.067 & 0.018 & 4.62 & $1.95\,\pm\,0.38$ & 0.63 \\
V10 & 0.212 & 0.096 & 0.045 & 0.037 & 4.55 & $3.64\,\pm\,0.21$ & 2.71 \\
V11 & 0.194 & 0.032 & 0.102 & 0.042 & 6.07 & $4.88\,\pm\,0.20$ & 2.66 \\
V13 & 0.222 & 0.046 & 0.143 & 0.111 & 5.03 & $4.37\,\pm\,0.23$ & 3.40 \\
V14 & 0.182 & 0.100 & 0.109 & 0.075 & 5.06 & $4.22\,\pm\,0.30$ & 0.95 \\
V16 & 0.082 & 0.505 & 0.309 & 0.588 & 4.06 & $3.16\,\pm\,0.56$ & 0.94 \\
V17 & 0.170 & 0.116 & 0.152 & 0.022 & 4.82 & $3.86\,\pm\,0.20$ & 6.08 \\
V19 & 0.144 & 0.157 & 0.054 & 0.101 & 3.73 & $2.74\,\pm\,0.75$ & 0.44 \\
\enddata
\end{deluxetable*}

\begin{deluxetable}{lcccccl}
\footnotesize
\tablecaption{NGC~4147:  Inferred physical parameters for ab-type variables}
\tablecolumns{7}
\tablenum{12}
\tablehead{
\colhead{ID} 
&\colhead{$D_m$}
&\colhead{[Fe/H]}
&\colhead{$M_V$}
&\colhead{$\left< \bmv\right>$}
&\colhead{$\left< \vmi\right>$}
&\colhead{Note}
}
\startdata
V1  & 46.1 & --1.42 & 0.80 & 0.304 & 0.449 \\
V2  & 162. & --2.30 & 0.76 & 0.310 & 0.457 \\
V6  & 3.21 & --1.29 & 0.76 & 0.348 & 0.506 \\
V7  & 3.93 & --1.34 & 0.82 & 0.331 & 0.482 \\
V12 & 46.4 & --1.29 & 0.81 & 0.306 & 0.452 & all modern data \\
V12 & 164. & --1.13 & 0.19 & 0.223 & 0.349 & AF04 only \\
\enddata
\end{deluxetable}

\begin{deluxetable}{lcccc}
\footnotesize
\tablecaption{NGC~4147:  Inferred physical parameters for c-type variables}
\tablecolumns{5}
\tablenum{13}
\tablehead{
\colhead{ID} 
&\colhead{Mass}
&\colhead{$\log L$}
&\colhead{$\log T_{\hbox{\footnotesize eff}}$}
&\colhead{$Y$}
}
\startdata
V3  & 0.624 & 1.674 & 3.870 & 0.282 \\
V4  & 0.683 & 1.717 & 3.865 & 0.267 \\
V8  & 0.771 & 1.720 & 3.866 & 0.263 \\
V10 & 0.568 & 1.728 & 3.861 & 0.270 \\
V11 & 0.436 & 1.699 & 3.862 & 0.286 \\
V13 & 0.509 & 1.752 & 3.856 & 0.267 \\
V14 & 0.493 & 1.699 & 3.863 & 0.282 \\
V16 & 0.660 & 1.780 & 3.855 & 0.252 \\
V17 & 0.554 & 1.743 & 3.858 & 0.266 \\
V19 & 0.625 & 1.666 & 3.872 & 0.284 \\
\enddata
\end{deluxetable}

We are reluctant to give undue emphasis to the physical parameters derived from
Fourier decomposition because (i)~the RRc relations are based on formulae that
appear to be incapable, for instance, of simultaneously providing masses and 
luminosities (\ie, they incorrectly predict the period-mean density 
relation, Catelan 2004b), and (ii)~the $D_m$ criterion which is used to
refine the RRab sample is also suspect, since it fails to effectively eliminate
Blazhko variables (\eg, Cacciari \etal\ 2005).  Nevertheless, accepting the
results in Table~13 at face value, they imply a mean mass for the RRc variables
of 0.59$\,$M${}_\odot$, a mean luminosity of 1.72$\,L_\odot$, and a mean
effective temperature of 7300$\,$K.  We can compare these results with those
compiled by Corwin \etal\ for several clusters.  The mean effective temperature
appears consistent with the cluster metallicity (indeed, with any [Fe/H] value
in the range --1.5 to --1.9).  The same conclusion applies to the mean mass,
although the luminosity is most consistent with an abundance close to the
metal-rich end of this range.  Among the clusters listed by Corwin \etal, these
physical properties place \ngc{4147} close to the boundary dividing the Oo~I
class from Oo~II---again, between M$\,$3 and M$\,$55.

In the case of the RRab stars (Table~12), we are hampered by the fact that only
two of them satisfy the nominal acceptance criterion, $D_m < 5$.  From these we
find a mean metallicity [Fe/H] = --1.32 on the Jurcsik (1995) metallicity scale
($\approx -1.54$ on the Zinn \& West scale), and an absolute visual magnitude of
0.79.  The inferred metallicity is, of course, somewhat high compared to
published values and compared to our photometric indicators discussed above. 
The derived absolute magnitude seems only marginally brighter (by
0.02$\,$mag--0.03$\,$mag) than what has been found by the same method for
metal-intermediate Oo~I clusters, but fainter than what was found for M$\,$55
(Oo~II) by some 0.1$\,$mag.  Again, we stress that these conclusions are
rendered more than normally uncertain by the fact that they are derived from
only two seemingly acceptable RRab stars, at least one of which---V6---appears to
show a varying amplitude.  

\subsection{Possible Second-Overtone Variables and the Period-Luminosity Relation in $I$}

Fig.~26 above also reveals the presence of a peak in the RR Lyrae period
distribution at a period $P_{\rm f} \approx 0.37$~d, corresponding to
short-period variables with periods $P \approx 0.27-0.28$~d. As discussed by
Alcock \etal\ (1996) and Clement \& Rowe (2000), this is the period range
characterizing candidate {\it second-overtone\/} RR~Lyrae (RRe) stars.
However, Kov\'acs (1998) argues that such variables may be more naturally
explained as the short-period wing of the RRc distribution.

One possible way to constrain the RRe possibility is to check the position
of the candidate second-overtone stars in the (fundamentalized)
period-luminosity diagram.  In Fig.~27 we show such a diagram, based on
our $I$-band mean magnitudes (\S5.5 above) and the ``modern'' periods (\S5.2),
with the pulsation type indicated for each star. Even though the candidate
RRe stars have been fundamentalized as if they were actually first-overtone
pulsators, they still seem to follow the trend defined by the RRc and
RRab stars, thus suggesting that they may indeed be more
straightforwardly explained as short-period RRc stars; see also Catelan
2004b for a similar argument in the cases of IC$\,$4499 (= C1452-820)
and M$\,$92 (= \ngc{6341}\ = C1715+432).  Note also that, in contrast with the
very small amplitudes that characterize candidate RRe stars (Clement \&
Rowe 2000), the three stars with the shortest periods in Fig.~27 do not appear
to show unusually small amplitudes, thus further supporting the likelihood that
they are simply short-period RRc stars.  

The sloping solid line in Fig.~27 is the predicted theoretical relation for a
metallicity $Z = 0.0005$, corresponding to a ${\rm [Fe/H]} = -1.84$ for an
assumed $\alpha$-element enhancement of $[\alpha/{\rm Fe}] = +0.3$ (see Salaris 
\etal\ 1993).  This theoretical result is based on Eqs.~(1) and
(2) and Table~9 in Catelan \etal\ (2004). A distance modulus
$(m-M)_V = 16.39$~mag has been assumed (Harris 1996; readers will recall that
Harris's tabulated value for $V_{HB}$ in \ngc{4147} agreed extremely well with
ours), corresponding to $(m-M)_0 = 16.33\,$mag and (as indicated in the plot)
$(m-M)_I = 16.36$~mag. As can be seen, there is excellent agreement between the
distance modulus tabulated in the Harris catalog and the Catelan \etal\
theoretical calibration.  This result can be compared with that obtained on the
basis of the same models from the ZAHB estimate provided in \S4.2. There we
found that $V_{\rm ZAHB} = 17.02$, whereas the Catelan et al. models give
$M_V^{\rm ZAHB} = 0.62$. This implies a distance modulus $(m-M)_V = 16.40$, in
perfect agreement with the Harris (1996) catalog and with the analysis of the
$I$-band period-luminosity relation.

\subsection{Non-Variable Stars near the Instability Strip}

Table~7 above also lists positions for seven stars in the field of \ngc{4147}
that appear to be non-varying even though they lie very near the
instability strip in the ground-based data.  They were selected by the criteria 

\begin{center}
$r < 200$\sec,\\
$16.80 < V < 17.20$, and\\
$0.400 < \bmi < 1.000.$
\end{center}

\noindent These possibly constant stars are also marked in Fig.~21.  They
have been numbered C1 through C7 in order of increasing right ascension, that
is, from right to left in the finding chart.  Stars C1, C2, and C7 are clearly
visible in the figure but C3 through C6 are in the crowded cluster center. 
As mentioned above, stars V5, V9, and V15 are most likely not variables.  These,
then, are three more apparently non-varying stars near the instability strip. 
We list the photometric indices of these stars in Table~14.  It seems
that apparently constant stars C2, V5, and V15, on the one hand, and variable
stars V2, V3, and V19, on the other (Table~9), delimit the blue edge of the
instability strip quite precisely at \bmv~=~0.17 or $(\bmv)_0 = 0.15$, with an
uncertainty of $\pm0.01$ or less.  These stars are all quite reasonably
uncrowded, with $\hbox{\it sep\/} > 3$.

\begin{deluxetable*}{lccccc}
\footnotesize
\tablecaption{NGC~4147:  Photometry for possible constant stars 
near the instability strip}
\tablecolumns{6}
\tablenum{14}
\tablehead{
\colhead{ID} 
&\colhead{$V$} 
&\colhead{\bmi} 
&\colhead{\bmv} 
&\colhead{\vmr} 
&\colhead{\vmi} 
}
\startdata
\multicolumn{6}{c}{Ground-based data} \\
C1  & 16.933$\pm$0.0009 & 0.996$\pm$0.0034 & 0.399$\pm$0.0030 & 0.293$\pm$0.0017 & 0.597$\pm$0.0019 \\
C2  & 16.983$\pm$0.0018 & 0.428$\pm$0.0044 & 0.172$\pm$0.0040 & 0.100$\pm$0.0026 & 0.256$\pm$0.0032 \\
C3  & 17.031$\pm$0.0047 & 0.610$\pm$0.0120 & 0.208$\pm$0.0084 & 0.185$\pm$0.0068 & 0.402$\pm$0.0108 \\
C4  & 16.829$\pm$0.0043 & 0.441$\pm$0.0120 & 0.118$\pm$0.0110 & 0.123$\pm$0.0082 & 0.323$\pm$0.0078 \\
C5  & 17.122$\pm$0.0035 & 0.638$\pm$0.0146 & 0.185$\pm$0.0121 & 0.223$\pm$0.0071 & 0.453$\pm$0.0095 \\
C6  & 16.938$\pm$0.0061 & 0.497$\pm$0.0156 & 0.134$\pm$0.0119 & 0.159$\pm$0.0088 & 0.363$\pm$0.0133 \\
C7  & 16.900$\pm$0.0014 & 0.904$\pm$0.0039 & 0.357$\pm$0.0038 & 0.260$\pm$0.0022 & 0.547$\pm$0.0023 \\
V5  & 16.961$\pm$0.0032 & 0.482$\pm$0.0092 & 0.187$\pm$0.0077 & 0.127$\pm$0.0041 & 0.295$\pm$0.0067 \\
V9  & 17.017$\pm$0.0015 & 0.305$\pm$0.0059 & 0.107$\pm$0.0042 & 0.068$\pm$0.0027 & 0.198$\pm$0.0046 \\
V15 & 16.981$\pm$0.0028 & 0.454$\pm$0.0060 & 0.173$\pm$0.0053 & 0.123$\pm$0.0040 & 0.281$\pm$0.0049 \\
V18 & 13.929$\pm$0.0065 & 2.982$\pm$0.0170 & 1.444$\pm$0.0127 & 0.838$\pm$0.0096 & 1.538$\pm$0.0146 \\
\multicolumn{6}{c}{WFPC2 data} \\
C1  & 16.898$\pm$0.0213 &  --- &   0.321$\pm$0.0321 &  --- &  --- \\
C3  & 17.478$\pm$0.0357 &  --- &   0.022$\pm$0.0419 &  --- &  --- \\
C4  & 17.194$\pm$0.0261 &  --- &   0.066$\pm$0.0371 &  --- &  --- \\
C5  & 17.699$\pm$0.0193 &  --- &   0.008$\pm$0.0289 &  --- &  --- \\
C6  & 17.422$\pm$0.0238 &  --- &   0.044$\pm$0.0368 &  --- &  --- \\
V5  & 16.906$\pm$0.0452 &  --- &   0.172$\pm$0.0520 &  --- &  --- \\
V15 & 17.012$\pm$0.0484 &  --- &   0.169$\pm$0.0558 &  --- &  --- \\
\enddata
\end{deluxetable*}

The WFPC2 results suggest that stars C3, C4, C5, and C6 are really normal blue
horizontal-branch stars that have been measured too red from the ground due to
blending with redder subgiants or turnoff stars.  (The reader is reminded that
the photometric errors assigned by ALLFRAME to stars in crowded regions are
lower limits.)  Stars C1 and C7 lie far enough from the cluster core
to be easily measurable from the ground and they appear to be uncrowded
($\hbox{\it sep} = 7.6$ and 4.5, respectively).  They both lie
rather near the red edge of the instability strip: Cacciari \etal\ (2005) find
at least some RR~Lyrae stars redder than $\bmv\,=\,0.40$ in M$\,$3.  C1 and C7
are close enough to the cluster that they are quite probably members:  the
surface density of such blue stars at this apparent magnitude level is very
small.  This may be confirmed by examination of Fig.~5, which represents an area
roughly 25 times larger than the 400\sec-diameter area where we carried out this
search.  In fact, C1 and C7 both lie within a maximum distance of 73\sec\ from
the adopted cluster center; this circle represents about 0.5\% of the
photometric survey area.  

But are these stars really near the instability strip, or have they been placed
there because of photometric errors, such as might be caused by blending?  It is
possible that the Two Micron All Sky Survey (2MASS) Point Source Catalog can
help us with this question.  We have searched in that catalog for detections
near our measured positions for the variable candidates and the constant stars
near the instability strip.  We also conducted the search in the reverse sense: 
for every 2MASS catalog entry within 6\min\ of our computed cluster center, we
have searched for the best possible cross-identification within our catalog. 
When the two cross-matches agreed, we concluded we had a possible
identification; when they disagreed, there was never any ambiguity about which
cross-match was more probable, based upon the relative separations and the
implied (optical)--(infrared) colors.  

Table~15 lists the results of this process.  For each provisional
cross-identification within a maximum match-up tolerance of 4\Sec0, it lists our
star name; the minutes and seconds of right ascension and the arcminutes and
arcseconds of declination for the proposed 2MASS counterpart; its infrared
magnitudes; inferred \imj\ and \vmk\ colors; and the angular distance
between the optical and infrared positions.  Obviously, the $V$ and $I$
magnitudes have been taken from our work, and the {\it JHK\/} magnitudes from
the 2MASS Point Source Catalog.  The uncertainties of the 2MASS magnitudes are
large, and not all of our stars have 2MASS counterparts; evidently these objects
are very near the 2MASS detection limit, and in fact the actual errors of the
infrared measurements may be larger than indicated due to the patchy underlying
cluster light.  We shall make an effort not to overinterpret the 2MASS data.

\begin{figure}[t]
  \figurenum{28}
  \plotone{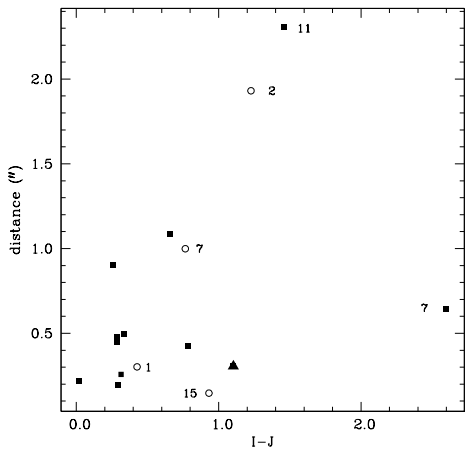}
\caption{A diagram testing the plausibility of our proposed
identifications between our catalog of objects within 6\min\ of the center
of \ngc{4147}, and entries in the Two Micron All Sky Survey lying within the
same area.  For each proposed identification, the separation between the
optical and infrared positions (vertical axis) is plotted against the
inferred \imj\ color.  The large triangle represents star V18, which lies
near the tip of the cluster giant branch.  Filled squares represent
possible identifications of catalogued variable stars, empty circles
represent possible identifications of constant stars near the instability
strip.  Labeled points are discussed in the text.}
      \label{Fig28}
\end{figure}

Fig.~28 plots the apparent separation on the sky between our star and
the closest 2MASS point source against the inferred \imj\ color.  The large
triangle near the bottom of the figure represents V18, which lies near the tip
of the \ngc{4147} giant branch.  Apart from this one star, all the other objects
in the diagram are supposed to be in or near the instability strip---\ie, much
bluer than V18.  Therefore, assuming that our cross-identification of {\it
this\/} object is correct, then it would be unreasonable to expect that any of
the other stars should be redder than it {\it if\/} the provisional
cross-identification between the optical object and the infrared catalog entry
is also correct.  On this basis, we can probably say with confidence that the
two provisional cross-identifications where the optical and infrared positions
differ by more then 1\Sec2 are both incorrect, since their inferred \imj\ colors
are redder than that of V18.  These are variable candidate V11 and possible
constant star C2; the 2MASS indices probably do not refer to these objects, and
can not tell us anything new about them.  The infrared sources are sufficiently
distant from the optical detections that even blending between them is probably
not a serious issue.

Among the stars whose positions agree to less than 1\Sec2, we have labeled four:
the putatively stable stars C1, C7, and V15 (empty circles) and the
RRab star V7 (filled square).  Clearly the inferred color for V7 is
anomalous.  In reversing the search, \ie, in looking for optically detected
objects near 2MASS sources rather than looking for 2MASS sources near our
variable stars, we did find that V7 is in fact the brightest, closest optical
object near the infrared detection.  It lies 7\Sec7 distant from our computed
position of the cluster center, and is contained within the WFPC2 coverage of
\ngc{4147}.  Within two arcseconds of its position there are 26 other objects 
with $V < 22.2$ in the WFPC2 detection catalog;  the closest and brightest of
these (defined as the one that produces the most severe contamination in
one-arcsecond seeing conditions) is a subgiant/turnoff star with $V=19.81$ and
$\bmv=0.47$.  In terms of the fraction of photons delivered to the position of
V7's centroid on the sky in one-arcsecond seeing, this companion is fainter than
V7 by 3.2$\,$mag in the $V$ photometric bandpass; correspondingly, it would
contaminate V7 at a 5\%\ level in $V$ if its contribution were unrecognized.  The other
companions would each contaminate V7 at a level of less than 1.6\% each.  Even
if all 26 of the companions were added together, and even allowing for the
fact that they are probably all cooler than V7 and, hence, not so faint---relatively
speaking---in the infrared bandpasses, it is not clear that the superposition of
these stars with V7 could produce the $J$-band magnitude of the 2MASS detection. 
The 2MASS source likely represents, rather, the integrated photometry of an
ensemble of stars considerably larger than 4\sec\ in diameter surrounding the
position of V7.  At this point we have no evidence that the putatively stable
stars, C1, C2, and V15, show any evidence of severe crowding like that affecting
V7.

\begin{deluxetable*}{lccccccccccc}
\footnotesize
\tablecaption{NGC~4147:  2MASS infrared photometry for candidate variables and possible constant stars near the instability strip}
\tablecolumns{12}
\tablenum{15}
\tablehead{
\colhead{ID} 
&\colhead{RA--12\hr} & \colhead{Dec--18\deg} 
&\colhead{$J$} & \colhead{$\sigma_J$}
&\colhead{$H$} & \colhead{$\sigma_H$}
&\colhead{$K$} & \colhead{$\sigma_K$}
&\colhead{\imj} 
&\colhead{\vmk} 
&\colhead{r(\sec)}
}
\startdata
\multicolumn{12}{c}{Variable stars} \\
V1  & 09~59.37 & 31~48.4 & 16.28 & 0.10 & 16.33 & 0.24 & 15.98 & 0.23 & 0.29 & 1.02 & 0.20 \\
V2  & 10~04.94 & 32~04.5 & 16.56 & 0.12 & 16.14 & 0.21 & 16.42 & 0.35 & 0.02 & 0.61 & 0.22 \\
V3  & 10~04.45 & 31~58.5 & 16.03 & 0.08 & 15.84 & 0.16 & 15.78 & 0.18 & 0.66 & 1.22 & 1.09 \\
V4  & 10~06.39 & 32~49.8 & 16.30 & 0.14 & 14.49 &  ---  & 14.47 &  ---  & 0.32 & 2.50 & 0.26 \\
V7  & 10~06.61 & 32~39.8 & 13.75 &  ---  & 14.79 & 0.25 & 13.24 &  ---  & 2.60 & 3.56 & 0.64 \\
V10 & 10~03.74 & 31~47.9 & 16.30 & 0.10 & 15.72 & 0.15 & 15.90 & 0.22 & 0.29 & 1.09 & 0.48 \\
V11 & 10~05.51 & 31~52.2 & 16.10 & 0.09 & 16.02 & 0.200 & 15.95 & 0.23 & 0.34 & 0.91 & 0.50 \\
V14 & 10~06.94 & 32~32.5 & 15.03 & 0.16 & 15.27 & 0.28 & 15.16 & 0.24 & 1.46 & 1.78 & 2.31 \\
V16 & 10~07.32 & 32~39.9 & 15.35 & 0.12 & 15.26 & 0.15 & 15.43 & 0.18 & 0.78 & 1.14 & 0.43 \\
V17 & 10~10.64 & 34~50.9 & 16.26 & 0.10 & 16.13 & 0.21 & 15.64 &  ---  & 0.29 & 1.29 & 0.45 \\
V18 & 10~05.61 & 32~11.4 & 11.29 & 0.03 & 10.55 & 0.03 & 10.41 & 0.02 & 1.10 & 3.52 & 0.31 \\
V19 & 10~21.92 & 35~01.5 & 16.47 & 0.12 & 16.38 & 0.21 & 15.72 &  ---  & 0.25 & 1.31 & 0.90 \\
\multicolumn{12}{c}{Constant stars} \\
C1  & 10~01.77 & 32~00.2 & 15.91 & 0.08 & 15.57 & 0.12 & 15.80 & 0.19 & 0.43 & 1.13 & 0.30 \\
C2  & 10~04.11 & 32~40.8 & 15.50 & 0.07 & 15.23 & 0.09 & 15.27 & 0.13 & 1.23 & 1.71 & 1.93 \\
C7  & 10~08.04 & 32~10.7 & 15.59 & 0.08 & 15.37 & 0.12 & 15.16 & 0.14 & 0.77 & 1.74 & 1.00 \\
V15 & 10~06.99 & 32~24.9 & 15.77 & 0.18 & 14.53 &  ---  & 14.48 &  ---  & 0.93 & 2.50 & 0.15 \\
\enddata
\end{deluxetable*}

\begin{figure}[t]
  \figurenum{29}
  \plotone{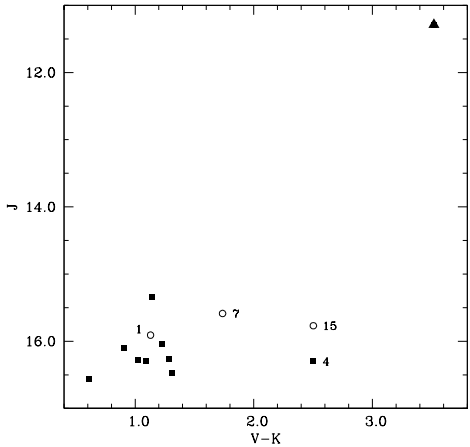}
\caption{An optical-infrared $(\vmk,J)$ color-magnitude
diagram for \ngc{4147}\ stars provisionally identified in the 2MASS
Point Source Catalog.  Point types are as in Fig.~27, and labeled points
are discussed in the text.}
      \label{Fig29}
\end{figure}

Fig.~29 is a (\vmk,$J$) color-magnitude diagram for the same stars
excepting V7, V11, and C2, since it is clear from the evidence of Fig.~28
that the infrared indices can not be assumed to apply meaningfully to these
objects.  As before, the large filled triangle represents the star near
the giant-branch tip, V18.  Four other objects have been labeled:  variable
candidate V4 (solid square), and allegedly constant stars C1, C7, and V15.
The optical colors appear to place V15 at the blue edge of the cluster's 
instability strip, but its infrared indices imply a much cooler temperature.
The star lies 12\Sec7 distant from our adopted cluster center, and there are
13 other WFPC2 detections within 2\Sec0, the closest and brightest of which
contaminates V15 at the 1\% level in one-arcsecond seeing.  As with V7, the
2MASS flux must represent a patch of sky appreciably larger than 4\sec\ in
diameter.  Variable candidate V4 does not appear in the WFPC2 field coverage
even though it lies within 17\sec\ of the cluster center; it fell in the crack
between detectors PC1 and WFC2.  Constant-star candidate C1 gives
every indication of being correctly identified in the 2MASS catalog, and in
infrared colors it is surrounded on all sides by RR~Lyrae stars, although it
is somewhat redder than they are in the optical indices.  Candidate C7 is a
slightly more ambiguous case, in that the identity between the optical and
infrared catalogs is a bit less secure.  If the identification is correct,
however, the optical and infrared colors suggest that it lies to the red of the
instability strip, and is therefore a likely red horizontal-branch star in the
cluster.  Taken together, C1 and C7 on the one hand, and V6 and V12 on the
other, appear to require the red edge of the instability strip in \ngc{4147} to
lie at $\bmv = 0.36$ or $(\bmv)_0 = 0.34$ with an uncertainty not worse than
0.04$\,$mag.  

Here we summarize the comparison of the 2MASS infrared photometry with our
optical photometry.  We conclude that the provisional matches for C2 and V11 are
incorrect:  their separations on the sky are relatively large, and their
infrared colors are unreasonably red for their optical colors.  The infrared
counterparts to V7, V15, and probably V4 are either spurious or, more likely,
rendered useless by confusion, because their apparent infrared fluxes are
unphysically large for stellar photospheres corresponding to their optical
colors.  In the case of V7 and V15, it is possible to use the WFPC2 images to
estimate the level of confusion with nearby stars, and in each case it is
difficult to account for the infrared flux merely by summing the predicted
contributions of the stars detected within a few arcseconds:  either the
confusion circle must be somewhat larger than 4\sec, or there must be additional
infrared sources not visible in the WFPC2 images, or the 2MASS photometric
errors for these stars must be larger than thought ($\sim 1\,$mag).  The
remaining entries in Table~15 are plausible cross-identifications on both
astrometric and photometric grounds.  Encouraged by this comparison that these
remaining stars have been well photometered, we are able to place quantitative
constraints on the colors of the blue and red edges of the instability strip. 
No constant star has been shown to lie definitively within the instability strip
on the basis of both optical {\it and\/} infrared colors, although some 
stars---C1 and C7, for instance---are very near or in the instability strip in
optical {\it or\/} infrared colors but not both.  These conclusions are
weaker than might otherwise be the case, due to the low mass and relative
remoteness of \ngc{4147}: its horizontal-branch magnitude appears to lie very
near the 2MASS detection limit, the total number of horizontal-branch stars is
small, and confusion is an issue for many of them because the angular extent of
the cluster is also small.  Similar optical-infrared analyses of closer and
larger clusters may well be more informative than this one has been.  

\section{DISCUSSION}

\subsection{Oosterhoff Classification} 

Castellani \& Quarta (1987) classified \ngc{4147}\ as Oosterhoff (1939) Type I
(Oo~I), making it one of the Galactic globular clusters with the bluest HBs ever
to be thus classified, and also the most metal-poor Oo~I cluster. As discussed
by Contreras \etal\ (2005), such a classification seems inconsistent with the
current theoretical paradigm, which holds that RR~Lyrae stars in Oo~I globular
clusters (comparatively metal-rich systems with predominantly red or
intermediate HBs and $\left<P_{ab}\right> \sim 0.55\,$d) are relatively
unevolved objects, whereas those in Oo~II globulars (metal-poor clusters with
predominantly blue HBs and $\left<P_{ab}\right> \sim 0.65\,$d) are evolved
from a position on the blue ZAHB (\eg, Lee \etal\ 1990; Clement \&
Shelton 1999).  However, when this potential conflict with the evolutionary
interpretation was identified, the possibility was raised that at least some of
the RR~Lyrae periods reported in the literature for \ngc{4147}\ were in fact
incorrect (Clement 2000).  While Catelan (2005) has shown that the Oosterhoff
dichotomy---as defined by the mean period of the RRab variables---is
present even among Galactic globular clusters with $N_{ab} \geq 5$, it is
important to check the position of the variables in the Bailey
(period-amplitude) diagram to place firmer constraints on the Oosterhoff status
of a cluster or galaxy (Catelan 2004b and references therein).

\begin{figure*}[t]
  \figurenum{30}
  \plotone{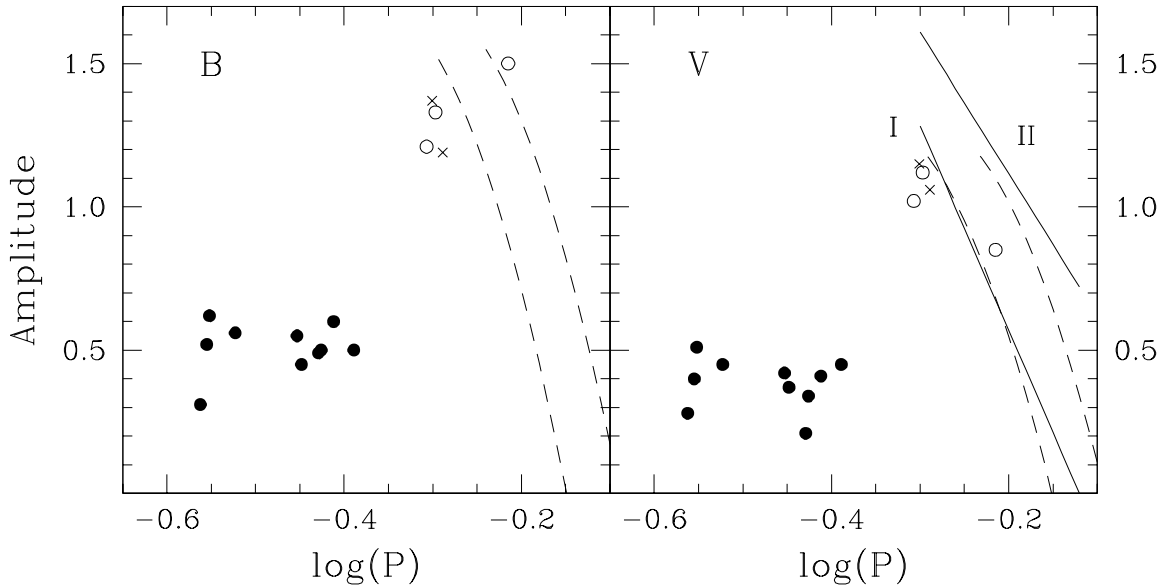}
\caption{The traditional Bailey diagram representing the total
amplitude of variation in the $B$ (left) and $V$ (right) photometric bandpasses
plotted against the logarithm of period, for ab-type (crosses for variables
with apparently constant amplitudes, open circles for those with variations
in amplitude and/or phase) and c-type RR~Lyrae variables in \ngc{4147}.  Curves
represent the fiducial ridge-lines for ab variables on Oosterhoff I- and II-type
clusters according to Clement \& Rowe (2000; solid) and Cacciari \etal\ (2005;
dashed).}
      \label{Fig30}
\end{figure*}

The two panels of Fig.~30 are versions of the traditional Bailey
diagram relating the logarithm of the period to the amplitude in the (in this
case) $B$ (left) and $V$ (right) photometric bandpasses.  The data for this
diagram come from our Tables~8 and 10 above: we have used the ``modern'' periods
and the amplitudes derived from the merger of our data with those of AF04.  
Filled circles represent the c-type (first overtone) pulsators with 
periods less than 0.45$\,$d; crosses and empty circles are for the ab-type
(fundamental mode) variables, with the empty circles designating the stars
with Blazhko-like amplitude variations in the available data.  The
solid curves in the right-hand panel of Fig.~28 are the standard lines for
RRab stars in M$\,$3 (prototype Oosterhoff class I) and $\omega$~Centauri
(Oosterhoff class II) from Clement \& Rowe (2000); the dashed curves in both
panels are the standard sequences for Oo~I and II proposed by Cacciari \etal\ 
(2005), with Oo~I and Oo~II being the left and right curves,
respectively.  The one star that stands near the Oo~II sequence in the $B$-band
Bailey diagram is the Blazhko-like star V6, which in these data has an anomalously
high amplitude in $B$.  In the $V$-band data---which are much more extensive
because we are able to include the data of AF04 with our own---the perceived
amplitude of this star is more moderate.  In the aggregate, however, the {\it
ab}-type variables of \ngc{4147} clearly lie closer to the Oo~I relations
despite the cluster's comparatively low metal abundance and predominantly blue
horizontal branch.  

The present analysis and that of AF04 have confirmed the observation of
Clement (2000) that many of Newburn's (1957) periods were uncertain and probably
incorrect;  in fact, essentially {\it all\/} the periods that Newburn flagged as
uncertain have turned out to be seriously imprecise or wrong. However,
despite these necessary revisions the Oo~I classification of the cluster stands
as a firm result.  Our analysis gives a $\langle P_{\rm ab}\rangle = 0.525$~d for
the five confirmed RRab variables, and $\langle P_{\rm c}\rangle = 0.339$~d for the
ten RRc (including the candidate RRe's). These values
are both clearly consistent with an Oo~I classification, and incompatible with
an Oo~II type.  

We note again that our analysis of the morphology of the evolved sequences in
the CMD of \ngc{4147} suggests that its metal abundance may be somewhat higher
than previously thought:  higher than the abundance of M$\,$55 and thus
intermediate between that cluster (Oo~II) and M$\,$3 (Oo~I).  The observed color
of the main sequence in \ngc{4147} leads to a similar conclusion {\it
provided\/} we accept that the interstellar reddening toward M$\,$55 is
considerably higher than the value listed in Harris's compilation catalog, but
very close to the value given by the Schlegel \etal\ reddening maps.  The
properties tentatively inferred from the Fourier decomposition of the RR Lyrae
light curves of \ngc{4147} also suggest a greater similarity to the variables of
M$\,$3 than to those of M$\,$55.

An intriguing aspect related to the Oo~I status of the cluster is its very large
RRc number fraction, $f_{\rm c} = 0.667$, which stands in sharp contrast with
the much smaller values ($f_{\rm c} \ltsim 0.3$) typically found in Oo~I
globular clusters with intermediate or predominantly red HB types. In fact, the
RRc number fraction seems to fall much more in line with the predominantly blue
HB of the cluster. The same combination of predominantly blue HB morphology,
large RRc number fraction and Oo~I status has recently also been confirmed for
the Galactic globular cluster M62 (\ngc{6266} = C1658--300) by Contreras et al. 
(2005).  We therefore favor the interpretation that \ngc{4147} is intermediate
in metallicity between M$\,$3 and M$\,$55, and close to whatever conceptual
boundary it is that separates Oo~I clusters from Oo~II; it is predominantly on
the Oo~I side of that boundary, but with some extreme properties compared to
other Oo~I clusters, such as its comparatively low metallicity, mainly blue
horizontal branch, and high fraction of RRc-type variables.  

\subsection{Broader significance} 

NGC$\,$4147 has recently surfaced from a long period of relative oblivion to a
renewed interest, due to its claimed association with the Sagittarius dwarf
spheroidal (dSph) satellite of the Milky Way, which appears to be merging with
the main body of the Galaxy and supplying the latter with its stars and globular
clusters (Bellazzini \etal\ 2003a, 2003b, and
references therein).  The present detailed analysis of the color-magnitude
diagram and variable-star properties of the cluster may help us understand its
place within the framework of formation models of the Galactic halo. 

Examination of the cluster's color-magnitude diagram confirms the presence of a
predominantly blue HB and a fairly low metallicity, ${\rm [Fe/H]} \sim -1.8$~dex
on the Zinn \& West (1984) scale.  This value is consistent with that tabulated
by Harris (1996), although we note that the cluster's metallicity has been the
subject of some controversy. Zinn \& West estimated the metallicity of the
cluster on the basis of the integrated-light photometric parameter $Q_{39}$,
finding $\hbox{\rm [Fe/H]} = -2.01$. They noted that the previous analysis of
the flux distribution from red giants in the cluster by Bell \& Gustafsson
(1983) had provided a metallicity ${\rm [M/H]} = -1.59$~dex, so the value that
Zinn \& West finally adopted for the cluster was a compromise: ${\rm [Fe/H]} =
-1.80 \pm 0.26$~dex. More recently, Suntzeff \etal\ (1988) have used several
metallicity-sensitive spectral indices to obtain ${\rm [Fe/H]} = -1.85$~dex for
the cluster. The value currently given in the Harris catalog, ${\rm [Fe/H]} =
-1.83$~dex, appears to correspond to a straight average of the Zinn \& West and
Suntzeff \etal\ values.  AF04, on the basis of Fourier decomposition of the
fundamental-mode (ab-type) RR Lyrae variables in the cluster, arrived at a much
higher metallicity, namely ${\rm [Fe/H]} = -1.22 \pm 0.31$~dex.  Our own Fourier
decomposition results also support a fairly high metallicity for the cluster,
namely ${\rm [Fe/H]} \approx -1.32$~dex (based on two RRab variables with
relatively small $D_m$ values). However, it should be remembered that this value
is provided in the Jurcsik metallicity scale, and translates to ${\rm
[Fe/H]} = -1.54$~dex on the Zinn \& West scale.  Our interpretation of the
CMD morphology of the cluster favors an [Fe/H] value on this scale lower than
that of M$\,$3 (\ie, $< -1.57$) but higher than that of M$\,$55 (\ie, $\gtsim
-1.81$); perhaps a compromise value near --1.7 may lie within the confidence
intervals of the various photometric, spectroscopic, and pulsational analyses
of the stars in the cluster.

A similar discrepancy between photometric and Fourier-based metallicity values
has recently been discussed by Nemec (2004) in the case of the metal-poor
globular cluster \ngc{5053}\ (= C1313+179).  

In contrast with previous photometric studies of the cluster, we have been able
to examine a wide field that has provided an essentially complete assessment of
the cluster's HB.  Furthermore, we have also been able to reach about 2.5~mag
below the TO point, thus being able to establish---for the first time---a secure
age for \ngc{4147}\ relative to other, better-studied globular clusters. Our
results indicate that, compared to similar objects in the studies of Rosenberg
\etal\ (1999) and VandenBerg (2000), \ngc{4147}\ has a completely normal age. 
(In the course of this exposition, we have noted that the $\Delta V_{\rm
TO}^{\rm HB}$ value adopted by VandenBerg from previous studies of M$\,$55 is
probably overestimated by $> 0.1\,$mag.) If the cluster is indeed associated
with the Sagittarius dSph, this suggests that the latter was able to form
globular clusters over an extended period of time, since at least for
Palomar~12, which people have also associated with the Sagittarius dwarf
(Da Costa \& Armandroff 1995; Dinescu \etal\ 2000), there appears to be a
consensus regarding its relative youth (Gratton \& Ortolani 1988; Stetson \etal\
1989; Chaboyer \etal\ 1996; Rosenberg et al. 1999; VandenBerg
2000; Salaris \& Weiss 2002).  While the variation in mean globular cluster age
with metallicity remains somewhat debatable, it seems fairly certain that
Pal~12 is at least 25\%--33\% younger than \ngc{4147}.  

As concerns the RR~Lyrae variable stars, together with AF04 we have been able to
correct the problematic period values to which Clement (2000) called attention,
thus confirming---on the basis of the average ab- and c-type periods,
as well as the Bailey (period-amplitude) diagram for the
RRab's---Castellani \& Quarta's (1987) classification of the cluster as
type Oo~I. Unlike what is typically found in Oo~I clusters, \ngc{4147} has a
predominantly blue HB.  Interestingly, the c-to-ab number ratio is more
in line with the cluster's blue HB type than it is with its Oo~I classification.
This all suggests that the stars in the instability strip are relatively unevolved
from the ZAHB, but concentrated toward the blue side, unlike traditional Oo~I
clusters where the variables are near the ZAHB but uniformly distributed or
concentrated to the red side of the instability strip, and also unlike
traditional Oo~II clusters, where the ZAHB is populated primarily to the
blueward of the instability strip, and many stars do not become pulsationally
unstable until after considerable luminosity and temperature evolution.  A
similar phenomenon has been recently seen in the case of M62 (Contreras \etal\
2005). For a critical discussion of the role played by evolutionary effects in
producing Oo~II clusters, the reader is referred to Pritzl et al. (2002), and
a discussion of appropriate nomenclature is to be found in Catelan (2004a;
last paragraph on his p.~411).

Other Sagittarius-related globulars have been studied by Salinas \etal\ (2005)
and Cacciari \etal\ (2002); the combined results suggest that
those globular clusters that have been notionally associated with the
Sagittarius dwarf galaxy have fairly unusual RR~Lyrae pulsation characteristics
when compared with the remainder of the Galactic globular cluster system.  This
suggests (see also Catelan 2004b, 2005) that the Sagittarius merger is not truly
representative of the typical process that has led to the formation of the
present-day Galactic halo: the latter's oldest stellar populations would have
looked quite different if Sagittarius-like protogalactic fragments were
primarily responsible for its assembly.

\acknowledgments

PBS is very grateful to Michael~Bolte, Howard~Bond, Alfred~Rosenberg, and
Nicholas~Suntzeff, who generously donated some of the data that were used
in this study.  He continues to appreciate the Canadian Astronomy Data
Centre, the Isaac Newton Group Archive, and the ESO Science Archive, who
provided data for this and other ongoing studies.  Much of the writing of this
paper took place while PBS was enjoying the hospitality of the Osservatorio
di Roma at Monte Porzio Catone.

MC warmly thanks Armando~Arellano~Ferro for kindly providing his light curves in
machine-readable format; Carla~Cacciari and Christine~Clement for providing
their Oo~I and Oo~II lines in the Bailey diagram; and Don~VandenBerg for
interesting discussions.  Support for MC was provided by Proyecto FONDECYT
Regular No.~1030954.  

HAS thanks the National Science Foundation for support under grant AST~02-05813.

\clearpage

\LongTables

\begin{deluxetable*}{rrrrr}
\footnotesize
\tablecaption{Photometric transfer sequence for \ngc{4147} WFPC2 data: astrometry}
\tablecolumns{5}
\tablenum{2}
\tablehead{
\colhead{ID} &\colhead{X} &\colhead{Y} 
&\colhead{RA} &\colhead{Dec} \\
\colhead{} & \colhead{\sec} & \colhead{\sec} &
\colhead{2000.0} & \colhead{2000.0}
}
\startdata
 5481 & --194.6 &   65.1 & 12~10~00.11 & +18~32~27.5 \\
 5544 & --190.8 &   83.1 & 12~10~00.38 & +18~32~45.5 \\
 5587 & --186.8 &   76.3 & 12~10~00.65 & +18~32~38.6 \\
 5603 & --185.8 &   46.1 & 12~10~00.73 & +18~32~08.5 \\
 5649 & --183.2 &  112.2 & 12~10~00.91 & +18~33~14.5 \\
 5659 & --182.6 &   41.5 & 12~10~00.95 & +18~32~03.9 \\
 5723 & --178.7 &   66.7 & 12~10~01.23 & +18~32~29.1 \\
 5827 & --173.7 &   11.6 & 12~10~01.58 & +18~31~34.0 \\
 5851 & --172.3 &   78.0 & 12~10~01.67 & +18~32~40.4 \\
 5901 & --170.0 &   45.2 & 12~10~01.84 & +18~32~07.5 \\
 5976 & --166.9 &  119.3 & 12~10~02.05 & +18~33~21.6 \\
 6009 & --165.7 &   99.1 & 12~10~02.14 & +18~33~01.4 \\
 6026 & --164.7 &   91.3 & 12~10~02.21 & +18~32~53.7 \\
 6041 & --163.8 &   33.1 & 12~10~02.27 & +18~31~55.5 \\
 6066 & --162.7 &   89.3 & 12~10~02.35 & +18~32~51.7 \\
 6115 & --161.2 &   78.7 & 12~10~02.46 & +18~32~41.1 \\
 6123 & --160.7 &   41.8 & 12~10~02.49 & +18~32~04.2 \\
 6126 & --160.7 &  101.6 & 12~10~02.49 & +18~33~03.9 \\
 6132 & --160.5 &   49.3 & 12~10~02.50 & +18~32~11.7 \\
 6136 & --160.3 &  111.3 & 12~10~02.52 & +18~33~13.7 \\
 6139 & --160.3 &   29.1 & 12~10~02.52 & +18~31~51.5 \\
 6203 & --157.8 &  120.9 & 12~10~02.69 & +18~33~23.3 \\
 6260 & --155.9 &   36.8 & 12~10~02.83 & +18~31~59.2 \\
 6275 & --155.3 &  109.8 & 12~10~02.87 & +18~33~12.2 \\
 6311 & --153.8 &  110.2 & 12~10~02.97 & +18~33~12.5 \\
 6312 & --153.8 &  112.7 & 12~10~02.97 & +18~33~15.1 \\
 6402 & --150.6 &   82.9 & 12~10~03.20 & +18~32~45.3 \\
 6406 & --150.5 &    5.4 & 12~10~03.21 & +18~31~27.8 \\
 6487 & --147.8 &   19.3 & 12~10~03.40 & +18~31~41.7 \\
 6499 & --147.4 &   93.7 & 12~10~03.42 & +18~32~56.0 \\
 6516 & --146.5 &  109.5 & 12~10~03.49 & +18~33~11.9 \\
 6543 & --145.4 &   59.0 & 12~10~03.57 & +18~32~21.3 \\
 6579 & --144.5 &    1.0 & 12~10~03.63 & +18~31~23.4 \\
 6612 & --143.5 &  131.6 & 12~10~03.70 & +18~33~34.0 \\
 6651 & --142.3 &   95.5 & 12~10~03.79 & +18~32~57.9 \\
 6655 & --142.0 &  145.3 & 12~10~03.81 & +18~33~47.6 \\
 6665 & --141.5 &   81.5 & 12~10~03.84 & +18~32~43.9 \\
 6774 & --138.2 &    9.6 & 12~10~04.07 & +18~31~32.0 \\
 6797 & --137.7 &   78.9 & 12~10~04.11 & +18~32~41.3 \\
 6900 & --135.0 &   27.6 & 12~10~04.30 & +18~31~50.0 \\
 6913 & --134.7 &  146.3 & 12~10~04.31 & +18~33~48.7 \\
 6923 & --134.5 &  135.2 & 12~10~04.33 & +18~33~37.5 \\
 6936 & --134.2 &  120.2 & 12~10~04.35 & +18~33~22.5 \\
 6956 & --133.8 &   49.7 & 12~10~04.38 & +18~32~12.1 \\
 7061 & --131.6 &   36.5 & 12~10~04.54 & +18~31~58.8 \\
 7107 & --130.8 &   51.8 & 12~10~04.59 & +18~32~14.2 \\
 7194 & --128.7 &  105.8 & 12~10~04.74 & +18~33~08.2 \\
 7345 & --125.6 &  114.5 & 12~10~04.96 & +18~33~16.8 \\
 7399 & --124.6 &  159.0 & 12~10~05.03 & +18~34~01.3 \\
 7489 & --122.6 &  119.3 & 12~10~05.17 & +18~33~21.7 \\
 7553 & --121.3 &   68.9 & 12~10~05.26 & +18~32~31.3 \\
 7558 & --121.1 &   89.5 & 12~10~05.27 & +18~32~51.9 \\
 7563 & --121.0 &   67.4 & 12~10~05.28 & +18~32~29.8 \\
 7634 & --119.6 &   76.7 & 12~10~05.38 & +18~32~39.1 \\
 7669 & --118.9 &  150.6 & 12~10~05.43 & +18~33~53.0 \\
 7672 & --118.8 &   93.9 & 12~10~05.44 & +18~32~56.3 \\
 7693 & --118.2 &  141.1 & 12~10~05.48 & +18~33~43.5 \\
 7699 & --118.1 &  138.4 & 12~10~05.49 & +18~33~40.7 \\
 7723 & --117.6 &   69.9 & 12~10~05.52 & +18~32~32.3 \\
 7734 & --117.3 &   60.8 & 12~10~05.54 & +18~32~23.2 \\
 7742 & --117.1 &  118.8 & 12~10~05.55 & +18~33~21.2 \\
 7807 & --115.6 &  100.8 & 12~10~05.66 & +18~33~03.2 \\
 7830 & --115.1 &  127.0 & 12~10~05.70 & +18~33~29.4 \\
 7861 & --114.5 &   96.5 & 12~10~05.74 & +18~32~58.9 \\
 7887 & --113.9 &  122.0 & 12~10~05.78 & +18~33~24.4 \\
 7917 & --113.3 &   79.7 & 12~10~05.82 & +18~32~42.1 \\
 7941 & --112.9 &   60.3 & 12~10~05.85 & +18~32~22.7 \\
 7982 & --112.1 &  103.0 & 12~10~05.91 & +18~33~05.4 \\
 8009 & --111.6 &   98.4 & 12~10~05.95 & +18~33~00.8 \\
 8195 & --107.5 &  110.2 & 12~10~06.24 & +18~33~12.6 \\
 8199 & --107.4 &  142.8 & 12~10~06.24 & +18~33~45.2 \\
 8223 & --107.0 &  137.9 & 12~10~06.27 & +18~33~40.3 \\
 8277 & --106.0 &  123.0 & 12~10~06.34 & +18~33~25.4 \\
 8328 & --105.0 &  106.5 & 12~10~06.41 & +18~33~08.9 \\
 8347 & --104.6 &  119.5 & 12~10~06.44 & +18~33~21.9 \\
 8360 & --104.3 &   57.5 & 12~10~06.46 & +18~32~19.9 \\
 8625 &  --99.0 &   87.1 & 12~10~06.83 & +18~32~49.4 \\
 8654 &  --98.5 &   82.4 & 12~10~06.86 & +18~32~44.8 \\
 8687 &  --97.7 &   85.4 & 12~10~06.92 & +18~32~47.8 \\
 8741 &  --96.6 &   62.4 & 12~10~07.00 & +18~32~24.8 \\
 8748 &  --96.5 &   81.0 & 12~10~07.01 & +18~32~43.4 \\
 8766 &  --96.1 &  146.7 & 12~10~07.03 & +18~33~49.1 \\
 8773 &  --95.9 &   81.9 & 12~10~07.05 & +18~32~44.3 \\
 8870 &  --93.9 &  122.7 & 12~10~07.19 & +18~33~25.0 \\
 8873 &  --93.8 &  106.5 & 12~10~07.20 & +18~33~08.9 \\
 8888 &  --93.5 &  114.3 & 12~10~07.21 & +18~33~16.7 \\
 8999 &  --91.2 &   72.7 & 12~10~07.38 & +18~32~35.1 \\
 9128 &  --88.2 &   99.1 & 12~10~07.59 & +18~33~01.5 \\
 9321 &  --83.6 &  120.8 & 12~10~07.91 & +18~33~23.1 \\
 9485 &  --79.4 &  118.7 & 12~10~08.21 & +18~33~21.1 \\
 9647 &  --74.8 &  113.2 & 12~10~08.53 & +18~33~15.6 \\
\enddata
\end{deluxetable*}

\begin{deluxetable*}{rrrrrrrrr}
\footnotesize
\tablecaption{Photometric transfer sequence for \ngc{4147} WFPC2 data: photometry}
\tablecolumns{9}
\tablenum{3}
\tablehead{
\colhead{ID} &\colhead{$\left<V\right>$} &\colhead{$\sigma$} 
&\colhead{$\left<B\right>$} &\colhead{$\sigma$} &\colhead{$\Delta V$} &
\colhead{$\sigma$} &\colhead{$\Delta B$} &\colhead{$\sigma$}
}
\startdata
 5481 &  19.657 &  0.004 &  20.205 &  0.005 & --0.035 &  0.047 & --0.002 &  0.055 \\
 5544 &  19.337 &  0.003 &  19.921 &  0.006 & --0.066 &  0.056 & --0.054 &  0.056 \\
 5587 &  18.840 &  0.003 &  19.462 &  0.004 & --0.049 &  0.080 & --0.013 &  0.039 \\
 5603 &  20.318 &  0.005 &  20.723 &  0.017 & --0.042 &  0.092 &   0.001 &  0.052 \\
 5649 &  16.688 &  0.001 &  17.454 &  0.003 & --0.028 &  0.054 & --0.009 &  0.032 \\
 5659 &  20.125 &  0.005 &  20.550 &  0.007 & --0.048 &  0.088 & --0.032 &  0.047 \\
 5723 &  20.474 &  0.006 &  20.872 &  0.009 &   0.042 &  0.066 &   0.103 &  0.071 \\
 5827 &  16.535 &  0.001 &  17.329 &  0.003 &   0.031 &  0.051 &   0.001 &  0.033 \\
 5851 &  16.426 &  0.001 &  17.236 &  0.003 &   0.031 &  0.029 &   0.047 &  0.031 \\
 5901 &  17.300 &  0.002 &  17.296 &  0.004 &   0.084 &  0.051 &   0.015 &  0.026 \\
 5976 &  17.857 &  0.002 &  18.518 &  0.003 &   0.034 &  0.035 &   0.005 &  0.048 \\
 6009 &  19.160 &  0.003 &  19.748 &  0.005 & --0.049 &  0.040 & --0.058 &  0.050 \\
 6026 &  20.140 &  0.006 &  20.524 &  0.018 & --0.093 &  0.063 & --0.070 &  0.058 \\
 6041 &  19.935 &  0.004 &  20.371 &  0.008 &   0.063 &  0.066 & --0.009 &  0.047 \\
 6066 &  17.049 &  0.002 &  17.111 &  0.004 &   0.030 &  0.030 &   0.014 &  0.031 \\
 6115 &  18.657 &  0.002 &  19.274 &  0.004 &   0.025 &  0.034 &   0.062 &  0.036 \\
 6123 &  17.199 &  0.002 &  17.172 &  0.006 &   0.115 &  0.051 &   0.023 &  0.026 \\
 6126 &  18.278 &  0.005 &  18.910 &  0.008 & --0.028 &  0.039 & --0.034 &  0.063 \\
 6132 &  19.489 &  0.004 &  20.061 &  0.012 &   0.000 &  0.062 &   0.008 &  0.042 \\
 6136 &  20.366 &  0.006 &  20.750 &  0.012 &   0.042 &  0.056 & --0.069 &  0.062 \\
 6139 &  18.928 &  0.004 &  19.541 &  0.004 &   0.052 &  0.060 & --0.054 &  0.074 \\
 6203 &  17.161 &  0.003 &  17.181 &  0.005 & --0.004 &  0.030 & --0.036 &  0.031 \\
 6260 &  17.142 &  0.002 &  17.150 &  0.004 &   0.058 &  0.052 &   0.013 &  0.032 \\
 6275 &  19.975 &  0.012 &  20.385 &  0.009 & --0.099 &  0.052 & --0.077 &  0.064 \\
 6311 &  20.123 &  0.010 &  20.524 &  0.013 & --0.054 &  0.056 & --0.070 &  0.064 \\
 6312 &  17.394 &  0.002 &  18.100 &  0.005 & --0.019 &  0.033 &   0.018 &  0.032 \\
 6402 &  16.962 &  0.001 &  17.709 &  0.003 &   0.047 &  0.029 &   0.065 &  0.031 \\
 6406 &  18.199 &  0.002 &  18.835 &  0.005 &   0.032 &  0.053 & --0.053 &  0.075 \\
 6487 &  20.364 &  0.006 &  20.760 &  0.013 &   0.043 &  0.068 & --0.036 &  0.088 \\
 6499 &  19.194 &  0.008 &  19.810 &  0.025 & --0.105 &  0.043 & --0.103 &  0.057 \\
 6516 &  20.144 &  0.005 &  20.528 &  0.008 & --0.034 &  0.052 & --0.062 &  0.067 \\
 6543 &  15.942 &  0.001 &  16.808 &  0.003 &   0.110 &  0.050 &   0.066 &  0.026 \\
 6579 &  20.180 &  0.005 &  20.586 &  0.014 &   0.001 &  0.081 & --0.094 &  0.056 \\
 6612 &  20.135 &  0.005 &  20.519 &  0.008 & --0.020 &  0.053 &   0.028 &  0.029 \\
 6651 &  16.776 &  0.002 &  17.545 &  0.005 &   0.042 &  0.036 &   0.014 &  0.044 \\
 6655 &  15.766 &  0.001 &  16.728 &  0.003 &   0.043 &  0.024 & --0.032 &  0.011 \\
 6665 &  17.505 &  0.002 &  17.484 &  0.005 &   0.017 &  0.031 & --0.018 &  0.032 \\
 6774 &  18.874 &  0.003 &  19.497 &  0.005 &   0.028 &  0.056 & --0.053 &  0.039 \\
 6797 &  16.875 &  0.002 &  17.379 &  0.004 &   0.023 &  0.030 &   0.096 &  0.031 \\
 6900 &  17.842 &  0.002 &  18.514 &  0.005 &   0.060 &  0.052 &   0.026 &  0.034 \\
 6913 &  19.905 &  0.005 &  20.329 &  0.006 & --0.055 &  0.041 &   0.040 &  0.041 \\
 6923 &  20.178 &  0.006 &  20.570 &  0.007 &   0.005 &  0.047 & --0.020 &  0.058 \\
 6936 &  19.718 &  0.006 &  20.227 &  0.011 &   0.046 &  0.040 &   0.043 &  0.043 \\
 6956 &  17.168 &  0.003 &  17.901 &  0.007 &   0.074 &  0.051 &   0.005 &  0.029 \\
 7061 &  17.840 &  0.002 &  18.502 &  0.008 &   0.024 &  0.052 & --0.007 &  0.030 \\
 7107 &  16.967 &  0.002 &  17.109 &  0.003 &   0.077 &  0.051 &   0.071 &  0.026 \\
 7194 &  15.284 &  0.001 &  16.238 &  0.002 &   0.033 &  0.027 & --0.020 &  0.010 \\
 7345 &  19.879 &  0.005 &  20.336 &  0.010 &   0.029 &  0.043 &   0.083 &  0.049 \\
 7399 &  18.192 &  0.002 &  18.888 &  0.005 & --0.060 &  0.088 & --0.033 &  0.041 \\
 7489 &  18.473 &  0.003 &  19.103 &  0.004 & --0.060 &  0.058 & --0.014 &  0.021 \\
 7553 &  17.547 &  0.005 &  18.220 &  0.011 & --0.100 &  0.044 & --0.081 &  0.047 \\
 7558 &  15.675 &  0.002 &  16.560 &  0.003 &   0.001 &  0.036 &   0.045 &  0.037 \\
 7563 &  16.904 &  0.004 &  17.647 &  0.009 & --0.029 &  0.039 & --0.021 &  0.045 \\
 7634 &  16.673 &  0.003 &  17.243 &  0.004 &   0.012 &  0.030 &   0.103 &  0.033 \\
 7669 &  18.558 &  0.003 &  19.179 &  0.004 & --0.077 &  0.073 & --0.002 &  0.028 \\
 7672 &  16.919 &  0.002 &  16.993 &  0.005 &   0.041 &  0.025 &   0.042 &  0.013 \\
 7693 &  19.233 &  0.004 &  19.822 &  0.005 &   0.004 &  0.051 &   0.012 &  0.038 \\
 7699 &  20.382 &  0.007 &  20.787 &  0.009 &   0.006 &  0.066 &   0.024 &  0.062 \\
 7723 &  17.006 &  0.004 &  17.030 &  0.008 &   0.063 &  0.024 &   0.042 &  0.044 \\
 7734 &  15.711 &  0.002 &  16.565 &  0.004 & --0.009 &  0.038 & --0.054 &  0.044 \\
 7742 &  19.711 &  0.004 &  20.219 &  0.007 & --0.029 &  0.056 & --0.067 &  0.045 \\
 7807 &  18.960 &  0.005 &  19.606 &  0.007 & --0.079 &  0.077 & --0.022 &  0.052 \\
 7830 &  19.493 &  0.003 &  20.054 &  0.006 &   0.006 &  0.045 & --0.012 &  0.040 \\
 7861 &  18.702 &  0.005 &  19.322 &  0.007 & --0.008 &  0.031 &   0.019 &  0.048 \\
 7887 &  17.234 &  0.002 &  17.237 &  0.005 &   0.037 &  0.025 &   0.002 &  0.013 \\
 7917 &  16.538 &  0.003 &  17.252 &  0.005 & --0.060 &  0.046 & --0.049 &  0.043 \\
 7941 &  17.451 &  0.004 &  17.457 &  0.008 & --0.057 &  0.053 & --0.049 &  0.046 \\
 7982 &  18.130 &  0.004 &  18.774 &  0.011 & --0.003 &  0.028 &   0.013 &  0.023 \\
 8009 &  19.053 &  0.004 &  19.625 &  0.009 & --0.062 &  0.038 & --0.021 &  0.031 \\
 8195 &  19.823 &  0.006 &  20.299 &  0.008 & --0.028 &  0.053 & --0.025 &  0.068 \\
 8199 &  18.506 &  0.005 &  19.132 &  0.009 & --0.015 &  0.075 &   0.001 &  0.038 \\
 8223 &  18.667 &  0.003 &  19.291 &  0.007 &   0.006 &  0.031 &   0.066 &  0.031 \\
 8277 &  20.225 &  0.006 &  20.646 &  0.010 & --0.033 &  0.044 & --0.060 &  0.055 \\
 8328 &  16.482 &  0.002 &  17.280 &  0.003 &   0.001 &  0.034 &   0.065 &  0.016 \\
 8347 &  19.891 &  0.006 &  20.339 &  0.018 & --0.059 &  0.140 & --0.011 &  0.049 \\
 8360 &  16.968 &  0.003 &  17.092 &  0.008 & --0.021 &  0.037 &   0.015 &  0.047 \\
 8625 &  17.231 &  0.003 &  17.933 &  0.009 &   0.009 &  0.046 &   0.016 &  0.047 \\
 8654 &  17.883 &  0.005 &  18.711 &  0.013 & --0.019 &  0.040 & --0.112 &  0.048 \\
 8687 &  17.343 &  0.003 &  17.350 &  0.008 &   0.038 &  0.033 &   0.036 &  0.034 \\
 8741 &  16.982 &  0.003 &  17.154 &  0.005 & --0.021 &  0.056 &   0.011 &  0.045 \\
 8748 &  18.266 &  0.006 &  18.913 &  0.014 & --0.098 &  0.047 & --0.100 &  0.044 \\
 8766 &  18.584 &  0.003 &  19.203 &  0.008 & --0.015 &  0.038 & --0.057 &  0.055 \\
 8773 &  17.885 &  0.005 &  18.550 &  0.014 & --0.011 &  0.048 & --0.019 &  0.051 \\
 8870 &  19.929 &  0.004 &  20.361 &  0.006 &   0.004 &  0.051 &   0.026 &  0.053 \\
 8873 &  19.785 &  0.008 &  20.268 &  0.008 & --0.020 &  0.075 & --0.022 &  0.096 \\
 8888 &  19.566 &  0.006 &  20.127 &  0.011 &   0.055 &  0.045 &   0.037 &  0.042 \\
 8999 &  16.961 &  0.003 &  17.146 &  0.007 &   0.063 &  0.051 &   0.101 &  0.046 \\
 9128 &  17.477 &  0.003 &  17.440 &  0.007 &   0.003 &  0.046 & --0.027 &  0.025 \\
 9321 &  20.109 &  0.013 &  20.508 &  0.022 & --0.061 &  0.056 & --0.121 &  0.053 \\
 9485 &  19.848 &  0.004 &  20.299 &  0.007 & --0.004 &  0.051 & --0.020 &  0.053 \\
 9647 &  19.996 &  0.006 &  20.412 &  0.007 & --0.023 &  0.061 & --0.040 &  0.041 \\
\enddata
\end{deluxetable*}


\begin{thebibliography}{}

\bibitem[]{} Alcaino,~G. 1975, \aaps, 22, 193 

\bibitem[]{} Alcock,~C., \etal\ \ 1996, \aj, 111, 1146

\bibitem[]{} Arellano~Ferro,~A., Ar\'evalo, ~M.~J., L\'azaro,~C.,
Rey,~M., Bramich,~D.~M., \& Giridhar,~S. 2004, RMxAA, 40, 209 (AF04) 

\bibitem[]{} Auri\`ere,~M., \& Lauzeral,~C. 1991, \aap, 244, 303 

\bibitem[Baade 1930]{baa30} Baade,~W. 1930, AN, 239, 353 

\bibitem[]{} Bedin,~L.~R., \etal\ 2000, \aap, 363, 159

\bibitem[]{} Bell,~R.~A., \& Gustafsson,~B. 1983, \mnras, 204, 249 

\bibitem[]{} Bellazzini,~M., Ferraro,~F.~R., \& Ibata,~R. 2003a, \aj, 125, 188

\bibitem[]{} Bellazzini,~M., Ibata,~R., Ferraro,~F.~R., \& Testa,~V. 2003b, \aap, 405, 577 

\bibitem[]{} Bolte,~M., Hesser,~J.~E., \& Stetson,~P.~B. 1993, \apjl, 408, L89

\bibitem[]{} Buonanno, R. 1993, in ASP Conf. Ser. 48, The Globular Cluster-Galaxy
        Connection, eds.\ G.~H.~Smith \& J.~P.~Brodie (San Francisco: ASP), 131

\bibitem[]{}  Buonanno, R., Corsi, C. E., Bellazzini, M., Ferraro, F. R., \&
Fusi~Pecci, F. 1997, \aj, 113, 706 

\bibitem[]{} Buonanno,~R., Corsi,~C.~E., \& Fusi~Pecci,~F. 1989, \aap, 216, 80 

\bibitem[]{} Cacciari,~C., Bellazzini,~M., \& Colucci,~S. 2002, in IAU Symp. 207, 
  Extragalactic Star Clusters, ed.\ D.~Geisler, E.~K.~Grebel, \& D.~Minniti
  (San Francisco: ASP), 168 

\bibitem[]{} Cacciari,~C., Corwin,~T.~M., \& Carney,~B.~W. 2005, \aj, 129, 267 

\bibitem[]{} Carney,~B., Fullton,~L., \& Trammell,~S. 1991, \aj, 101, 1699 

\bibitem[]{} Carney,~B., Latham,~D.~W., \& Laird,~J.~B. 1990, \aj, 99, 572 

\bibitem[]{} Carretta,~E., \& Gratton,~R.~G. 1997, \aaps, 121, 95 (CG97) 

\bibitem[]{} Cassisi,~S., \& Salaris,~M. 1997, \mnras, 285, 593 

\bibitem[]{} Castellani,~M., Castellani,~V., \& Cassisi,~S. 2005, \aap, in press
(astro-ph/0504470) 

\bibitem[]{} Castellani,~V., \& Quarta,~M.~L. 1987, \aaps, 71, 1

\bibitem[]{} Catelan, M. 1992, \aap, 261, 443

\bibitem[]{} Catelan, M. 2004a, \apj, 600, 419

\bibitem[]{} Catelan, M. 2004b, in ASP Conf. Ser. 310, Variable Stars in the
Local Group, ed.\ D.~W.~Kurtz \& K.~Pollard (San Francisco: ASP), 113

\bibitem[]{} Catelan,~M. 2005, in ASP Conf. Ser., Resolved Stellar Populations,
ed.\ D.~Valls-Gabaud \& M.~Ch\'avez (San Francisco: ASP), in press
(astro-ph/0507464) 

\bibitem[]{} Catelan,~M., Bellazzini,~M., Landsman,~W.~B., Ferraro,~F.~R., 
Fusi~Pecci,~F., \& Galleti,~S. 2001, \aj, 122, 3171

\bibitem[]{} Catelan, M., Pritzl, B. J., \& Smith, H. A. 2004, \apjs, 154, 633

\bibitem[]{} Chaboyer,~B., Demarque,~P., \& Sarajedini,~A. 1996, \apj, 459,
558 

\bibitem[Christian \etal\ 1985]{chr85} Christian,~C.~A., \etal\ 1985, \pasp, 97,
363  

\bibitem[Clement 2000]{cle00} Clement,~C. 2000, in ASP Conf.\ Ser.\ vol.\ 220,
Amateur--Professional Partnerships in Astronomy, ed.\ J.~R.~Percy \&
J.~B.~Wilson, (San Francisco: ASP), 93 

\bibitem[]{} Clement,~C.~M., \& Rowe,~J. \ 2000, \aj, 120, 2579 

\bibitem[]{} Clement,~C.~M., \& Shelton,~I. 1999, \apjl, 515, L85

\bibitem[]{} Clement,~C.~M., \etal\ \ 2001, \aj, 122, 2587 

\bibitem[]{} Contreras,~R., Catelan,~M., Smith,~H.~A., Pritzl,~B.~J., \&
Borissova,~J. 2005, \apjl, 623, L117

\bibitem[]{} Corwin,~T.~M., Catelan,~M., Smith,~H.~A., Borissova,~J.,
Ferraro,~F.~R., \& Raburn,~W.~S. 2003, \aj, 125, 2543 

\bibitem[]{} Da~Costa,~G.~S., \& Armandroff,~T.~E. 1995, \aj, 109, 2533

\bibitem[Davis 1917]{dav17} Davis,~H. 1917, \pasp, 29, 260 

\bibitem[]{} Dinescu,~D.~I., Majewski,~S.~R., Girard,~T.~M., \& Cudworth,~K.~M. 2000, \aj, 
  120, 1892

\bibitem[]{} Djorgovski,~S., \& King,~I.~R. 1986, \apjl, 305, L61 

\bibitem[]{} Ferraro,~F.~R., \etal\ 1999, \aj, 118, 1738 

\bibitem[]{} Gratton,~R.~G., \& Ortolani,~S. 1988, \aaps, 73, 137

\bibitem[]{} Harris,~W.~E. 1996, \aj, 112, 1487 

\bibitem[]{} Holtzman,~J.~A., \etal\ 1995, \pasp,  107,  1065 

\bibitem[]{} Jurcsik,~J. 1995, AcA, 45, 653

\bibitem[]{} Jurcsik,~J., \& Kov\'acs,~G. 1996, \aap, 312, 111 

\bibitem[]{} Kov\'acs, G. 1998, in ASP Conf.\ Ser.\ 135, A Half Century of
Stellar Pulsation Interpretations: A Tribute to Arthur~N.~Cox, ed.\
P.~A.~Bradley \& J.~A.~Guzik (San Francisco, ASP), 52

\bibitem[Landolt 1992]{lan92} Landolt,~A.~U. 1993, \aj, 104, 340 

\bibitem[]{} Lee,~S.-W. 1977, \aaps, 29, 1 

\bibitem[]{} Lee,~Y.-W., Demarque,~P., \& Zinn,~R. 1990, \apj, 350, 155  

\bibitem[]{} Mackey,~A.~D., \& van~den~Bergh,~S. 2005, \mnras, in press
(astro-ph/0504142)

\bibitem[]{} Mandushev,~G.~I., Fahlman,~G.~G., Richer,~H.~B., \& Thompson,~I.~B.
1996, \aj, 112, 1536 

\bibitem[]{} Mannino,~G. 1957, MmSAI, 28, 285 

\bibitem[Monet+ 1998]{mon98} Monet,~D.~G., \etal\ 1998, USNO-A2.0 (Flagstaff: US
Naval Obs.), CD-ROM 

\bibitem[]{} Nemec,~J.~M. 2004, \aj, 127, 2185 

\bibitem[]{} Newburn,~R.~L. 1957, \aj, 62, 197 

\bibitem[]{} Oosterhoff,~P.~Th. 1939, Observatory, 62, 104 

\bibitem[]{} Peterson,~C.~J. 1986, \pasp, 98, 1258

\bibitem[]{} Piotto,~G., \etal\ 2002, \aap, 391, 945

\bibitem[]{} Piotto, G., \etal\ 2004, \apjl, 604, L109

\bibitem[]{} Pritzl,~B.~J., Smith,~H.~A., Catelan,~M., \& Sweigart,~A.~V. 2002,
\aj, 124, 949

\bibitem[]{} Pryor,~C., \& Meylan,~G. 1993, in Structure
and Dynamics of Globular Clusters, eds.~S.~G.~Djorgovski and G.~Meylan,
ASP Conf.\ Ser. (San Francisco: ASP), 50, p. 357 

\bibitem[]{} Rathbun,~P., \& Smith,~H.  1997, \pasp, 109, 1128

\bibitem[]{} Rood,~R.~T., \& Crocker,~D.~A. 1989, in IAU Colloq.\ 111, The Use
of Pulsating Stars in Fundamental Problems of Astronomy, ed.\ E.~G.~Schmidt
(Cambridge: Cambridge University Press), 218

\bibitem[]{} Rosenberg,~A., Saviane, I., Piotto,~G., \& Aparicio,~A. 1999,
\aj, 118, 2306 

\bibitem[]{} Sabbi, E., Ferraro, F. R., Sills, A., \& Rood, R. T. 2004, \apj,
617, 1296

\bibitem[]{} Salaris, M., Chieffi, A., \& Straniero, O. 1993, \apj, 414, 580

\bibitem[]{} Salinas,~R., Catelan,~M., Smith,~H.~A., Pritzl,~B.~J., \&
Borissova,~J. 2005, in preparation 

\bibitem[]{} Sandage,~A.~R., \& Walker,~M.~F. 1955, \aj, 60, 230 

\bibitem[]{} Schlegel, D.~J., Finkbeiner, D.~P., \& Davis, M.~1998, ApJ, 500,
525 

\bibitem[]{} Simon,~N.~R., \& Clement,~C.~M, 1993, \apj, 410, 526 

\bibitem[Stetson 1979]{ste79} Stetson,~P.~B. 1979, \aj, 84, 1056 

\bibitem[Stetson 1987]{ste87} Stetson,~P.~B. 1987, \pasp, 99, 191 

\bibitem[Stetson 1990]{ste90}  Stetson,~P.~B. 1990, \pasp, 102, 932 

\bibitem[Stetson 1993]{ste93} Stetson,~P.~B. 1993, in IAU Coll.\ 136, Stellar
Photometry --- Current Techniques and Future Developments, ed.\ C.~J.~Butler \&
I.~Elliot (Cambridge: Cambridge Univ.\ Press), 291 

\bibitem[Stetson 1994]{ste94} Stetson,~P.~B. 1994, \pasp, 106, 250 

\bibitem[Stetson 1998]{ste98} Stetson,~P.~B. 1998, \pasp, 110, 1448 

\bibitem[Stetson 2000]{ste2000} Stetson,~P.~B. 2000, \pasp, 112, 925 

\bibitem[]{} Stetson,~P.~B. 2005, accepted for publication in \pasp 

\bibitem[Stetson \etal\ 1998]{pbs98} Stetson,~P.~B. \etal\ 1998, \apj, 508, 491 

\bibitem[Stetson \etal\ 1999]{pbs99} Stetson,~P.~B. \etal\ 1999, \aj, 117, 247 

\bibitem[Stetson \etal\ 2003]{ste03} Stetson,~P.B., Bruntt,~H., \& Grundahl,~F. 2003, \pasp, 115, 413 

\bibitem[]{} Stetson,~P.~B., McClure,~R.~D., \& VandenBerg, ~D.~A. 2004, \pasp,
116, 1012 

\bibitem[]{} Stetson,~P.~B., VandenBerg,~D.~A., Bolte,~M., Hesser,~J.E., \&
Smith,~G.~H. 1989, \aj, 97, 1360

\bibitem[]{} Suntzeff,~N.~B., Kraft,~R.~P., \& Kinman,~T.~D. 1988, \aj, 95, 91 

\bibitem[]{} Trager,~S.~C., Djorgovski,~S., \& King,~I.~R. 1993, in Structure
and Dynamics of Globular Clusters, eds.~S.~G.~Djorgovski and G.~Meylan,
ASP Conf.\ Ser. (San Francisco: ASP), 50, p. 347 

\bibitem[]{} VandenBerg,~D.~A. 2000, \apjs, 129, 315 

\bibitem[van den Bergh \& Mackey 2004]{vdb04} van~den~Bergh,~S., \& Mackey,~A.~D.
2004, \mnras, 354, 713 

\bibitem[]{} Wang,~J.~J., Chen,~L., Wu,~Z.~Y., Gupta,~A.~C., \& Geffert,~M. 2000, \aaps, 
  142, 373

\bibitem[]{} Zinn,~R., \& West,~M.~J. 1984 \apjs, 55, 45 

\end{thebibliography}
\end{document}